\documentclass[showpacs,preprintnumbers,amsmath,amssymb,APSl,prd,nofootinbib,superscriptaddress, twocolumn]{revtex4-2}

\usepackage{dcolumn}
\usepackage{bm}
\usepackage{ifpdf}
\usepackage{hyperref}
\usepackage{bm}
\usepackage{xcolor,color,graphicx,graphics}
\usepackage[spanish,english]{babel}
\usepackage[latin1]{inputenc}
\usepackage[OT1]{fontenc}
\usepackage{latexsym,amssymb,amsmath,amsfonts}
\usepackage{makeidx}
\usepackage{epsfig,subfigure}
\usepackage{natbib}
\usepackage{epstopdf}
\usepackage{mathrsfs}
\hypersetup{colorlinks=true, linkcolor=blue, citecolor=green}
\usepackage{enumerate}
\usepackage{xcolor}
 \usepackage{multirow}

\definecolor{red}{rgb}{1,0,0}

\def\+{^\dagger}

\def\<{\leftarrow}
\def\>{\rightarrow}

\def\({\left(}
\def\){\right)}

\def\sc{\mathop{\rm sc}\nolimits}
\def\arcsinh{\mathop{\rm arcsinh}\nolimits}

\newcommand{\bi}{\begin{itemize}} 				\newcommand{\ei}{\end{itemize}}
\newcommand{\benu}{\begin{enumerate}} 		\newcommand{\enu}{\end{enumerate}}
\newcommand{\bd}{\begin{dinglist}{0}}     \newcommand{\ed}{\end{dinglist}}
\newcommand{\bfig}{\begin{figure}[htbp]}  \newcommand{\efig}{\end{figure}}
        			
\newcommand{\bc}{\begin{center}} 				  \newcommand{\ec}{\end{center}}
\newcommand{\be}{\begin{equation}} 				\newcommand{\ee}{\end{equation}}
\newcommand{\bsub}{\begin{subequations}}  \newcommand{\esub}{\end{subequations}}
\newcommand{\ben}{\begin{eqnarray}} 			\newcommand{\een}{\end{eqnarray}}
\newcommand{\ba}[1]{\begin{array}{#1}} 		\newcommand{\ea}{\end{array}}
\newcommand{\bea}{\begin{equation}\begin{array}{rcl}}
\newcommand{\eea}{\end{array}\end{equation}}

\renewcommand*{\thefootnote}{\fnsymbol{footnote}}

\begin{document}
\title{Shadows and photon rings of regular black holes and geonic horizonless compact objects}

	\author{Gonzalo J. Olmo} \email{gonzalo.olmo@uv.es}
\affiliation{Departamento de F\'{i}sica Te\'{o}rica and IFIC, Centro Mixto Universidad de Valencia - CSIC.
Universidad de Valencia, Burjassot-46100, Valencia, Spain}
\affiliation{Universidade Federal do Cear\'a (UFC), Departamento de F\'isica, Campus do Pici, Fortaleza - CE, C.P. 6030, 60455-760 - Brazil}
\author{Jo\~ao Lu\'is Rosa} \email{joaoluis92@gmail.com}
\affiliation{Institute of Physics, University of Tartu, W. Ostwaldi 1, 50411 Tartu, Estonia}
\affiliation{University of Gda\'{n}sk, Jana Ba\.{z}y\'{n}skiego 8, 80-309 Gda\'{n}sk, Poland}
\author{Diego Rubiera-Garcia}    \email{drubiera@ucm.es} \thanks{Corresponding author}
\affiliation{Departamento de F\'isica Te\'orica and IPARCOS,
	Universidad Complutense de Madrid, E-28040 Madrid, Spain}
	\author{Diego S\'aez-Chill\'on G\'omez} \email{diego.saez@uva.es}
\affiliation{Department of Theoretical Physics, Atomic and Optics, and IMUVA, \\ 
University of Valladolid, Paseo Bel\'en, 7, 47011 Valladolid, Spain}

\date{\today}
\begin{abstract}
The optical appearance of a body compact enough to feature an unstable bound orbit, when surrounded by an accretion disk, is expected to be dominated by a luminous ring of radiation enclosing a central brightness depression typically known as the shadow. Despite observational limitations, the rough details of this picture have been now confirmed by the results of  the EHT Collaboration on the imaging of the M87 and Milky Way supermassive central objects.  However, the precise characterization of both features - ring and shadow - depends on the interaction between the background geometry and the accretion disk, thus being a fertile playground to test our theories on the nature of compact objects and the gravitational field itself in the strong-field regime. In this work we use both features in order to test a continuous family of solutions interpolating between regular black holes and horizonless compact objects, which arise within the Eddington-inspired Born-Infeld theory of gravity, a viable extension of Einstein's General Relativity (GR). To this end we consider seven distinctive classes of such configurations (five black holes and two traversable wormholes) and study their optical appearances under illumination by a geometrically and optically thin accretion disk, emitting monochromatically with three analytic intensity profiles previously suggested in the literature. We build such images and consider the sub-ring structure created by light rays crossing the disk more than once and existing on top of the main ring of radiation. We discuss in detail the modifications as compared to their GR counterparts, the Lyapunov exponents of unstable nearly-bound orbits, as well as the differences between black hole and traversable wormholes for the three intensity profiles. In addition we use the claim by the EHT Collaboration on the radius of the bright ring acting (under proper calibrations) as a proxy for the radius of the shadow itself to explore the parameter space of our solutions compatible with such a result.
\end{abstract}

\maketitle

\newpage

\tableofcontents

\section{Introduction}

\subsection{Multimessenger quantum gravity era}

In the last few years, we have witnessed the beginning of a new era in gravitational physics triggered by the leap in the quantity, quality, and variety of observational data from different probes. This newly born discipline relies on several messengers (photons, neutrinos, cosmic rays and gravitational waves) from different sources, consequently being dubbed as multimessenger astronomy \cite{Addazi:2021xuf}. Equipped with its power, a new window is open to access to energy scales beyond those attainable by particle accelerators, allowing to probe deeper into phenomenological clues of the long-sought quantum theory of gravity. On a more conservative basis, it provides an opportunity to test the nature and properties of compact objects (neutron stars and black holes alike  \cite{Cardoso:2019rvt}) as well as to verify the consistency of the cosmological concordance model \cite{Bull:2015stt}.

Traditionally, the community has addressed the quantum gravity problem via either fundamental approaches (mostly string theory and loop quantum gravity) or via effective implementations known as modified theories of gravity \cite{DeFelice:2010aj,Nojiri:2017ncd}. Within the latter one also finds model-agnostic approaches, in which one either builds ad hoc specific solutions with the desired properties, or considers parameterized deviations from General Relativity (GR) solutions \cite{Johannsen:2011dh,Konoplya:2016jvv} hoping that both post-Newtonian and strong-field limit observations will allow to sufficiently constrain its coefficients. This procedure aims to extract useful lessons on the viable candidates to supersede GR expectations on its strong-field regime \cite{EventHorizonTelescope:2020qrl}. Either way, a pressuring topic in the community is how to bring putative Planck-scale effects associated to quantum gravity into scales that are directly testable by multimessenger astronomy, e.g. horizon or photon sphere scales (see \cite{Eichhorn:2022oma,Eichhorn:2022bbn} for some ideas at this regard). On the theoretical front, every such a theory should have something to say about the issue of space-time singularities  \cite{Senovilla:2014gza}, namely, the unavoidable loss of causal determinism and predictability of GR caused by the geodesic incompleteness within some of its most physically relevant solutions: black holes and early cosmological evolution.

\subsection{The metric-affine proposal}

Among the pool of frameworks and proposals to extend GR, this work is anchored in the hypothesis that allowing for a larger flexibility in the basic blocks of our gravitational theories, understood as a manifestation of a space-time imbued with geometrical properties, may contain some clues on how to deal with the issue of space-time singularities in (semi-)classical gravity. Such blocks are the metric - defining the notion of local distances, areas and volumes - and the affine connection - defining parallelism and thus appearing in the definition of covariant derivatives -. Disentangling them as {\it a priori}  independent entities opens up the world of the {\it metric-affine} foundation of gravity, in which an affine connection is split into its curvature, torsion, and non-metricity pieces (see \cite{BeltranJimenez:2019esp} for how to build gravitational theories based on each piece), while the field equations of the theory are obtained by independent variations of the action with respect to metric and connection. This needs to be accompanied by a recipe on how to connect the (curvature/energy/size) scales at which this conceptual modification may become relevant, with the GR-world at the scales which have been tested so far.

It turns out that the physics of solid state systems harbours useful lessons on how a continuum, macroscopic, regular geometrical-type world may emerge from a discrete, microscopic world filled with irregularities or {\it defects}.  The latter are any type of imperfection rupturing the array distribution of individual atoms in real crystals, which arise either as point-like (interstices/vacancies), one (dislocations) and two-dimensional (textures), and that unavoidably exist within any real material.  The presence of any such defects breaks the possibility of a pure Riemannian description in the transition to its continuum geometrical limit, requiring a larger flexibility of the encompassing geometry: non-metricity is associated to the point-like defects while torsion accounts for dislocations (check the book \cite{KittelBook} for a broad discussion of this connection). Moreover, defects have {\it dynamics}, i.e., they are able to move and interact, even to recombine with other defects to form larger ones or even to annihilate each other. This way, the metric-affine approach makes a natural implementation of an underlying {\it quantum foam} world yielding an emergent continuum space-time geometry embedded (in general) with curvature, non-metricity and torsion \cite{Lobo:2014nwa}. When translated into the gravitational context, it provides a natural connection between the above fundamental geometrical objects and the implementation of symmetry principles under unified gauge-type frameworks \cite{Cabral:2020fax}. Furthermore, it turns out that the metric-affine picture may resolve the problem of space-time singularities within black hole scenarios under different choices for the gravitational and matter sectors and without any violation of the energy conditions \cite{Olmo:2015dba,Olmo:2015axa,Bejarano:2017fgz,Menchon:2017qed}. It is thus natural to explore the phenomenological consequences of these theoretically well-grounded models.

\subsection{Shadow and photon ring observations}

The main aim of this work is to start a programme based on the systematic analysis of shadows and photon rings in ultra-compact objects within the metric-affine models mentioned above and their singularity-free solutions, and their capability to compete in the interpretation of real images with the canonical GR-based compact objects. In this sense, the observation in 2019, by the Event Horizon Telescope (EHT) \cite{EventHorizonTelescope:2019dse}, of the optical image of the central supermassive object at the center of the M87 galaxy, followed in 2022 by a similar observation at the center of our own Milky Way Galaxy (Sagittarius A$^*$  \cite{EventHorizonTelescope:2022wkp}), is allowing the field of ray-tracing and the analysis of photon trajectories in the vicinity of ultra-compact objects to blossom to the best of its potential \cite{Psaltis:2018xkc}. The EHT observations are founded in the well-tested idea that light trajectories near a massive body, which follow null geodesics in the corresponding space-time geometry, experience large deflections when approaching the critical points of its effective potential \cite{Bozza:2002zj}. The trajectories for which this deflection turns formally infinite can thus be dubbed (though this terminology is far from being universal)  as the {\it critical curve(s)}, the photon sphere being the corresponding spherically symmetric limit. This allows light rays to revolve several times around the central object before being released to asymptotic infinity to generate, in the observer's screen, a thin photon ring, consisting of strongly lensed radiation appearing on top of the direct emission generated by the disk itself.

Even though one would intuitively expect such a critical curve and its associated thin photon ring to define the outer edge of the central brightness depression - thus in this view the shadow fills entirely the critical curve \cite{Luminet:1979nyg,Falcke:1999pj} -, it turns out that the location of such an edge also depends on the interaction between the geometry of the black hole and the physics of the accretion disk providing the main source of illumination and, therefore, it is also critical to define the optical appearance of the object \cite{Lara:2021zth}. Two main features of the disk contribute to this: its optical thickness (whether it is transparent to its own radiation or not), and its geometrical shape: infinitesimally-thin, spherical, or any other intermediate thickness. The results of \cite{Vincent:2022fwj} consistently argue that, no matter how thick the disk may be as long as it is not completely spherical, the apparent size of the shadow can be strongly reduced from the size of the critical curve down to a minimum inner shadow limit defined by the geometry of the object (basically given by the apparent location of the event horizon \cite{Chael:2021rjo}). Furthermore, the photon ring itself in this situation is decomposed into an infinite number of self-similar rings with exponentially decreasing contributions to the total luminosity and approaching the inner shadow limit \cite{Gralla:2020srx}.  Finally, when the background geometry is generalized to include rotation - i.e. the Kerr solution - the critical curve degenerates into a photon shell of unstable bound geodesics \cite{Johnson:2019ljv} but the rough details of this picture hold (though significantly more involved in their description).

Such photon rings are a trademark of a given geometry, thus potentially harbouring a way to make robust tests of the Kerr hypothesis \cite{Gralla:2019drh,Wielgus:2021peu,Glampedakis:2021oie,Ayzenberg:2022twz}. While EHT observations may be successfully reproduced with a Kerr black hole supplied within General Relativistic Magnetic-Hydrodynamical (GRMHD) simulations of the accretion flow \cite{Vincent:2020dij}, many works in the literature have sought for modifications to GR canonical black holes via addition of new fields \cite{He:2021htq,Uniyal:2022vdu,Berry:2020ntz,Okyay:2021nnh,Wen:2022hkv,Guo:2021bhr,Zeng:2022pvb} and hairy black holes \cite{Gan:2021xdl,Gan:2021pwu}, horizonless compact objects such as  naked singularities \cite{Shaikh:2019hbm,Joshi:2020tlq}, black bounces \cite{Guerrero:2021ues,Guo:2021wid,Ghosh:2022mka}, boson stars \cite{Olivares:2018abq,Herdeiro:2021lwl,Rosa:2022tfv,Rosa:2022toh,Rosa:2023qcv}, thin-shell \cite{Guo:2022iiy} rotating \cite{Paul:2019trt} and asymmetric wormholes \cite{Peng:2021osd,Wang:2020emr} and other objects \cite{Eichhorn:2022oma,Rosa:2023hfm}. Images of modified black holes beyond GR have been also considered:  Gauss-Bonnet \cite{Zeng:2020dco}, asymptotic safety \cite{Held:2019xde}, noncommutative geometry  \cite{Zeng:2021dlj}, Einstein-\AE{}ther \cite{Liu:2021yev}, Horndeski theory  \cite{Afrin:2021wlj,Kumar:2021cyl}, quadratic gravity  \cite{Daas:2022iid}, or braneworlds \cite{Hou:2021okc}, to mention a few.

\subsection{This work: aims and goals}

The main goal of the present paper is to start the systematic analysis of the observational appearances of modified black holes and horizonless compact objects supported by metric-affine theories of gravity. To this end, in this work we shall consider a particularly gifted and well-behaved member of such theories, as given by the Eddington-inspired Born-Infeld proposal \cite{Banados:2010ix}, coupled to a standard electric (Maxwell) field. Depending on the interplay between the mass, electric charge, and the theory's parameter, the model holds both modified black holes as compared to their GR (Reissner-Nordstr\"om) counterparts) and several types of horizonless compact objects, thus allowing to discuss the corresponding optical appearances of different types of objects on an equal-footing. To this end, we shall consider the illumination of such objects by an optically and geometrically thin accretion disk, using three illumination profiles previously employed in the literature. Our main observables of interest are the shadow's size, and the features of the photon rings (number, relative luminosity, and location).

This work is organized as follows: in Sec. \ref{sec:II} we set the background geometries we shall work with throughout the paper, and discuss the different classes of configurations (modified black holes and horizonless objects). In Sec. \ref{sec:III} we discuss the main observables of interest, namely, the photon rings and the shadow's size, and motivate the particular values of mass, electric charge, and theory's parameter we shall take in order to capture every significant family of configurations (up to seven). With these choices in hand, we proceed in Sec. \ref{sec:IV} to build the optical appearances of such configurations under the hypotheses above for the accretion flow, and discuss the main qualitative and quantitative (using for the latter the Lyapunov exponents of nearly-bound orbits) differences between each class. We conclude in \ref{sec:V} with a recollection of the most succulent aspects of our results as well as with a critical discussion of the limitations and future upgrades of our approach and the place this work occupies in our long-term strategy for these metric-affine gravities.


\section{Background geometries} \label{sec:II}

\subsection{Geometry features}

We are interested in static, spherically symmetric geometries described in the set of spherical coordinates $\left(t,x,\theta,\phi\right)$ by a line element of the form
\begin{equation} \label{eq:SSS}
ds^2=-A(x)dt^2+B(x)dx^2+r^2(x)d\Omega^2,
\end{equation}
where $A\left(x\right)$ and $B\left(x\right)$ are well-behaved functions of $x$, $d\Omega^2=d\theta^2 +\sin^2 \theta d\phi^2$ is the surface element on the 2-sphere, and the notation $r^2(x)$ allows for a non-monotonic behavior of the radial function. Note that one can always perform a coordinate transformation to reabsorb the function $B(x)$ into the radial coordinate via $B(x)dx^2=dy^2$, resulting in a space-time with only two independent metric functions, as demanded by the spherical symmetry of the system. However, given the theoretical properties of the configurations considered in this work and the advantages brought by working with the explicit form of $r^2(x)$ in some specific models, we shall preserve the former description for the moment. Indeed, we are interested in electrically charged spherically symmetric space-times corresponding to solutions of a certain metric-affine theory of gravity dubbed as Eddington-inspired Born-Infeld gravity (EiBI) (the best account of this theory  can be found in the review paper \cite{BeltranJimenez:2017doy}, and we refer the reader there for an enhanced discussion of its properties), with action
\begin{eqnarray} \label{eq:EiBI}
\mathcal{S}_{EiBI}&=&\frac{1}{\kappa^2 \epsilon} \int d^4x\left(\sqrt{\vert g_{\mu\nu}+ \epsilon R_{\mu\nu}(\Gamma) \vert}-\lambda \sqrt{-g}\right) \nonumber \\
&+& \mathcal{S}_m(g_{\mu\nu},\psi_m)
\end{eqnarray}
Here $\kappa^2=8\pi G$ is Newton's constant in suitable units, $\epsilon$ is a (small) parameter with units of length squared, $R_{\mu\nu}(\Gamma)$ is the Ricci tensor of the affine connection $\Gamma \equiv \Gamma_{\mu\nu}^{\lambda}$, $g$ is the determinant of the space-time metric $g_{\mu\nu}$, vertical bars denote a determinant too, $\lambda$ is a constant parameter, $\mathcal S_m$ is the matter action, and $\psi_m$ collectively represents the matter fields. In the limit $\vert R_{\mu\nu} \vert \ll \epsilon^{-1}$ the theory boils down to GR plus a cosmological constant term given by $\Lambda=\frac{\lambda-1}{\kappa^2}$ (from now on we set $\lambda=1$ for asymptotically flat solutions) and the next-to-leading order term corresponds to quadratic gravity corrections. As we are working on the metric-affine approach, after applying the variational principle to the action, the corresponding field equations are obtained as two independent sets for the metric and for the affine connection; nonetheless, the latter can be removed in favour of additional contributions of the matter fields to the dynamics of the EiBI theory. This is a  general feature of a large class of metric-affine gravities (see \cite{Afonso:2018bpv} for an overview) in which deviations from GR dynamics are fuelled by new terms associated to the matter fields relevant only in the strong-field regime, whereas weak-field limit tests are naturally satisfied and we just need to be concerned with the corrections near the matter sources.

When coupled to a standard electromagnetic (Maxwell) field, the action (\ref{eq:EiBI}) admits a line element of the form (\ref{eq:SSS}) with analytical functions characterizing it. The latter were first found in Ref. \cite{Olmo:2013gqa} and later discussed in detail in Ref. \cite{Olmo:2015bya}, and we shall take here the main ingredients needed for our analysis below: the parameters characterizing the line element are the Schwarzschild radius $r_S=2M$, the charge radius $r_Q^2=\frac{\kappa^2 Q^2}{4\pi}$ (in this work a system of geometrized units for which $G=c=1$, which implies that $\kappa^2=8\pi$, and consequently $r_Q^2=2Q^2$, is employed), and a EiBI parameter conveniently rewritten as $\epsilon=-2l^2_{\epsilon}\neq 0$ (so we shall work in the negative branch of $\epsilon$ only) encoding the deviations from GR predictions. In this framework, the corresponding metric functions are given by
\begin{equation} \label{eq:AEiBI}
A(z)=\frac{1}{\Omega_+}\frac{(1-\delta_1 G(z))}{\delta_2 z\Omega_{-}^{1/2}} \quad;\quad  B(z)=\frac{1}{A(z)\Omega_{+}^2},
\end{equation}
with the following definitions and conventions: the objects $\Omega_{\pm}$, such that the inequality $g_{tt} \neq g^{xx}$ holds, are written in a dimensionless form via the introduction of a new variable $z=r/r_c$, where the constant $r_c$ is a normalised length-scale given by $r_{c}^2=l_{\epsilon}^2 r_Q^2$, leading to:
\begin{equation}
\Omega_{\pm}=1\pm\frac{1}{z^4};
\end{equation}
As for the function $G(z)$ characterizing the metric, it can be written in terms of an infinite series expansion as
\begin{equation}
G(z)=-\frac{1}{\delta_c}+\frac{\sqrt{z^4-1}}{2} \left(f_{3/4}(z)+f_{7/4}(z)\right),
\end{equation}
where the $f_{\lambda}(z)={_2}F_1\left[\frac{1}{2},\lambda,\frac{3}{2},1-z^4 \right]$ are hypergeometric functions; the constants $\delta_i$ introduced previously are defined as
\begin{equation}
\delta_1=\frac{1}{2r_S}\sqrt{\frac{r_Q^3}{l_{\epsilon}}} \quad;\quad  \delta_2=\frac{r_c}{r_S};
\end{equation}
and the constant $\delta_c \approx 0.572069$ emerges from the integration of the field equations to guarantee the compatibility of the asymptotic and central expansions of the metric. Finally, the non-monotonic function $r^2(x)$ is obtained via $x^2=r^2\Omega_{-}^{1/2}$, which can be solved exactly and provides, in the dimensionless notation $z \equiv r/r_c$, the following solution:
\begin{equation} \label{eq:zx}
z^2(x)=\frac{x^2+\sqrt{x^4+4}}{2},
\end{equation}
where the coordinate $x$ ranges in the interval $x\in(-\infty,+\infty)$, whereas the radial function $r$ satisfies $r \geq r_c$. The relationship between $r$ and $x$ can also be written in a differential form as (note that an implicit $r_c$ factor is present here)
\begin{equation} \label{eq:dxdr}
\frac{dx}{dz}=\pm \frac{\Omega_+}{\Omega_-^{1/2}}.
\end{equation}

The spherically symmetric geometry described by Eqs. (\ref{eq:AEiBI}) to (\ref{eq:zx}) may seem a bit involved, but is actually nothing else but an extension of the usual Reissner-Nordstr\"om (RN) solution of GR via $l_{\epsilon}$-corrections. Such corrections are important only in the strong-field regime, whereas in the weak-field regime (large distances, $r \gg r_c$), the metric components behave as
\begin{equation}\label{eq:metrlarg}
g_{tt} \approx - g^{rr} \approx -1 +\frac{1}{\delta_2 z} -\frac{\delta_1}{\delta_2 z^2} +\mathcal{O}\left(\frac{1}{z^4}\right),
\end{equation}
which, upon a restoration of the usual radial, mass, and electric charge quantities, corresponds precisely to the RN geometry. Let us recall that RN configurations feature horizons located at radii $r=r_\pm$ given by
\begin{equation}
r_\pm=M \pm\sqrt{M^2-Q^2}.
\end{equation}
In such a case, if $M^2>Q^2$ the geometry has two horizons, the inner horizon at $r=r_-$, and the event horizon at $r=r_+$, both converging into a single (degenerate) horizon for $M^2=Q^2$, while if $M^2<Q^2$ the space-time features a naked singularity. These latter configurations are physically problematic as they display a geodesic incompleteness character (visible to external observers) and a resultant lack of predictability and causal determinism, barred by cosmic censorship conjecture arguments and also by its incompatibility with current observations. 

The most dramatic deviation in the geometry considered here from its GR counterpart is the behaviour of the radial function $r^2(x)$. Indeed, while in the usual RN solutions this function is trivial, for this geometry it is given by Eq.(\ref{eq:zx}). In the positive branch of $l_{\epsilon}^2>0$, the radial function $r(x)$ does not approach zero in the interior region of the geometry; instead, it reaches a spherical surface of minimum area $S=4\pi r_c^2$ at $x=0$. As one approaches this surface from the asymptotically flat region (say) $x>0$, the ingoing sets of geodesics do not converge to a focal point, but instead re-expand to a new asymptotically flat region of the geometry with negative $x<0$ values (note that $r^2(x)$ remains positive in the two pieces of the space-time; in other words, $x \in (-\infty,+\infty)$ but $r^2>r_c^2$, or $z \geq 1$). This mechanism,  canonically interpreted in the literature as a {\it wormhole structure}, allows in the present case for the restoration of the geodesic completeness of the geometry. Note that this structure is present for the whole spectrum of mass and electric charge of the solutions considered, provided that $l_{\epsilon}^2  > 0$ and, perhaps more strikingly, despite the fact that curvature scalars generically diverge at the wormhole throat $r=r_c$ (for more details, see below).

However, for the sake of shadow/photon ring images, the wormhole structure plays a marginal role, particularly in those configurations featuring an event horizon. In such cases, it is more convenient to use Eq.(\ref{eq:dxdr}) to cast the line element in Eq.(\ref{eq:SSS}) as
\begin{equation}
ds^2=-A(r)dt^2+\frac{dr^2}{A(r)\Omega_{-}(r)}+r^2d\Omega^2,
\end{equation}
recalling that the radial coordinate $r$ is still bounded as $r \geq r_c$, where $r=r_c$ corresponds to the wormhole throat. When simulating the image of a shadow, whenever the throat is hidden behind an event horizon, the presence of a throat is irrelevant; however, the theory also encompasses horizonless compact objects and in such cases the simulations extend down to the throat (in this work we do not consider flows of light rays crossing the wormhole throat into our local universe patch).

\subsection{Classes of configurations}

There are several solutions of interest depending on the relative values of mass, electric charge and the parameter $l_{\epsilon}^2$ that defines the theory.  Recall that the zeroes of  $g^{rr}(r_\pm)=0$ provide the radii of the horizons (if any), $r=r_\pm$,  of every configuration; however, finding the explicit expressions for the radii $r_\pm$ is impossible for the metric above. Nonetheless, it is possible to determine analytically the number of horizons of each configuration via a combined analysis of the behaviour at the throat with the asymptotic behaviour given in Eq.(\ref{eq:metrlarg}) by considering the possible number of crossings with the surface  $g^{rr}(r_\pm)=0$.  In this sense, taking a series expansion of the metric components in the region near the wormhole throat, $z=1$, one obtains
\begin{eqnarray}
g_{tt}&  \approx &   \frac{1-\delta_1/\delta_c}{4\delta_2 \sqrt{z-1}} -\frac{1}{2}\left(1-\frac{\delta_1}{\delta_2}\right) + \mathcal{O}(z-1)^{1/2}, \label{eq:gttc} \\
g^{rr}&\approx & -\frac{1-\delta_1/\delta_c}{\delta_2 \sqrt{z-1}}+ 2\left(1-\frac{\delta_1}{\delta_2}\right) + \mathcal{O}(z-1)^{1/2}. \label{eq:grrc}
\end{eqnarray}
This way, we see that the leading-order terms in Eqs.(\ref{eq:gttc}) and  (\ref{eq:grrc}) determine both the nature (time-like or space-like) of the central region of the configurations near the wormhole throat, and the number and type of horizons, as controlled by the ratio $\delta_1/\delta_c$. Consequently, the configurations are split into three different general types:

\begin{itemize}

\item  Regular Schwarzschild-like black holes: When $\delta_1<\delta_c$ the central region is space-like and a single non-degenerate event horizon is present. Thus, the wormhole lurking in the innermost region of the geometry behind the event horizon is not visible to external observers.

\item Regular Reissner-Nordstr\"om-like solutions: When $\delta_1>\delta_c$ the central region is time-like and the global structure of the solution resembles the RN black hole geometry, splitting into three different possibilities:

\begin{itemize}

\item Regular two-horizon Reissner-Nordstr\"om-like solutions: Two non-degenerate (inner and event horizon, respectively) appear on each side of the geometry depending on a fairly complex interplay between $r_S$, $r_Q$ and $l_{\epsilon}$.

\item Regular extreme Reissner-Nordstr\"om-like black holes: This is the limiting case of the previous situation, in which both horizons merge into a degenerate one.

\item Regular horizonless Reissner-Nordstr\"om-like solutions: In such a case the horizons are absent and the solution describes a horizonless compact object. While the metric describing this object resembles the overcharged RN solution, this solution features a traversable wormhole with a throat radius $r=r_c$ instead of a naked singularity.

\end{itemize}

\item Regular, finite-curvature (FC) solutions: When $\delta_1=\delta_c$ then the leading-order coefficients in Eqs.(\ref{eq:gttc}) and (\ref{eq:grrc}) are constant. A consequence of this is that all curvature scalars turn out to remain finite at the throat (and everywhere). These solutions also split into three sub-cases:

\begin{itemize}

\item  Regular, finite-curvature traversable wormholes: When $\delta_2 > \delta_c$ no horizon is present and this configuration represents a traversable wormhole.

\item Regular, finite-curvature non-traversable wormholes: When $\delta_2 < \delta_c$ a horizon is present, rendering the wormhole as non-traversable.

\item Regular, finite-curvature transition case: When $\delta_2=\delta_c$ the horizon is exactly located at the throat itself. This case should thus be considered as non-traversable as well.

\end{itemize}

\end{itemize}
\begin{table}[t]
\begin{center}
\begin{tabular}{| c | c | c | c | c |}
\hline
\multicolumn{2}{ |c| }{Type} & $\delta_1$ & $\delta_2$ & Figure\\ \hline
\multicolumn{2}{ |c| }{Reg. Sch. BH} & $<\delta_c$ & - & \ref{fig:PSch} \\ \hline
\multirow{3}{*}{Reg. RN BH} & Two Horizons &\multirow{3}{*}{$>\delta_c$} & \multirow{3}{*}{-} & \ref{fig:PRN2} \\ 
 & Extreme & & & \ref{fig:PRN1} \\ 
& Horizonless & & & \ref{fig:PRN0}  \\ \hline
\multirow{3}{*}{FC solutions} & Transversable WH& \multirow{3}{*}{$=\delta_c$} & $>\delta_c$ & \ref{fig:PMk1} \\ 
&  Non-transversable WH &  & $<\delta_c$ &   \ref{fig:PMk1e}\\ 
& Transition case & & $=\delta_c$ & \ref{fig:FC0}\\ \hline
\end{tabular}
\caption{Summary of the configurations and the corresponding figure for the optical appearances below.}
\label{tab:summary}
\end{center}
\end{table}

A summary of these configurations is provided in Table \ref{tab:summary}. It should be emphasized that all the three types of configurations above, along with their several sub-classes, are regular according to the geodesic completeness criterion, which is the main concept ingrained in the singularity theorems \cite{Senovilla:2014gza}. This is a consequence of the presence of the wormhole throat at $r=r_{th}=r_c$, which in the present cases allows for the extensibility of every (time-like and null) geodesic able to reach the region beyond $x=0$ to the other side of the throat, as explicitly verified in \cite{Olmo:2015bya}. This happens despite the curvature scalars generically diverging at the wormhole throat (unless $\delta_1 = \delta_c$, as already stated), since this feature does not apparently entail the presence of any pathological behaviour, as has been shown through several routes, e.g. using tidal forces upon geodesic congruences plus causal transmission between the constituents of a body \cite{Olmo:2016fuc}, scattering of scalar waves off the wormhole \cite{Olmo:2015dba}, and completeness of accelerated observers' paths \cite{Olmo:2017fbc}.

\begin{center}
\begin{figure}[t]
\begin{center}
\includegraphics[width=8.6cm,height=6.0cm]{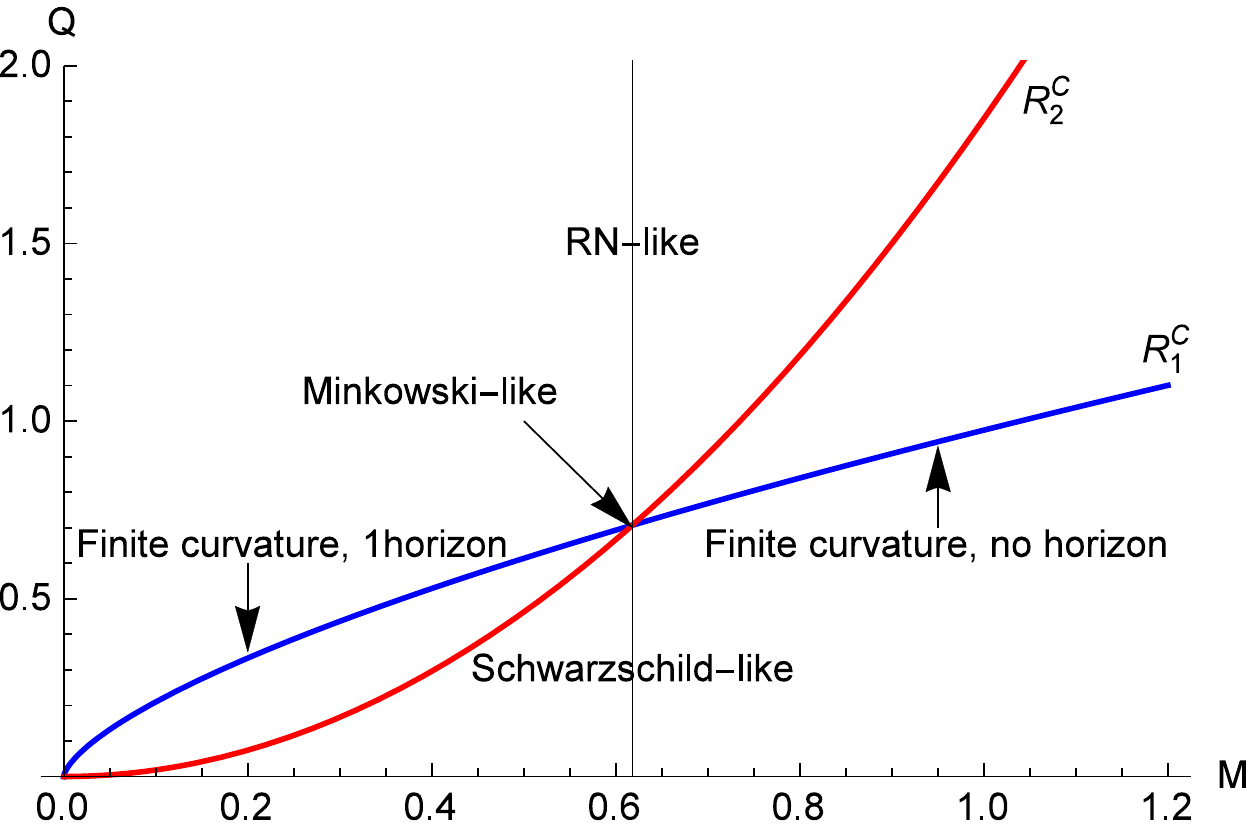}
\caption{The space of configurations for $l_{\epsilon}^2=1/4$ as given by the ratios $R_1^c$ and $R_2^c$ defined in Eqs.(\ref{eq:R1}) and (\ref{eq:R2}), respectively, and depicted in blue and red. For $R_1>R_1^c$ ($R_1<R_1^c$) we find Schwarzschild-like (RN-like) configurations. For those with $R_1=R_1^c$ (i.e. on the blue curve), those with $R_2>R_2^c$ (i.e. above the red line) have finite curvature everywhere but a single horizon, while those with  $R_2<R_2^c$ are finite-curvature, naked objects. The point at which $R_1^c=R_2^c$ meet, behaves as pure Minkowski. }
\label{fig:configurations}
\end{center}
\end{figure}
\end{center}

The absence of singularities in these solutions implies that, as opposed to the transition from RN black holes to naked singularities in GR (above $Q^2/M^2=1$), the cosmic censorship conjecture is unnecessary, as the transition from black hole solutions to horizonless traversable wormholes e.g. by changing the ratio $\delta_1/\delta_c$, does not uncloak any singularity. This ratio can be more conveniently expressed in terms of a critical cubic-charge-to-quadratic-mass ratio $R^1$ of the form
\begin{equation} \label{eq:R1}
R^1 =\frac{Q^3}{M^2},\qquad R_c^1 =Cl_{\epsilon},
\end{equation}
where $R_c^1$ is the critical value of the ratio $R^1$ and we defined the constant $C=\frac{16\delta_c^2}{2^{3/2}} \approx 1.8513$. Thus, RN-like configurations require $R^1>R_c^1$, Schwarzschild-like satisfy $R^1<R_c^1$, while FC solutions live on the line $R_1=R_c^1$. For the RN-like configurations, the transition between two-horizon black holes and horizonless compact objects cannot be found analytically for generic values of the parameters of the problem. As opposed to that, one can extract such an information in the particular cases in which $R^1=R_c^1$, since here the transition between the single-horizon and naked configurations occurs at $\delta_2=\delta_c$, which can also be expressed as a critical charge-to-quadratic-mass ratio $R^2$ of the form
\begin{equation} \label{eq:R2}
R^2=\frac{Q}{M^2},\qquad R_c^2=\frac{C}{2l_{\epsilon}},
\end{equation}
where $R_c^2$ is the critical value of the ratio $R^2$, so that naked configurations require $R^2>R_c^2$ in addition to the constraint on $R_c^1$ above, while a horizon is present otherwise. The space of configurations classified this way is depicted in Fig. \ref{fig:configurations} for the choice of EiBI parameter $l_{\epsilon}^2=1/4$ in the plane $\{M,Q\}$, where the divide between Schwarzschild-like and RN-like configurations is marked by the blue curve $R_1=R_1^c$, while the configurations with finite curvature correspond to those located {\it on} the blue curve. For the latter, the presence/absence of an event horizon depends on the value of its ratio $R^2$ as compared to the critical one $R_c^2$, i.e., above or below the red curve.

\section{Black hole imaging via null trajectories} \label{sec:III}

\subsection{Null geodesic equations}

Let us consider a null geodesic $g_{\mu\nu} k^{\mu}k^{\nu}=0$ with $k^{\mu}=\dot{x}^{\mu}$ a null vector representing the photon's wave number, and where a dot denotes a derivative with respect to the affine parameter.  For the line element given in Eq.(\ref{eq:SSS}) this equation reads (restricting the motion to the equatorial plane $\theta=\pi/2$ without loss of any generality due to spherical symmetry)
\begin{equation}
-A\dot{t}^2+B\dot{x}^2+r^2(x)\dot{\phi}^2=0.
\end{equation}
The spherical symmetry of the system also allows one to introduce two conserved quantities, $E=A\dot{t}$ and $L=r^2(x)\dot{\phi}$, corresponding to the energy and angular momentum per unit mass, respectively. With these definitions the previous equation can be rewritten as (a factor $L^2$ is re-absorbed by a redefinition of the affine parameter)
\begin{equation} \label{eq:geoeq}
AB\dot{x}^2=\frac{1}{b^2}-V(r),
\end{equation}
where we have introduced the impact parameter $b  \equiv L^2/E^2$, while the effective potential $V\left(r\right)$ is given by the expression
\begin{equation} \label{eq:Veff}
V(r(x))=\frac{A(x)}{r^2(x)}.
\end{equation}

In the process of generating images, it is more convenient to make use of the previously defined conserved quantities to rewrite the geodesic equation as the variation of the azimuthal angle with respect to the coordinate $x$ as
\begin{equation}
\frac{d\phi}{dx}=\mp \frac{b}{r^2(x)} \sqrt{\frac{AB}{1-\frac{Ab^2}{r^2(x)}}}.
\end{equation}
A light ray issued from the observer's screen with a given impact parameter $b$ approaching the object experiences a variation on its azimuthal angle given by the expression above with a negative sign. At the turning point, the square root on the right-hand-side vanishes, and the integration of the equation above continues with the positive sign. For the geometry of interest in this work, this equation reads
\begin{equation} \label{eq:nullphi}
\frac{d\phi}{dx}=\mp \frac{b}{\Omega_+ r^2(x)} \frac{1}{\sqrt{1-\frac{Ab^2}{r^2(x)}}}.
\end{equation}
Under this form of the geodesic equation the factor $\Omega_+$ varies between one (at asymptotic infinity) and two (at the wormhole throat), so its influence is minor at large distances, and will only manifest itself for the light rays flowing near the wormhole throat. The most important modifications with respect to GR solutions appear both via $A(r(x))$, which is quantitatively and qualitatively different from the RN solution of GR, and via $r^2(x)$, given its non-trivial behavior according to Eq.(\ref{eq:zx}).

Alternatively, one can also use Eq.(\ref{eq:dxdr}) to rewrite the previous equation in terms of the radial coordinate $r$ as
\begin{equation} \label{eq:geoeqr}
\frac{d\phi}{dr}=\mp \frac{b}{\Omega_{-}^{1/2} r^2} \frac{1}{\sqrt{1-\frac{Ab^2}{r^2}}}
\end{equation}
where in this case one must add the condition that the radial function is bounded by $r \geq r_c$. Now, $\Omega_{-}$ varies between one and zero, but the latter is only attained at the wormhole throat, while the photon sphere radius stands far from that region (recall that we are not considering radiation flow across the wormhole throat in the cases in which the wormhole is traversable). Thus this feature will be mostly irrelevant for black hole configurations, but it will acquire relevance in the horizonless ones.

\subsection{Photon sphere and photon rings}

A turning point $r=r_0$ in a light ray trajectory happens when the right-hand-side of Eq.(\ref{eq:geoeq}) vanishes, which corresponds to
\begin{equation}
b_0^2=\frac{r_0^2}{A_0}
\end{equation}
where $A_0 \equiv A(r_0)$. The smallest impact parameter, $b_c$, for which such a turning point exists, corresponds to the maximum of the effective potential, that is
\begin{equation} \label{eq:bc}
V_{eff}(r_{ps})=\frac{1}{b_c^2},\quad \frac{dV_{eff}}{dr}(r_{ps})=0,\quad \frac{d^2V_{eff}}{dr^2}(r_{ps})<0.
\end{equation}
The value $b_c$ is dubbed as the {\it critical impact parameter}, while the location of the maximum, $x=x_{ps}$ (or $r=r_{ps}$ in the $r$-representation), is sometimes referred to as the critical curve, though it is (perhaps) more popularly known as the {\it photon sphere} given the spherical symmetry of the system (in the rotating case, it forms instead an oblate finite-size region known as the photon shell). It corresponds to the locus of null unstable geodesics, i.e., light ray trajectories that asymptotically approach a critical point of the effective potential. Using the conditions above, from Eq.(\ref{eq:Veff}), the critical points are given by the solutions to the implicit equation
\begin{equation}
r'\vert_{x_{ps}}A(x_{ps})-\frac{1}{2}r(x_{ps})A'(x_{ps})=0
\end{equation}
where primes denote derivatives with respect to the variable $x$. Alternatively, one can rewrite this equation in terms of derivatives with respect to the radial function itself using the relation (\ref{eq:dxdr})) as
\begin{equation} \label{eq:critcur}
A(r_{ps})-\frac{1}{2}r_{ps} \frac{dA}{dr} \Big\vert_{r_{ps}}=0
\end{equation}
as long as $dr/dx \neq 0$, i.e., everywhere except at the wormhole throat.

We consider a setting in which the compact object is oriented face-on with respect to the asymptotic observer.  A light ray issued from asymptotic infinity and approaching the compact object in a trajectory with an impact parameter $b>b_c$ sees its trajectory gravitationally deflected with a given angle $\delta \phi$ before reaching the turning point given by a zero of the denominator of Eq.(\ref{eq:nullphi}), and afterwards is released back to asymptotic infinity. The closer the impact parameter to its critical value $b_c$ is, the larger the deflection angle becomes. Eventually, the bending is so large that the photon executes one or more half-turns around the compact object. These half-turns can be measured via the definition introduced in \cite{Gralla:2019xty}
\begin{equation}
n=\frac{\delta \phi+\pi}{2\pi}.
\end{equation}
Such a light ray crosses the equatorial plane just once if $1/2 \leq n <3/4$, where the lower bound corresponds to a completely non-deflected trajectory.  These trajectories yield the {\it direct emission} of the disk, which is the major contribution to the image of the compact object. Alternatively, and for the sake of this work, we shall use instead the number of crossings a light ray makes with the equatorial plane, labelled by an integer number $m \geq 1$ such that
\begin{equation}
\frac{m}{2}+\frac{1}{4} \leq n < \frac{m}{2}+\frac{3}{4}
\end{equation}
In this language the direct emission corresponds to $m=0$ (i.e. it departs from the disk itself instead of crossing it, and in such a case the lower bound in this expression is $1/2$ instead), while highly-bent trajectories ($m=1,2,\ldots$) produce a thin bright ring of radiation in the optical image of the object, dubbed as the {\it photon ring}, which tracks the location of the critical curve in the sky. For values of the impact parameter $b<b_c$, the light ray spirals down and reaches the event horizon. Due to this fact, a brightness depression is generically expected to be present in the central part of the image, at least in black hole solutions, usually known as the {\it shadow}. However, its outer edge does not necessarily coincide with the critical curve, since we did not take into account yet the features of the accretion disk providing the main source of illumination, something we shall tackle in Sec. \ref{sec:IV}.

In the usual RN solution of GR the critical impact parameter is given by the analytical expression
\begin{equation}
b_c =\frac{\left(3M \left(\sqrt{9M^2-8 Q^2}+3M\right)-4Q^2\right)}{\sqrt{2M\left(\sqrt{9M^2-8 Q^2}+3M\right)-4 Q^2}},
\end{equation}
while the radius of the photon sphere is given by
\begin{equation}
r_{ps}=\frac{3M+\sqrt{9M^2-8Q^2}}{2}.
\end{equation}
This implies that for any $M^2 \leq (9/8)Q^2$ the solution features a photon sphere which always lies outside the event horizon, if the latter exists, and thus it is accessible to photons. In particular, if the condition $Q^2<M^2\leq (9/8)Q^2$ is met, the solution features a photon sphere but not an event horizon: naked singularities in RN geometry may thus have a photon sphere. The modifications to the RN geometry introduced by the EiBI  solutions modify these two relevant quantities but do not allow one to obtain an explicit expression. Nonetheless, a numerical analysis on a case-by-case basis allows one to find these quantities for any configuration of interest.

\subsection{On the photon rings}

Both the critical curve and the shadow are relevant features in the theoretical characterization of black hole images. However, it is important to make a distinction between critical curve/photon sphere, and photon rings: while the latter are observable, the former is {\it not}. Indeed, while the critical curve is a unique theoretical feature of the background geometry alone, the actual location, separation, and luminosity of the photon ring(s) are dependent also on the features of the accretion disk mainly due to two reasons. First, the fact that the photon ring is actually an (infinite) composition of sub-rings corresponding to light rays that have executed $m$ half-turns around the black hole, provided that a mild number of astrophysical assumptions on the accretion flow hold: optical thinness (i.e. transparency to its own radiation) in the emission region, and departure from complete spherical shape (i.e. existence of gaps in the emission region) \cite{Vincent:2022fwj}. The second one is determined by the inner edge of the emission region: when the accretion flow is spherically symmetric, the shadow completely fills the critical curve, whereas in thin or thick disk scenarios the shadow is strongly reduced (and further darkened) to an inner shadow \cite{Chael:2021rjo}.  In the latter case, if the emission is extended all the way down to the event horizon, the shadow is essentially a lensed image of the event horizon though apparently augmented due to gravitational lensing effects. Under these two assumptions (optical thinness and non-spherical geometry of the disk), an infinite number of photon rings ($m=1,2,\ldots$) appear on the image, several possibly superimposed on top of the direct emission ($m=0$) and appearing inside the photon sphere radius.

\begin{figure*}[t]
\includegraphics[width=7.8cm,height=5.5cm]{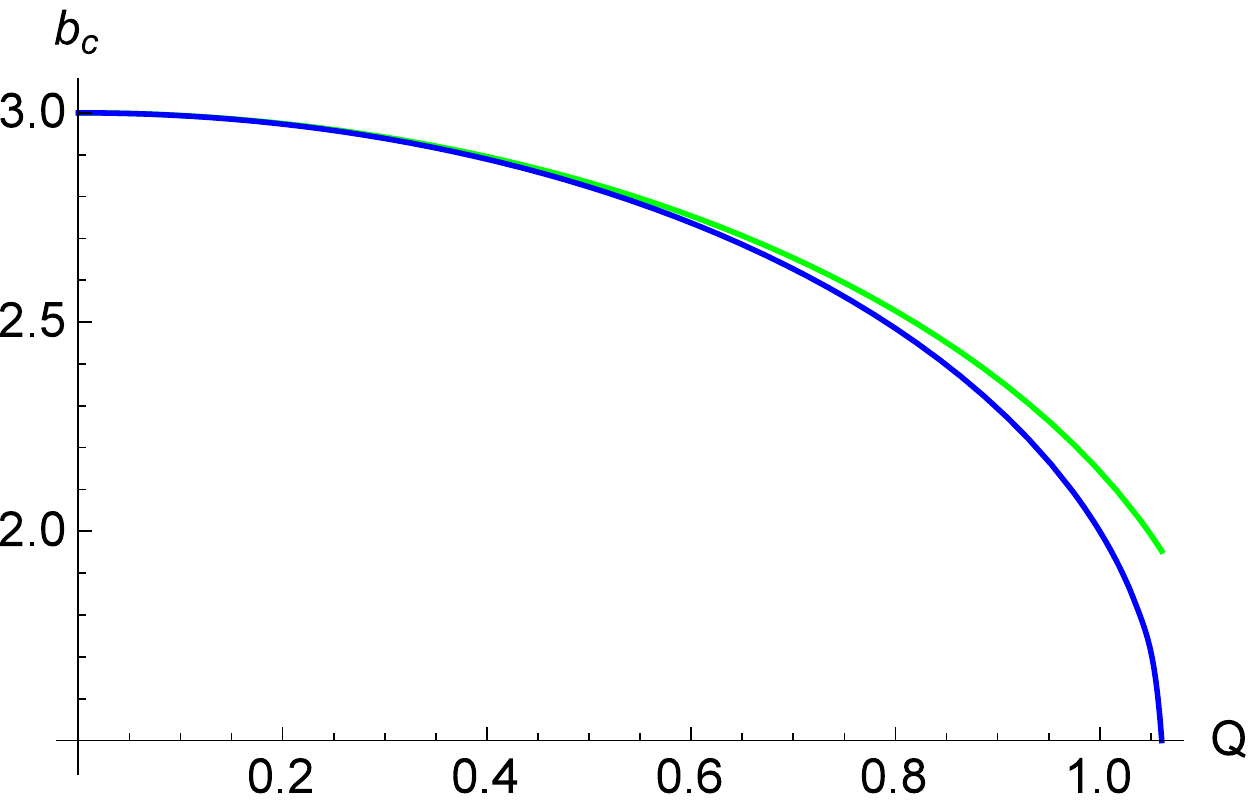}
\includegraphics[width=7.8cm,height=5.5cm]{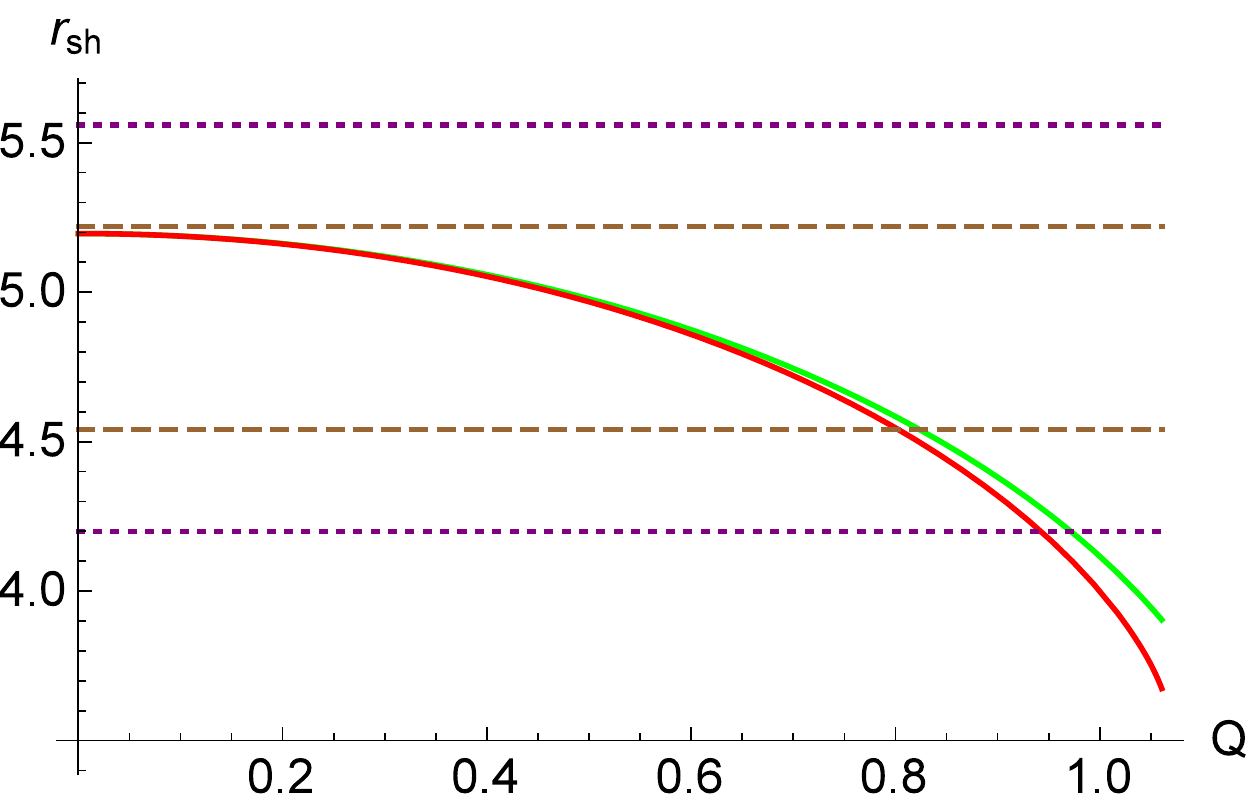}
\caption{The critical curve location (left) as given by the Eq.(\ref{eq:critcur}), and the shadow radius $r_{sh}$ (right) as given by Eq.(\ref{eq:sr}), for GR-RN solutions (blue and red, respectively); and for EiBI solutions with $l_{\epsilon}^2=1/4$ (green curves), as a function of the electric charge $Q$.  The set of straight dashed (brown) and dotted (purple) lines corresponds to the bounds on the shadow's size at $1\sigma$ and $2\sigma$, respectively, as reported by the EHT Collaboration \cite{EventHorizonTelescope:2022xqj}. In these plots, $M=1$.}
\label{fig:radsizePal}
\end{figure*}

\subsection{Size of the shadow}

The outer edge of the central brightness depression cannot be directly measured by the EHT Collaboration \cite{EventHorizonTelescope:2019ths} given the fact that its sensitivity makes impossible to extra dim features below $\sim 10 \%$ of the peak brightness. However, its simulated images of the supermassive object in the center of our Milky Way galaxy - Sgr A$^*$ - report the possibility of using the radius of the bright ring enclosing the object as a proxy of the shadow radius itself \cite{EventHorizonTelescope:2022xqj}, defined as the apparent boundary of the gravitationally lensed image of the photon sphere, given by the simple relation (for a more detailed analysis of analytical studies of the shadow's size see e.g. \cite{Perlick:2021aok}):
\begin{equation} \label{eq:sr}
r_{sh}=\frac{r_{ps}}{\sqrt{A(r_{ps})}} \ .
\end{equation}
This possibility is subject to two main requirements: i) the presence of a sufficiently numerous source of photons strongly lensed near the black hole event horizon; and ii) a (sufficiently) geometrically thick emission region which is optically thin at the EHT's wavelength operation window. While these bounds are only marginally affected by the details of the accretion flow beyond the two above, nonetheless a calibration factor must be introduced between the radius of the ring and the one of the shadow, addressing to which extent the former can be used as a proxy of the latter. This factor includes both theoretical and observational uncertainties on the modelling of the accretion flow and on the measurement of this bright radius, respectively.  For a Schwarzschild black hole, the shadow radius in Eq.(\ref{eq:sr}) yields (we restore units here) $r_{sh}=3\sqrt{3}GM/c^2$, therefore subtending (in the small angle approximation) an angular diameter  $\theta_{sh}=6\sqrt{3}GM/(c^2D)$, where $D$ is the distance between the observer and the source. Since for Sgr A$^*$ the ratio $M/D$ is known, the EHT Collaboration states that one can use these ingredients to find the following bounds on the shadow size
\begin{equation}
4.54 \lesssim r_{sh}/M \lesssim 5.22
\end{equation}
at $1\sigma$ and
\begin{equation}
4.20 \lesssim r_{sh}/M \lesssim 5.56
\end{equation}
at $2\sigma$. Note that the generalization of the previous bounds to extremal Kerr black holes modifies the result by a factor $\lesssim 7\%$ down \cite{Psaltis:2018xkc}.

To approach these results in an agnostic way (i.e. without judging their validity with regards to its assumptions), we shall consider its compatibility with our modified electrically charged  geometries (also using as a reference point the usual RN configurations in GR). For this purpose, we shall analyze the dependence of the radius of the shadow in the electric charge (for a fixed gravity parameter $l_{\epsilon}^2$ in the present case). This strategy was implemented in Ref. \cite{Vagnozzi:2022moj} for an overwhelming number of modified gravitational configurations inspired by either modifications to the matter fields or to the gravitational sector, though the one considered in this work is not included there.  The evolution of the critical curve radius and the shadow radius for both RN and EiBI solutions is depicted in Fig. \ref{fig:radsizePal} alongside with the bounds above, where we have taken a reference value of $l_{\epsilon}^2=1/4$ for convenience. In the usual RN case, for values of (taking $M=1$) $Q \gtrsim 0.8$ the shadow size is incompatible with the $1\sigma$ bound, and for $Q \gtrsim 0.95$ with the $2\sigma$ bound. This seems to discard naked RN singularities as viable models to describe the shadow of Sgr A$^{\star}$. As for the EiBI solutions, both the critical curve radius and the shadow radius are enhanced as compared to the RN solution, the size of the enhancement increasing with the electric charge, i.e., as the new gravitational dynamics encoded in the EiBI field are powered up by stronger contributions in the matter fields. In particular, this implies that constraints on the size of the shadow allow for slightly larger values of the electric charge in the EiBI version, and that this effect would become more pronounced for larger values of the EiBI parameter $l_{\epsilon}^2$.

\subsection{Selected configurations for generation of images}

In this section we shall fix the parameters of the configurations we are interested in working with.  It should be stressed that bounds on  $l_{\epsilon}^2$ (i.e. on $\epsilon$ do exist from several sources, e. g., light-by-light scattering \cite{BeltranJimenez:2021oaq} using ATLAS data, absence of anomalous beyond-standard model interactions in the LHC data \cite{Delhom:2019wir}, solar system  \cite{Casanellas:2011kf} and other astrophysical constraints \cite{Berti:2015itd}. These constraints, particularly those of astrophysical origin, contain some drawbacks on the modelization used in order to extract them. Furthermore, when considering optical appearances of EiBI configurations one must bear in mind the many unknowns in the physical features of the accretion disk and the difficulty of extracting reliable observational signatures, we have thus opted for setting the above value of $l_{\epsilon}^2=1/4$. This values allows for mild enough deviations with respect to the GR predictions which does not immediately enter into blatant contradiction with current shadow observations but, at the same time, entails the presence of configurations which strongly deviate from a qualitative and quantitative point of view in the shape of their metrics and effective potentials, following our discussion above on the global structure of the configurations for every $l_{\epsilon}^2 \neq 0$. This way, our subsequent analysis will allows us to see what predictions this model entails for the optical appearances of the different objects it harbours within, and leave for further works the actual grounding of this model to the current observational constraints from other sources.

For this choice, the seven classes of configurations can be classified according to the ratios $R_1^c$ and $R_2^c$ as depicted in Fig. \ref{fig:configurations}. The first ratio splits the Schwarzschild-like from the RN-like cases, with the finite-curvature (including the limiting Minkowski-like) cases corresponding to the curve $R=R_1^c$ itself. For the latter configurations, with a fixed charge-to-(quadratic) mass ratio, values of the mass below the critical $M_c \approx 0.6180$ yield solutions with a single horizon, while those with $M>M_c$ feature none. The transition between both cases yields a critical charge $Q_c \approx 0.7071$. Based on this scheme, we select the parameters $\{M,Q\}$ characterizing the seven qualitatively different classes of configurations according to the number and type of their causal structure (i.e. horizons). These are reported in Table \ref{Table:I} alongside their corresponding values for the four quantities of interest in the generation of shadow images: the radius of the throat, the radius of the event horizon, the radius of the critical curve, and the  value of the critical impact parameter.

\begin{center}
\begin{table}[t]
\begin{center}
\begin{tabular}{|c|c|c|c|c|c|c|c|}
\hline
  & Sch  & RN2  & RN1e  & RN0  & FC1  & FCt & FC0 \\ \hline
$M$ & 1  & 1  & 1  & 1  & 1  & 0.6180 & 0.55 \\ \hline
$Q$ & 0.5 & 1  & 1.0511  & 1.1  & 0.9745  & 0.7071  & 0.6542  \\ \hline
$r_t$ & 0.59  &  0.84  & 0.86   & 0.88   &  0.83   &  0.70   & 0.68   \\ \hline
$r_h$ & 1.87 & 1.31   &  1.05  & X  & 1.37  & 0.70   & X  \\ \hline
$r_{ps}$ & 2.83 & 2.14   & 1.98  & 1.76  & 2.20  & 1.21  & 1.07   \\ \hline
$b_c$ & 4.97  & 4.11   & 3.94   & 3.72  & 4.19  & 2.37  & 2.08    \\ \hline
\end{tabular}
\begin{center}
\caption{The values of the features of interest for generation of images for the chosen configurations representative of all the possible sub-cases: Schwarzschild-like black holes (Sch), RN-like with two (RN2), a single (extreme, RN1e) or none (RN0) horizons, and finite-curvature (FC) solutions with a single horizon (FC1) or none (FC0), including the transition case (FCt). The parameters $\{M,Q\}$ refer to the masses and charges of the configurations, while $\{r_t,r_h,r_{ps},b_c\}$ correspond to the throat's radius, event horizon radius, critical curve radius, and critical impact parameter, respectively.}
\label{Table:I}
\end{center}
\end{center}
\end{table}
\end{center}

\begin{figure}[t!]
\begin{center}
\includegraphics[width=8.6cm,height=6.0cm]{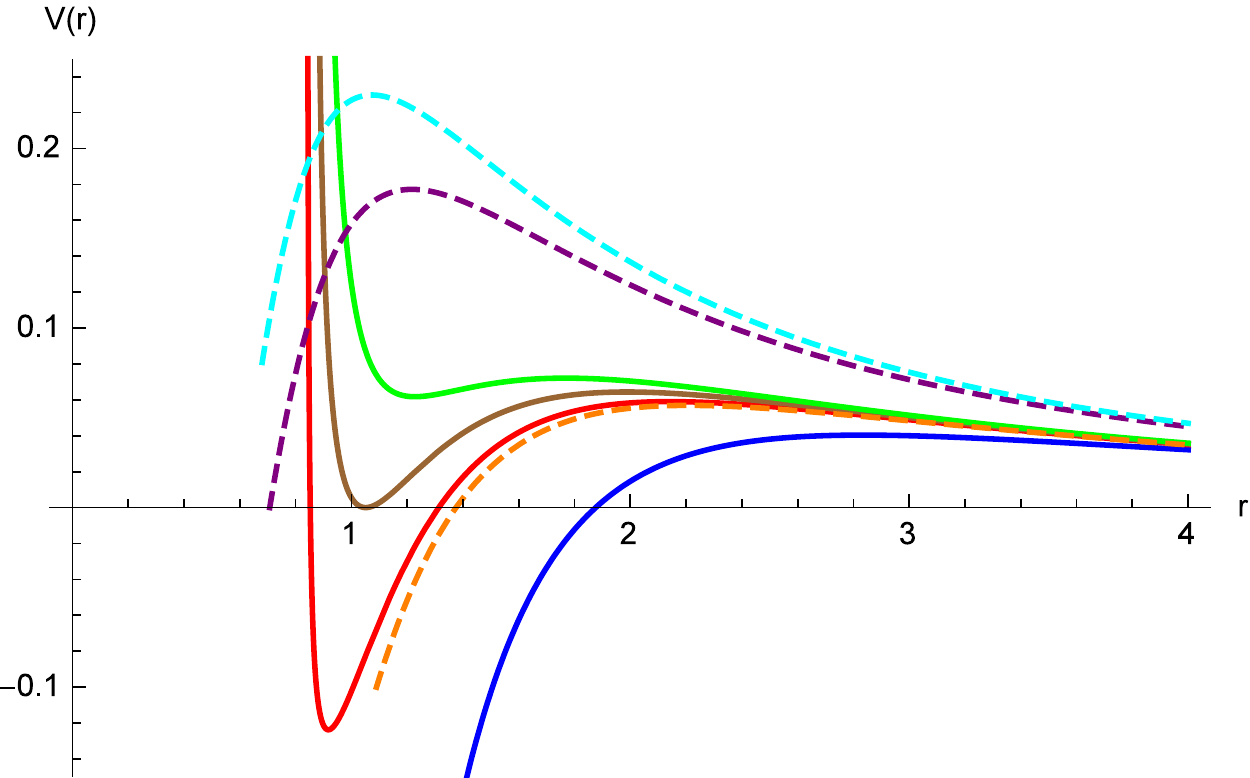}
\caption{The effective potential for the seven configurations reported in  Table \ref{Table:I}: Sch (blue), RN2 (red), RN1e (brown), RN0 (green), FC1 (dashed orange), FC0 (dashed cyan) and FCt (dashed purple).}
\label{fig:potential}
\end{center}
\end{figure}

A typical shadow image is the one of a black hole (e.g. Schwarzschild) encircled by a critical curve driving light trajectories with more than one half-turn around the central object, and contributing to create photon rings on top of the direct emission. In our case this corresponds to the Sch, RN2, RN1e,  and FC1 configurations of Table \ref{Table:I}, as depicted in Fig. \ref{fig:potential} for the effective potential (blue, red, brown and green curves, respectively). In addition we have a configuration, RN0, without event horizons but featuring a critical curve, a (stable) minimum of the potential (an anti-photon sphere), and an infinite slope of the potential at the center, all of them accessible to photons with impact parameters below the critical one (this is somewhat similar to the usual naked Reissner-Nordstr\"om configurations of GR \cite{Tsukamoto:2021fsz} but, in contrast, RN0 harbours no singularities). This mimics previous (ad hoc) toy-models studied in the field, such as in generalized black hole bounces \cite{Lobo:2020ffi}, where their toy-ness comes from the fact that they are postulated a priori as opposed to those found here, which arise naturally from resolution of the field equations of the EiBI theory. It should be mentioned that the presence of the pair of unstable/stable critical curves is a generic property of horizonless compact objects as long as they are compact enough \cite{Cunha:2017qtt}, and carry the burden of being capable to trigger the development of non-linear instabilities \cite{Cardoso:2014sna} (see \cite{Ghosh:2021txu} for an analysis of the stationary case) which would be pathological if the associated time scale is small enough (something to be analyzed on a case-by-case basis for every solution). We also consider a configuration, FC0, in which there is a critical curve, no horizon, but the potential is everywhere finite (extending down to the wormhole throat), and no stable anti-photon sphere is present. Finally, because the configuration FCt features a horizon located at the throat itself, we shall take it as a limiting case of the previous configurations, being of black hole-type. Thus, our selection of configurations offers several qualitatively different types of potentials, whose net effect in terms of their shadow images we want to study here.

Before going into that, it should be stressed that in comparing these configurations with their GR-RN counterparts, the nature of the horizons for each class may change. For instance, the RN2 configuration (having two non-extreme horizons) corresponds to an extreme black hole in GR, while the RN1e configuration is a naked singularity in GR instead of an extreme black hole. As for the RN0, FCt, and FC0 configurations, their counterparts in GR do not have photon spheres, so the latter are not of interest to us in this work.

\section{Imaging modified black holes and horizonless objects} \label{sec:IV}

\subsection{Ray-tracing procedure} \label{sec:IV.A}

A ray-tracing procedure is a tracking of all light trajectories that reach the observer, and that integrate backwards the gravitational deflection equation (\ref{eq:geoeqr}) for a bunch of impact parameters to find the location in the sky every such light ray came from. In doing so, an important variable to store is the location in the equatorial plane (as a function of the impact parameter) each light ray crossed it. This is so because for those light rays nearing in their impact parameter the photon sphere there may be more than one such a crossings, labelled by a function $r_m(b)$, where $m=1,2,\ldots$ belongs to a lensed image of order $m$ of the source (for the sake of this work we favor this language over the one of {\it lensing/photon ring emissions} introduced in \cite{Gralla:2019xty}), which in the main case of physical interest corresponds to an accretion disk ($m=0$ identified as the direct emission of the disk, as already mentioned).

In order to produce shadow images of sufficient visual quality but within manageable computational times, we ray-trace a bunch of light rays from the observer's screen, taken to be located at a far away distance of $r=1000M$.  We perform this integration in two steps. First, we cover a range of impact parameters between zero and three times the one of the critical curve using $\sim 500$ iterations. Secondly, we track much more closely the critical curve performing an additional set of $\sim 1500$ iterations within the range $b_c(1-0.5 \times 10^{-2})<b<b_c(1+0.5 \times 10^{-2})$ in order to finely capture the quick variations of luminosities on its restricted range of impact parameter values for the photon rings $m=1,2$. For the case of horizonless compact objects [see Sec. \ref{Sec:hopr}] in which the contributions of higher-order $m>2$ trajectories can become non-negligible (RN0 and to a lower extend FC0), we adapt this tracking of the critical curve in order to detect additional crossings of the light trajectories with the equatorial plane. In all cases, the transfer function $r=r_m(b)$ storing the information on every crossing of the light ray with the equatorial plane is required to be sufficiently smooth after all iterations have been carried out. For turning points, where the denominator of the deflection angle diverges, we set a precision of $1-b^2A(r)/r^2 <10^{-15}$ to stop the integration from asymptotic infinity towards this point, and select the opposite sign to continue with the integration from there back towards asymptotic infinity. For trajectories below the critical impact parameter value, this part of the integration never takes place and continues instead until hitting the event horizon in the black hole cases, or the throat in the wormhole cases.

We produce a list of the stored data on $r_m$ for all trajectories $m=0,1,2,\ldots$ and build separately their corresponding contributions to the luminosity of the object bearing in mind the gravitational redshift accounting from the trip of the photon from source to observer [see Eq.(\ref{eq:Iob}) below]: an interpolation function guarantees the smoothness of the corresponding profiles. One then obtains a collection of three functions which are subsequently merged to produce the total intensity function $I(b)$ after a number of additional effects are taken into account [discussed below in Sec. \ref{sec:ogep}]. 


\subsection{Optical, geometrical and emission properties of the accretion disk} \label{sec:ogep}

For the sake of our images we consider an optically and geometrically thin accretion disk assumed to be located on the equatorial plane and made up of particles moving on nearly geodesic, circular (Keplerian) motion \cite{Cunningham:1975zz}.  The optically-thin assumption means that the disk is transparent to its own radiation, and it is obviously needed for the purposes of this work since it is the one allowing photons to circle the central object several times, raising the possibility of having photon ring(s) on top of the direct emission. The geometrically-thin assumption implies that the effective source of emission comes only from an infinitely-thin surface extending up to a certain minimum radius (not necessarily the event horizon itself). Moreover, this must be supplemented with choices for the emissivity and intensity (absorptivity is effectively taken to zero via the optically thin assumption above) making up the disk, which are linked to each other via a radiative transfer equation.  Solving such equation, in general, is only accessible via GRMHD simulations for a pool of assumptions on such quantities \cite{Gold:2020iql}. 

Here we cut significantly the complexity of this process as follows: under the hypothesis that no absorption is present, by Liouville's theorem one finds that $I_{\nu}/\nu^3$ is conserved along the beam of radiation, that is \cite{Lindquist:1966igj}
\begin{equation}
\frac{I_{\nu_e}}{\nu_e^3}=\frac{I_{\nu_o}}{\nu_o^3}
\end{equation}
where the sub-indices $_e$ and $_o$ refers to the emission and observer's frames, respectively. This way, in the frame of the asymptotic observer this transforms into $I_{\nu_o}=g^3I_{\nu_e}$ where the factor $g=\nu_0/\nu_e$ becomes in the line element  (\ref{eq:SSS}) as $g=A^{1/2}$. Now, by assuming a monochromatic emission in the disk's frame then the specific intensity only depends on the radial distance, i.e., $I_{\nu_e} \equiv I(r)$. This way we can integrate over the full spectrum of frequencies to find \cite{Gralla:2019xty}
\begin{equation}
I^{ob}=\int d\nu_o  I_{\nu_o}=A^2(r)I(r) \ .
\end{equation}
In addition, one needs to bear in mind the subsequent intersections with the accretion disk, so one finally writes
\begin{equation} \label{eq:Iob}
I^{ob}_{total}=\sum_{m=0}^{i} A^2(r) I(r)
\end{equation}
where $i$ denotes the order of the photon ring beyond which the contributions to the luminosity become negligible ($i \leq 2$ for black hole configurations). This leaves as single free function $I(r)$ to characterize the emission of the disk. This function should be demanded a minimum number of properties such as its smoothness and to be mostly localized a few Schwarzschild radius units away from the event horizon. Typically one would expect the disk's inner edge to extend all the way down to the event horizon, to match current observations by the EHT Collaboration. Nonetheless, this scenario makes it hard to observe the photon ring sub-structure, since due to its relative dimness it is completely dazzled by the contribution of the direct emission (we shall further comment on this later on). To account for both scenarios, we consider a single emission profile given by the functional form proposed in \cite{Gralla:2020srx} (hereafter referred to as GLM models) and described by the unbounded Johnson Distribution: 
\begin{equation} \label{eq:Ichoice}
I(r;\gamma,\mu,\sigma)=\frac{e^{-\frac{1}{2} \left[\gamma +\arcsinh\left(\frac{r-\mu}{\sigma}\right)\right]^2}}{\sqrt{(r-\mu)^2+\sigma^2}}
\end{equation}
where $\{\gamma,\mu,\sigma\} $  are free functions of the modelling acting as follows: $\gamma$ drives the rate of growth of the intensity profile from infinity down to its peak, $\mu$ performs a translation of the intensity profile in order to shift the profile to a desired position, and $\sigma$ controls the dilation of the profile. This shape of the profile was originally developed  to match the results of GRMHD for rotating (Kerr) black holes in order to explore the structure of their photon rings, see e.g. \cite{Paugnat:2022qzy}, so we will simplify the problem a little bit given the spherical symmetry of our setting. In particular, the constant $\mu$ is related in such models either to the innermost circular time-like orbit (ISCO) radius or to the inner horizon of a Kerr black hole, and we shall take here their restrictions and match it to the non-rotating and uncharged (Schwarzschild) case. 

\begin{figure}[t!]
\begin{center}
\includegraphics[width=8.6cm,height=6.0cm]{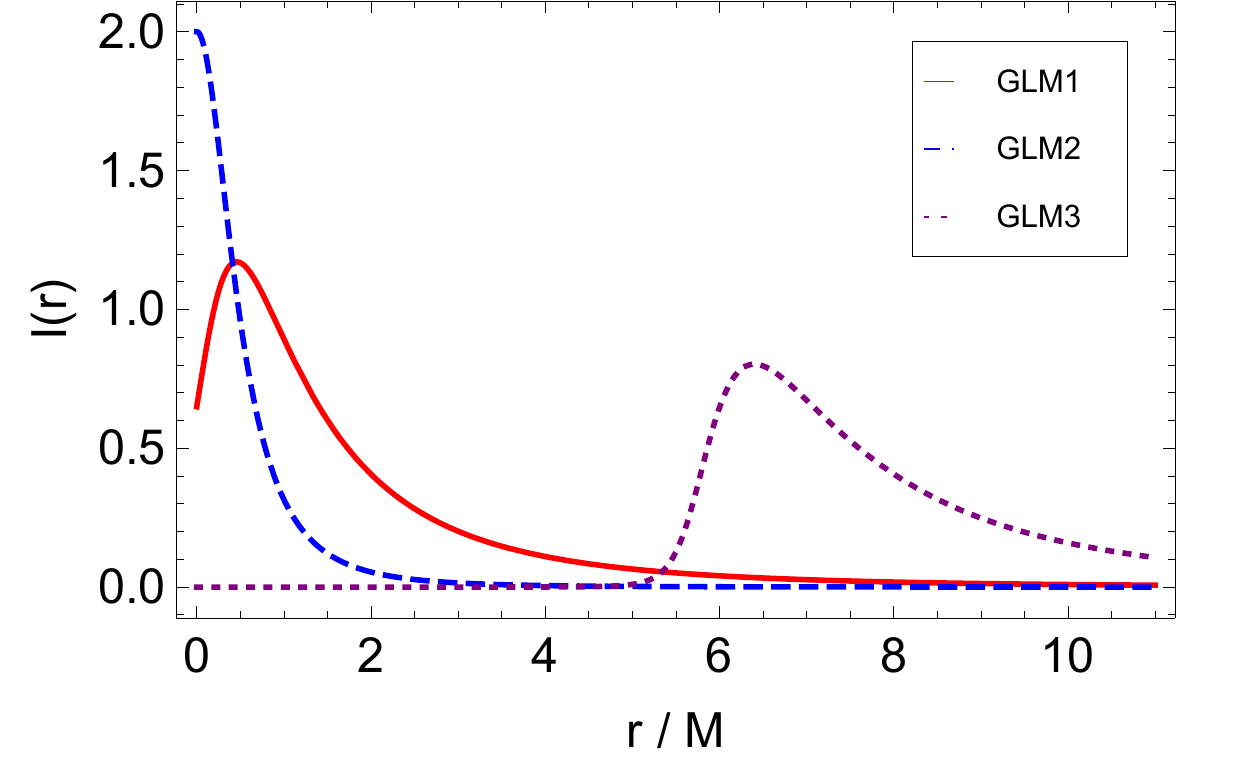}
\caption{Intensity profiles $I(r)$ (before normalization) as a function of the normalized radius $r / M$ of the three GLM models considered in Eqs.(\ref{eq:adGLM3}), (\ref{eq:adGLM1}) and (\ref{eq:adGLM2}).}
\label{fig:disks}
\end{center}
\end{figure}

This way, for the sake of this work we consider the following three models:
\begin{eqnarray}
\text{GLM3} &:& \gamma=-2,\mu=\frac{17M}{3}, \sigma=\frac{M}{2} \label{eq:adGLM3} \\
\text{GLM1} &:& \gamma=-\frac{3}{2},\mu=0, \sigma=\frac{M}{2} \label{eq:adGLM1} \\
\text{GLM2} &:& \gamma=0,\mu=0, \sigma=\frac{M}{2} \label{eq:adGLM2}
\end{eqnarray}
which closely resemble the ones of \cite{Gralla:2020srx} save by the fact that we halve the value of $\sigma$ in GLM3 for a better visualization of the corresponding photon rings. The intensity profiles of these GLM models are plotted in Fig. \ref{fig:disks}. The GLM3 model is aimed to probe the photon ring structure at a maximum of emission located near the ISCO radius of a Schwarzschild black hole: while such a radius obviously depends on charge (in the RN case), and also on the constant $l_{\epsilon}^2$ (in the present EiBI case), for the sake of comparison of photon ring images associated to a given geometry under the same conditions of the disk we shall take the same profile for all configurations (GR and EiBI alike), regardless of where exactly the ISCO is located for each of them. As for the GLM1 and GLM2 models, they probe the geometry all the way down to the event horizon or, in the naked case, to the wormhole throat. They are distinguished in the strength of their respective decay: GLM2 is the softer of the pair. This will have an impact in the distribution of their total luminosity. 

\begin{figure*}[t!]
\begin{center}
\includegraphics[width=5.9cm,height=4.4cm]{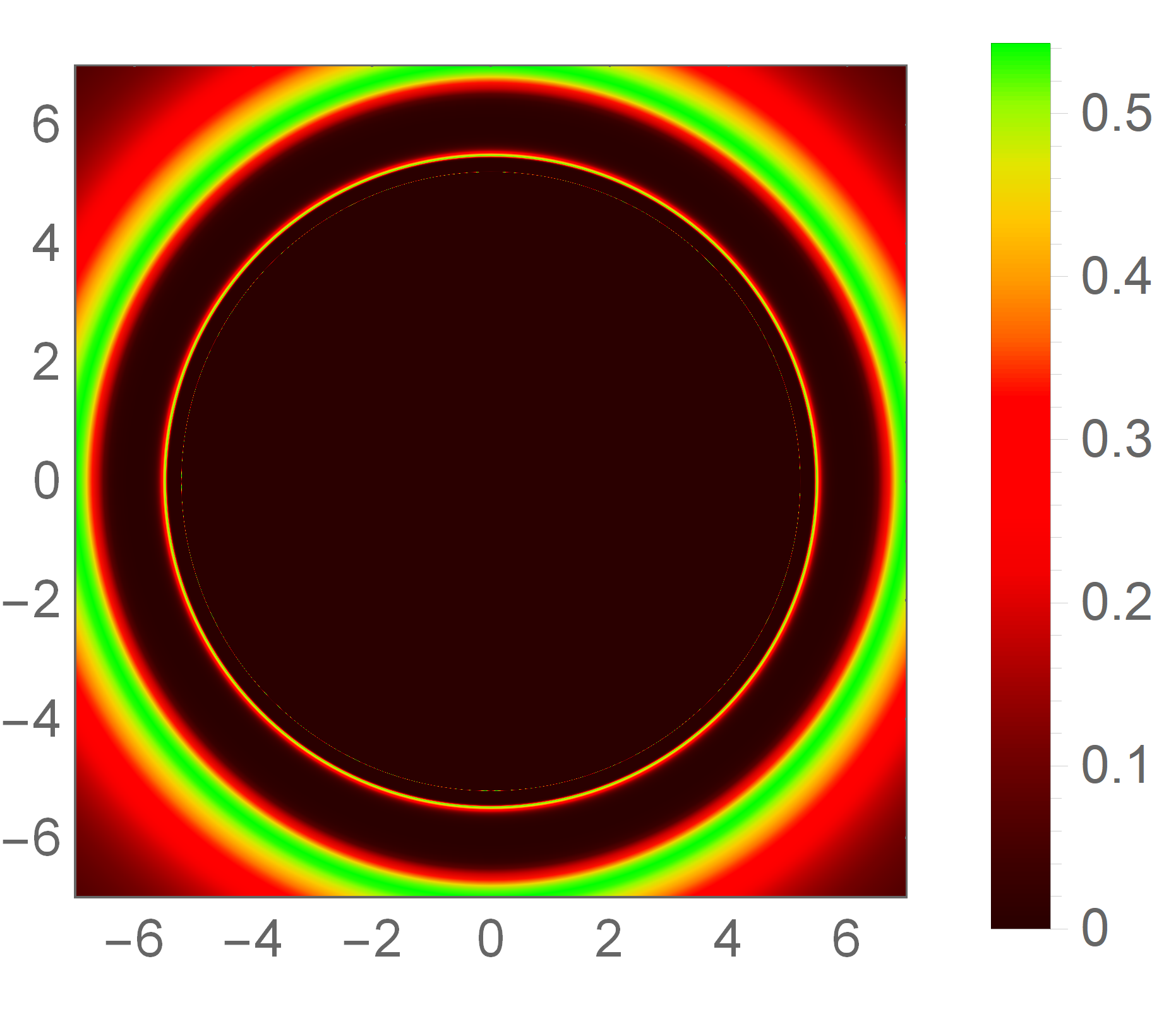}
\includegraphics[width=5.9cm,height=4.4cm]{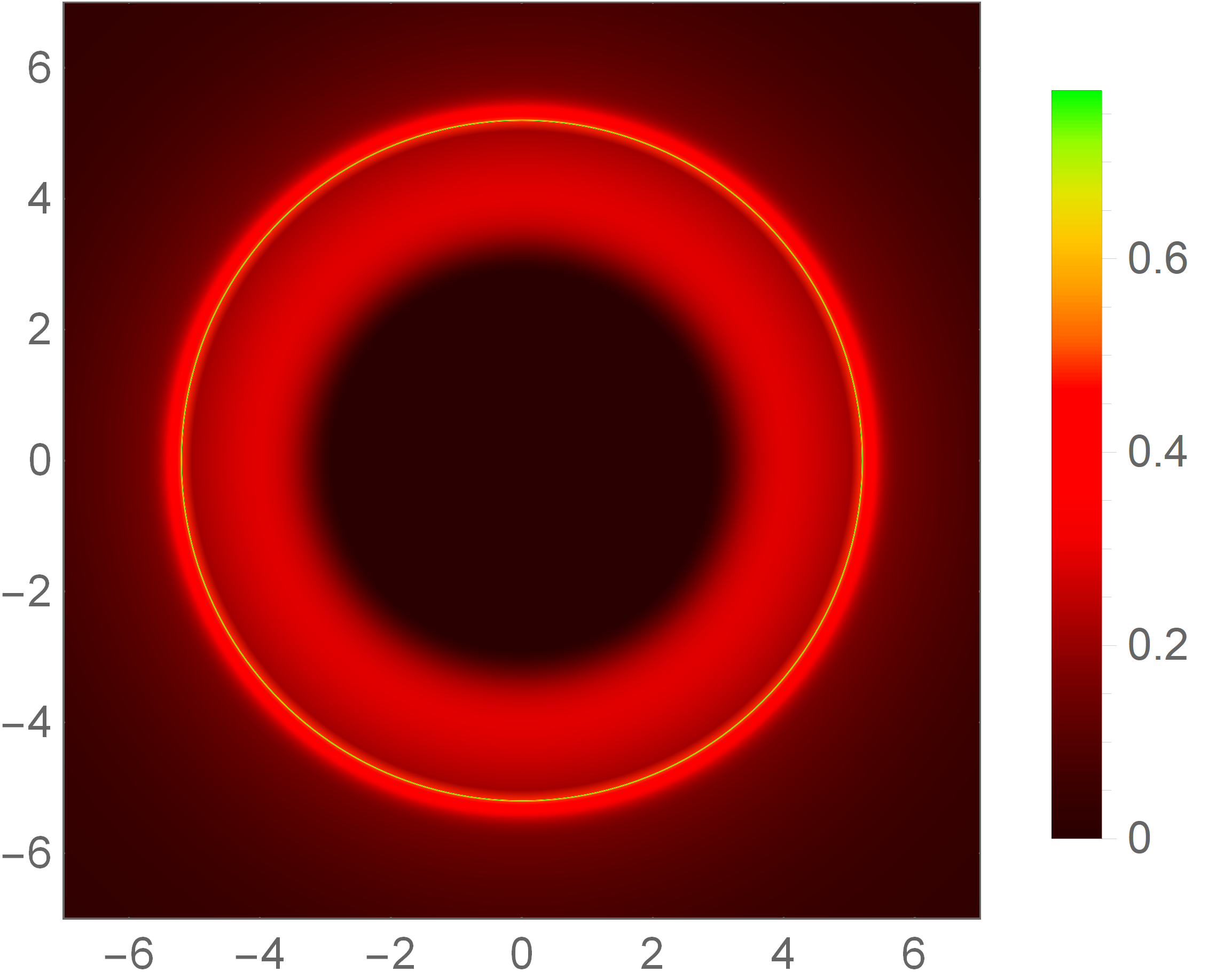}
\includegraphics[width=5.9cm,height=4.4cm]{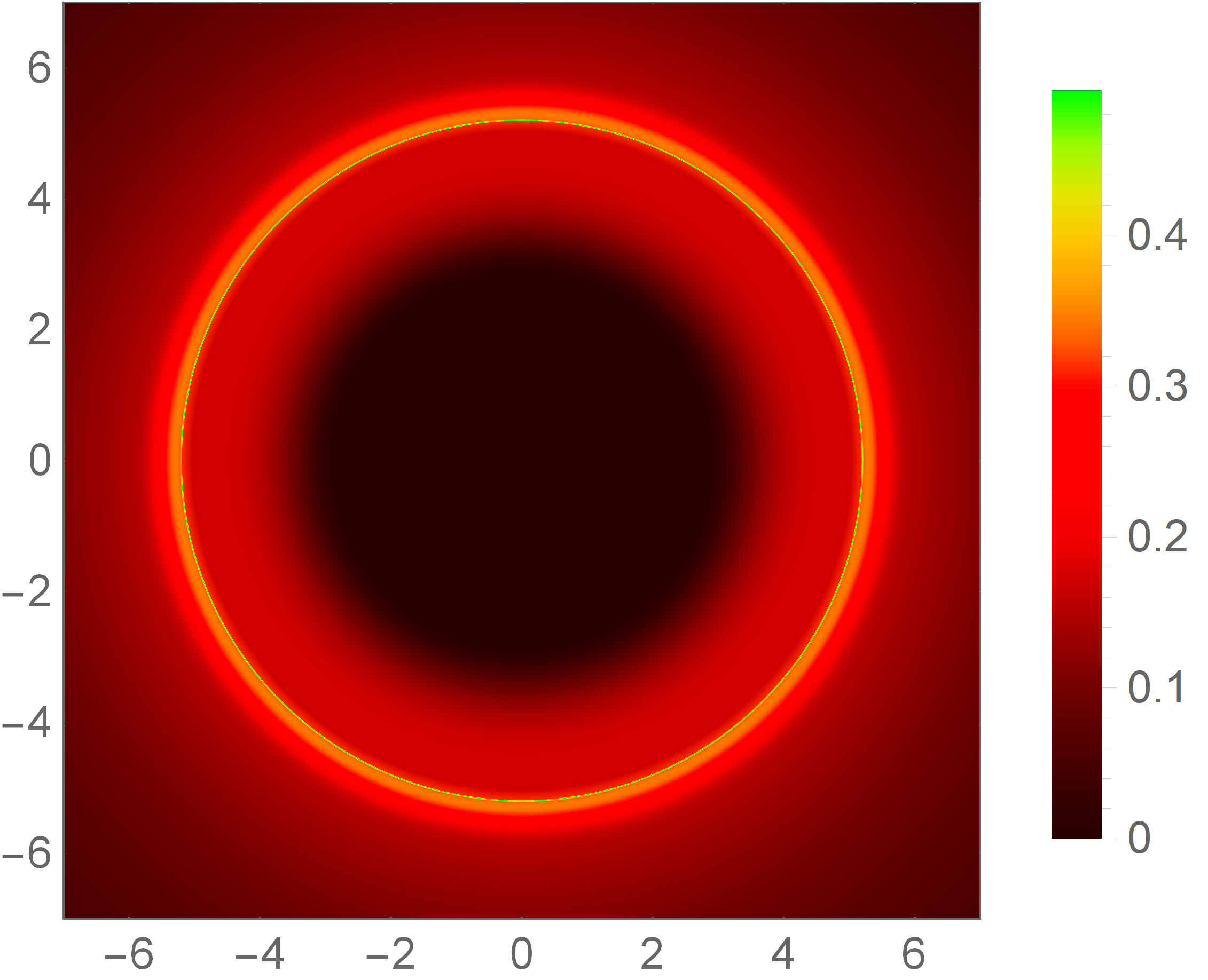}
\includegraphics[width=5.9cm,height=4.4cm]{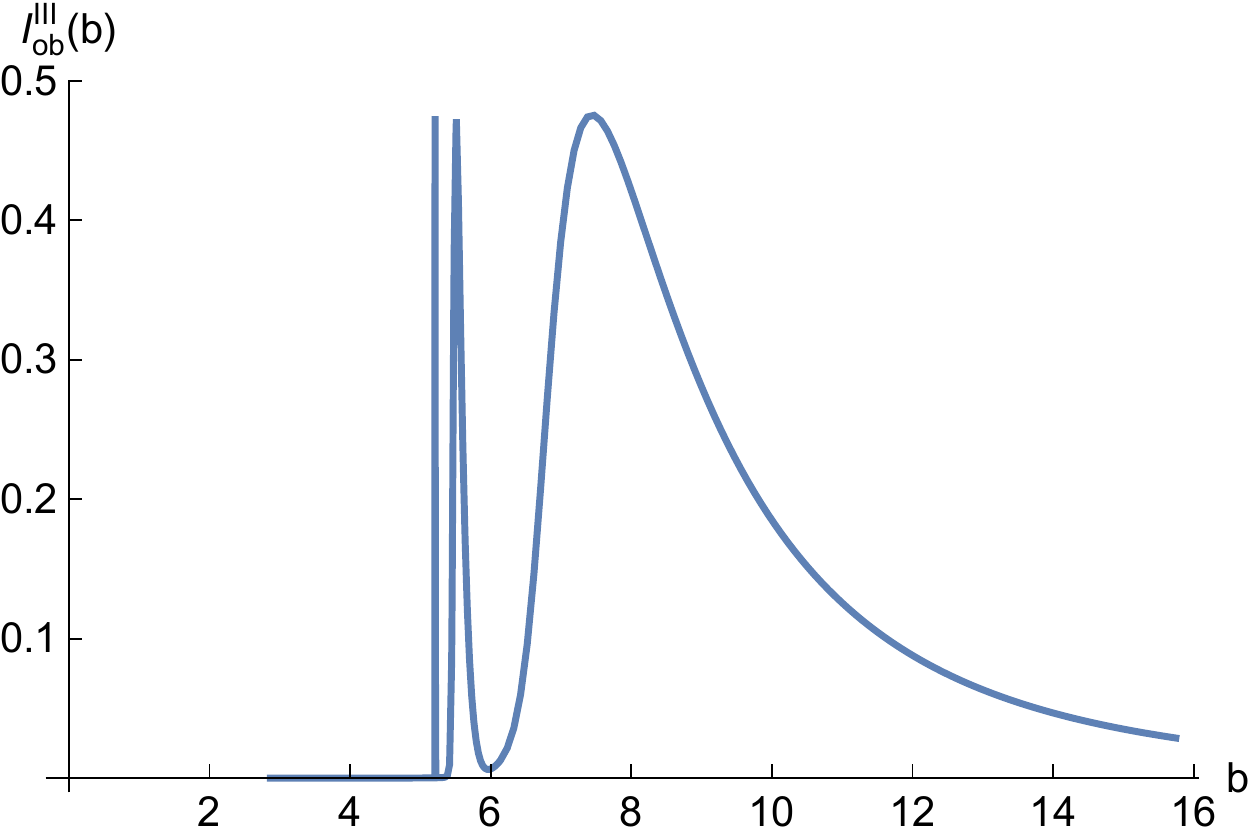}
\includegraphics[width=5.9cm,height=4.4cm]{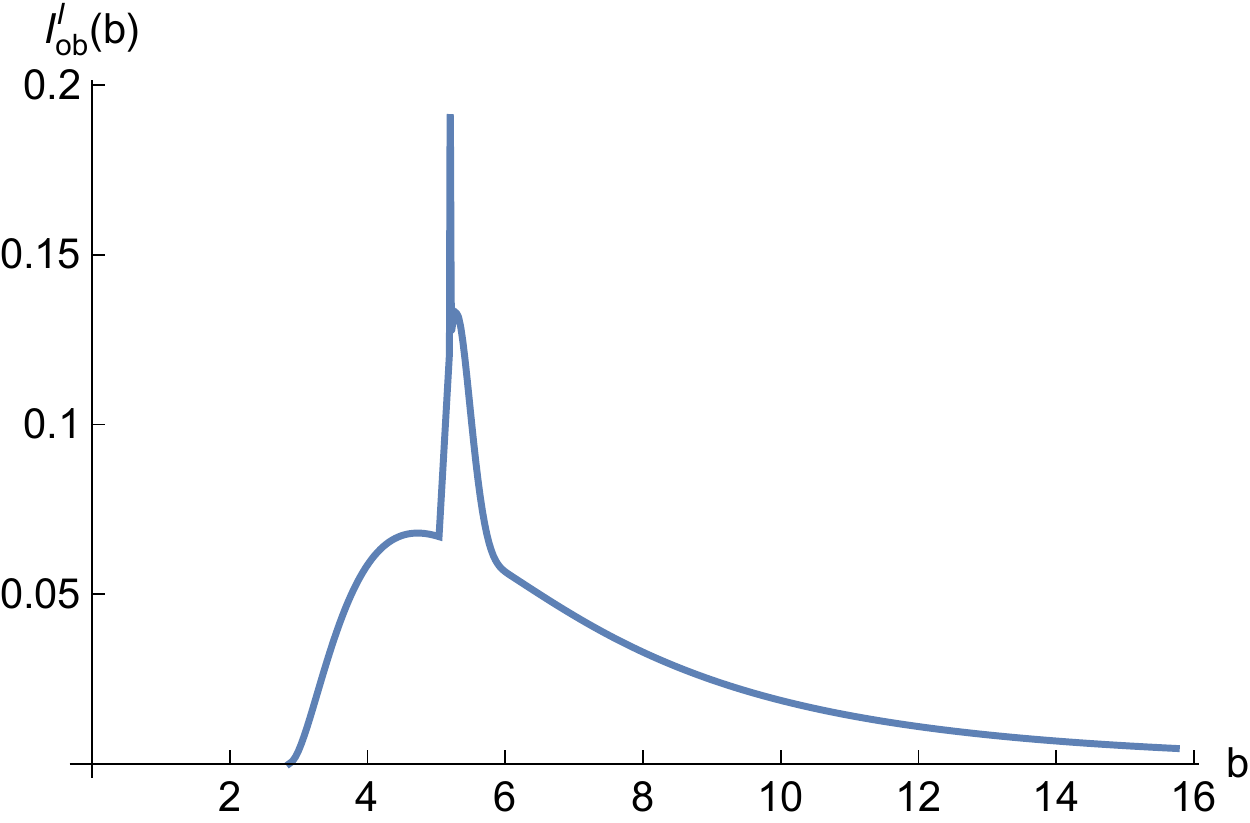}
\includegraphics[width=5.9cm,height=4.4cm]{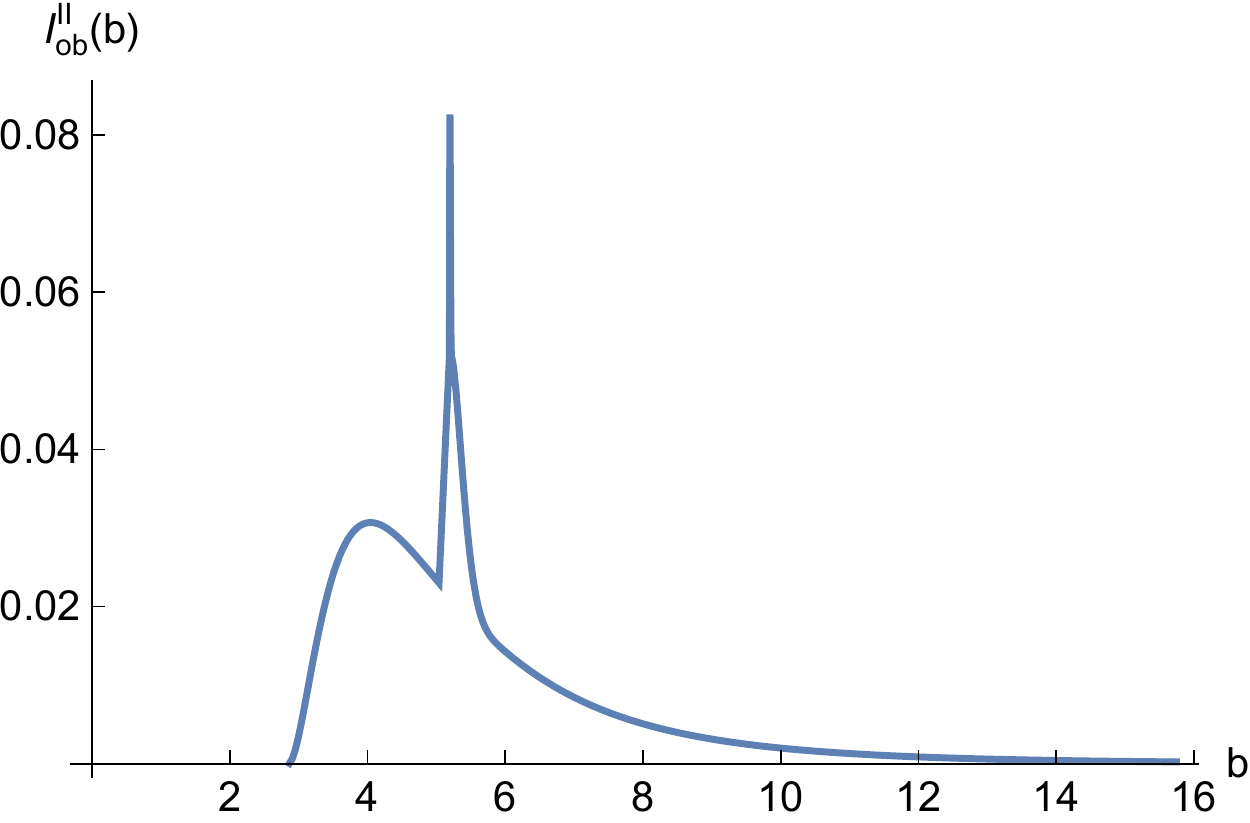}
\begin{center}
\caption{The optical appearance of a Schwarzschild black hole (top panel) in the GLM3 (\ref{eq:adGLM3}), GLM1 (\ref{eq:adGLM1}) and GLM2 (\ref{eq:adGLM2}) emission models (taking $M=1$) and the associated (normalized to their total values on each case) intensity for each of them (bottom panel), displaying the direct ($m=0$) and photon ring ($m=1,2$) emissions, respectively.}
\label{fig:Sch}
\end{center}
\end{center}
\end{figure*}

It should be stressed that the luminosity profiles GLM1/GLM2, for black hole configurations, should be restricted solely to the region outside the event horizon $r>r_h$. Since the location of the latter is different for every configuration, the peak luminosity will be different too. This way, in order to compare every configuration on as an equal-footing as possible, we normalize all emission profiles to their maximum values for each geometry, corresponding to their values at the event horizon for the GLM1/GLM2 models, and at its peak intensity for GLM3, which is the same for every configuration. Furthermore, on a second step we also normalize the data list relying the observed luminosity (\ref{eq:Iob}) to its total value (which is also different for every configuration). For future reference, we depict the application of these three models in their observed profiles under the conditions stated above, for the Schwarzschild geometry with $M=1$ in Fig. \ref{fig:Sch}. As expected, the optical appearance of the same model illuminated by different emission profiles may be very different, particularly at the level of the distribution of the luminosity over the direct and photon ring contributions, and the visibility of the latter (one ring clearly seen in the GLM3 model, but not so in the GLM1/GLM2 models).

\subsection{Generation of images}

\begin{figure*}[t!]
\begin{center}
\includegraphics[width=5.9cm,height=4.4cm]{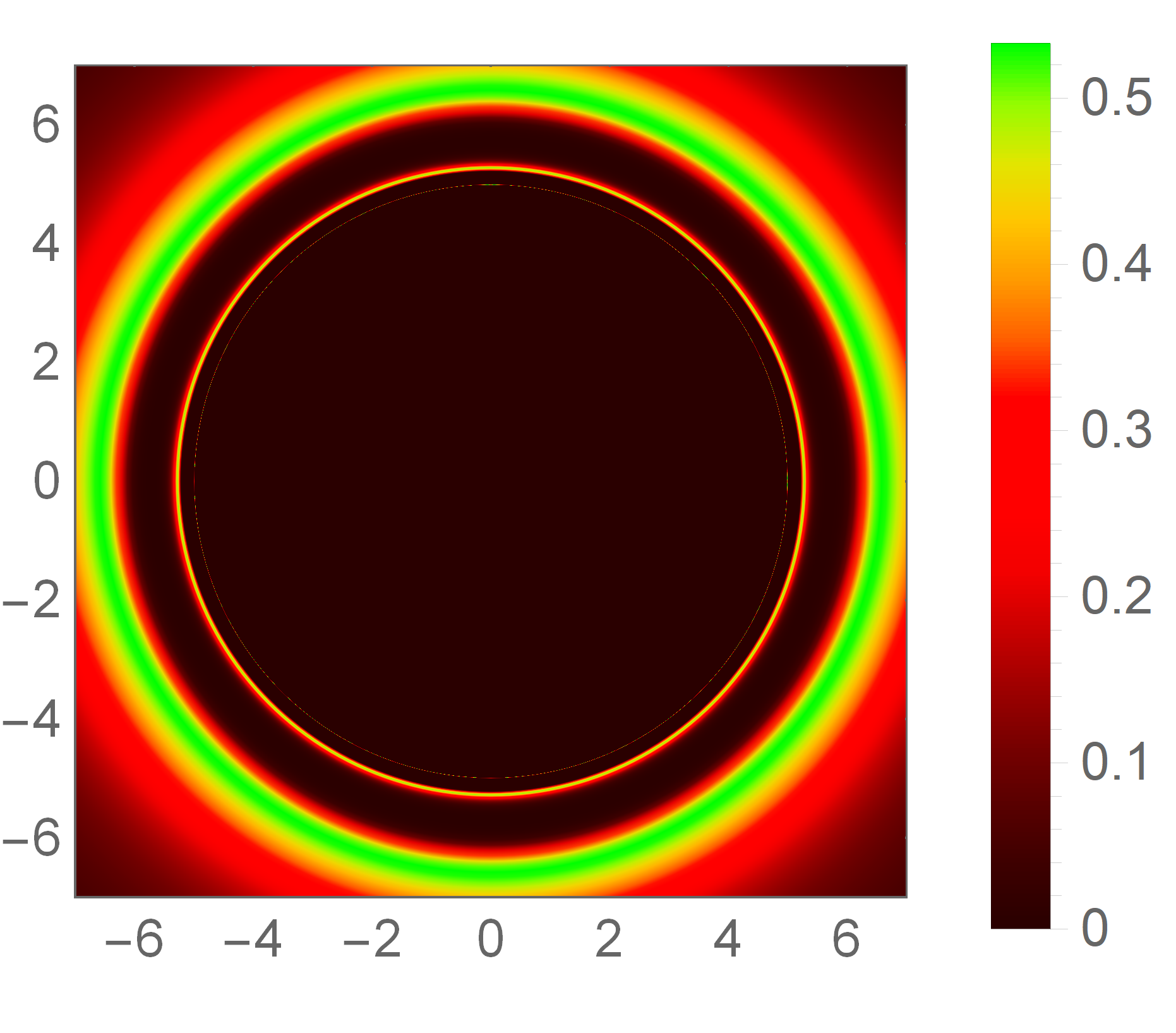}
\includegraphics[width=5.9cm,height=4.4cm]{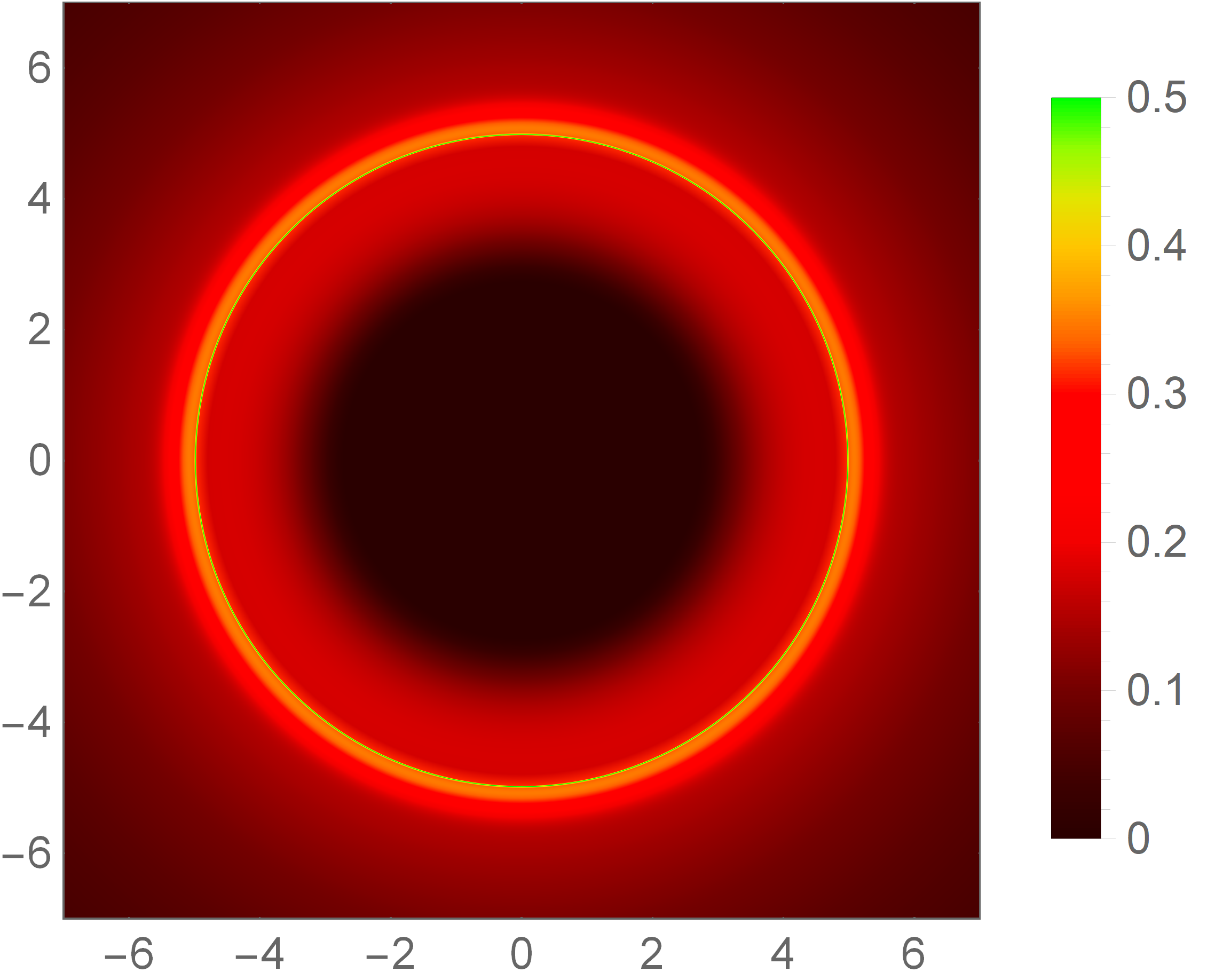}
\includegraphics[width=5.9cm,height=4.4cm]{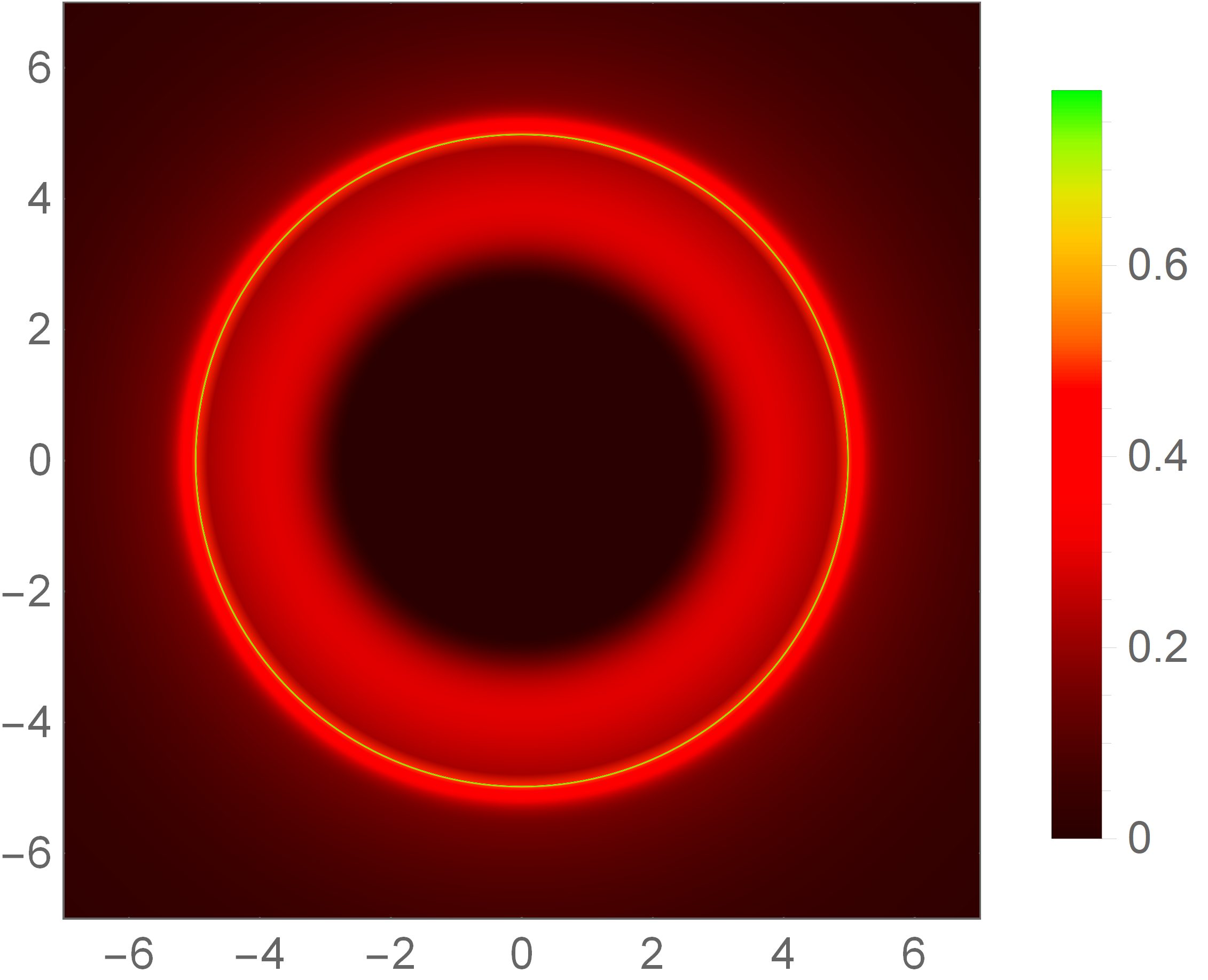}
\includegraphics[width=5.9cm,height=4.4cm]{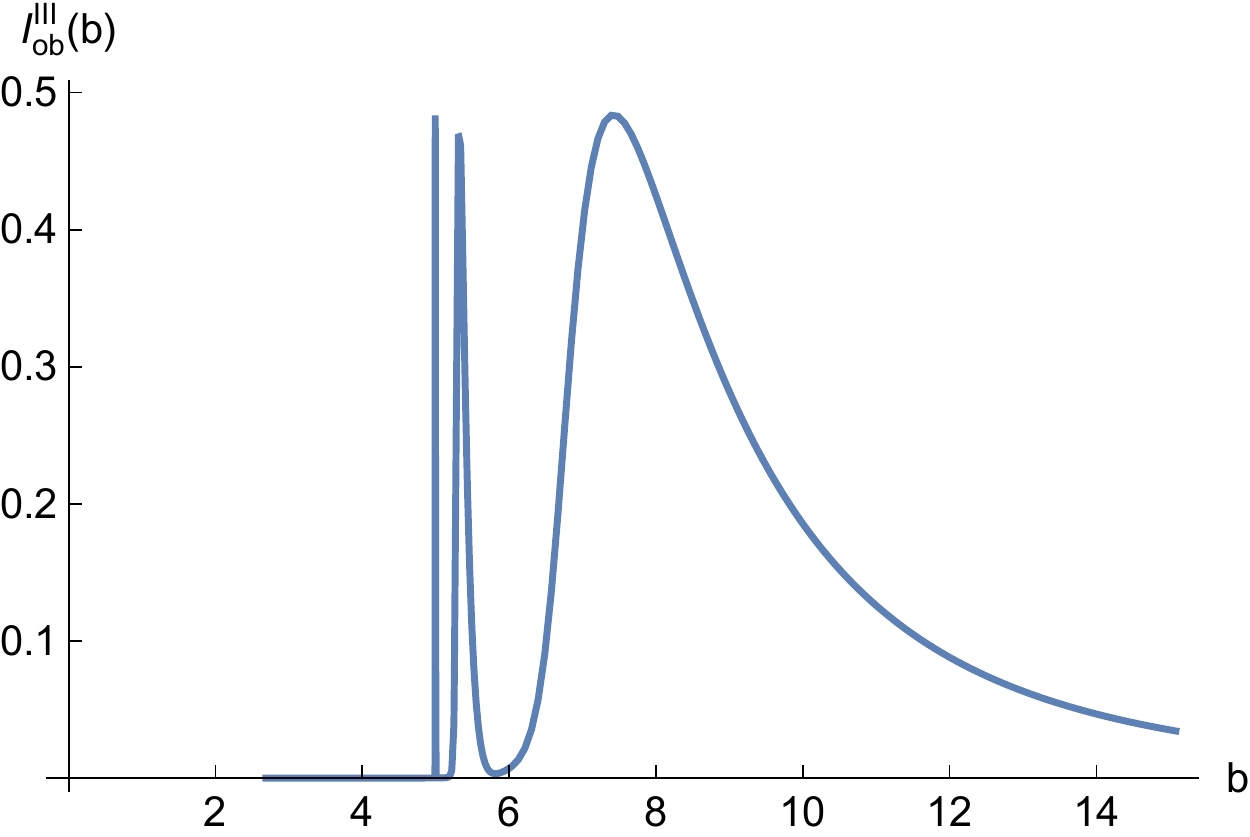}
\includegraphics[width=5.9cm,height=4.4cm]{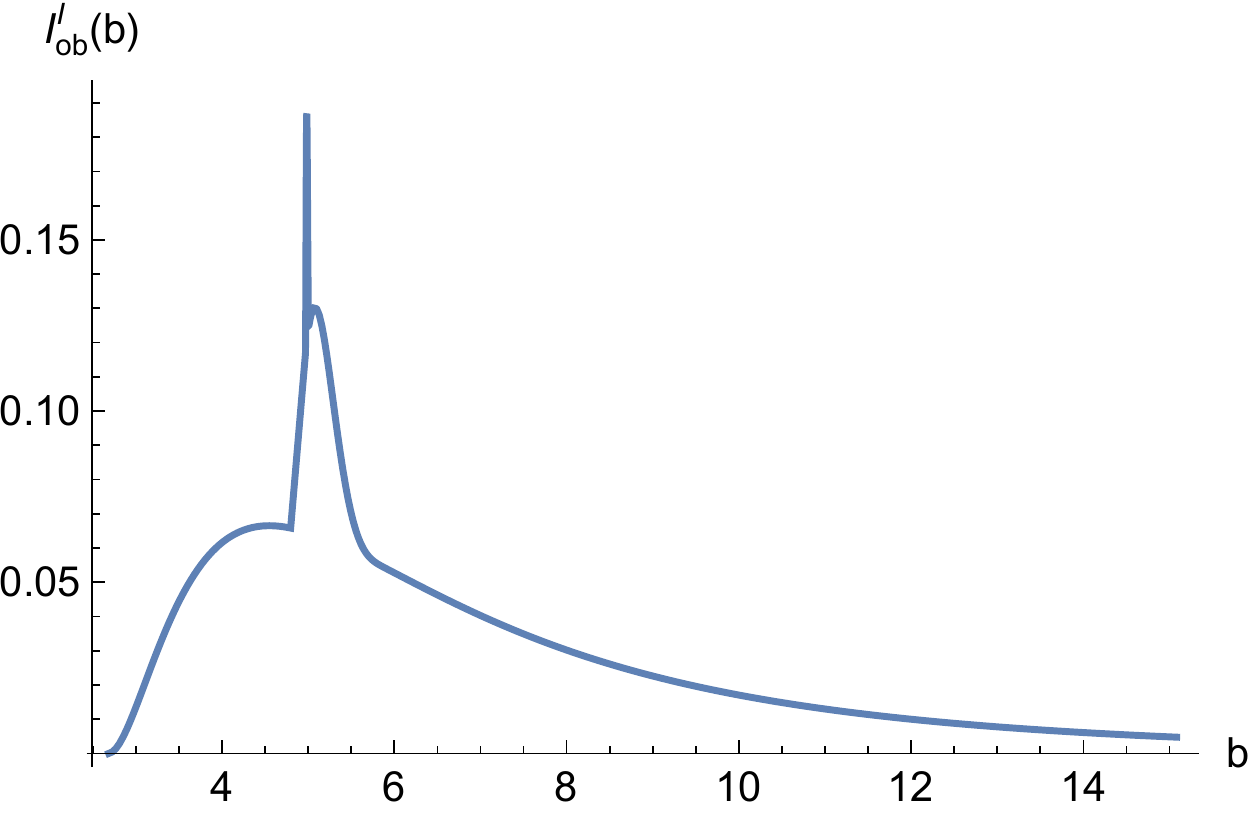}
\includegraphics[width=5.9cm,height=4.4cm]{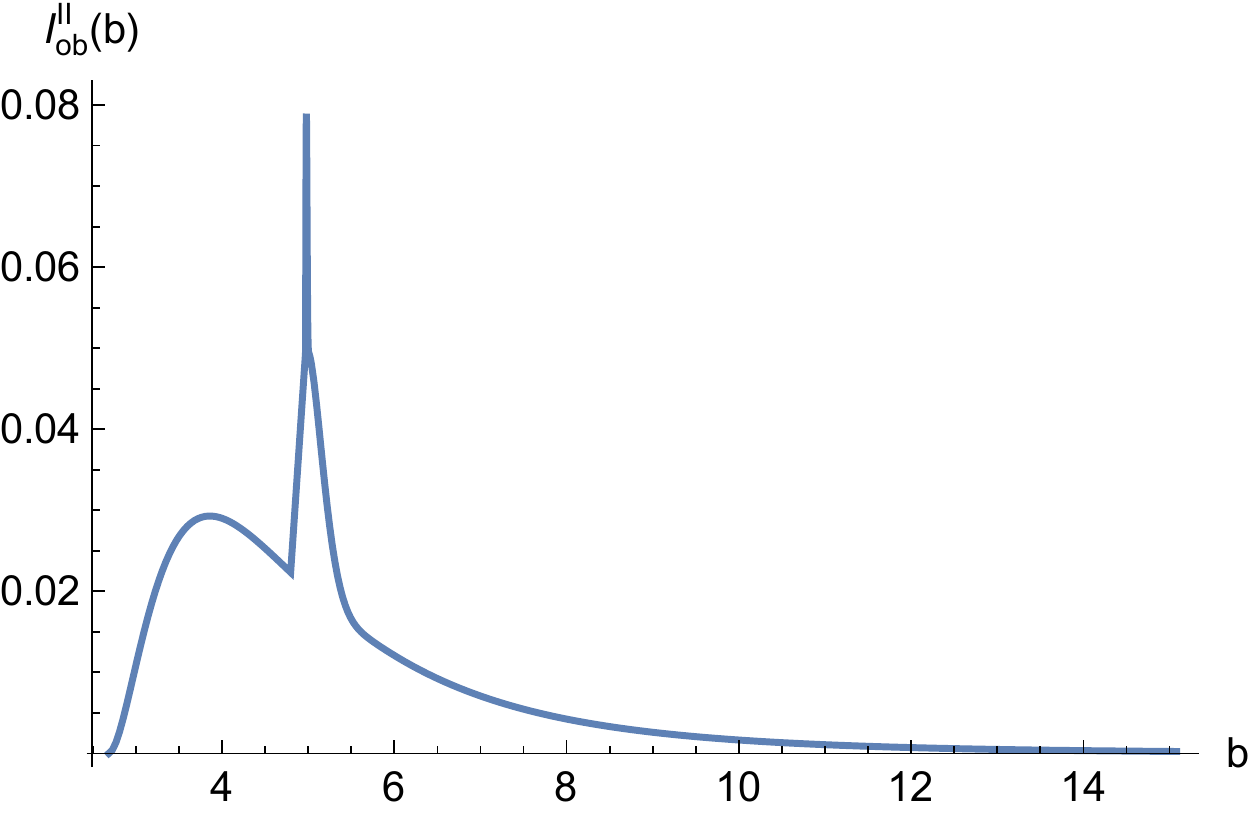}
\caption{The optical appearance of EiBI Sch configurations (top panel)  in the GLM3 (\ref{eq:adGLM3}), GLM1 (\ref{eq:adGLM1}) and GLM2 (\ref{eq:adGLM2}) emission models, and the associated intensity for each of them (bottom panel), displaying the direct ($m=0$) and photon ring ($m=1,2$) emissions,  respectively.}
\label{fig:PSch}
\end{center} 
\begin{center}
\includegraphics[width=5.9cm,height=4.4cm]{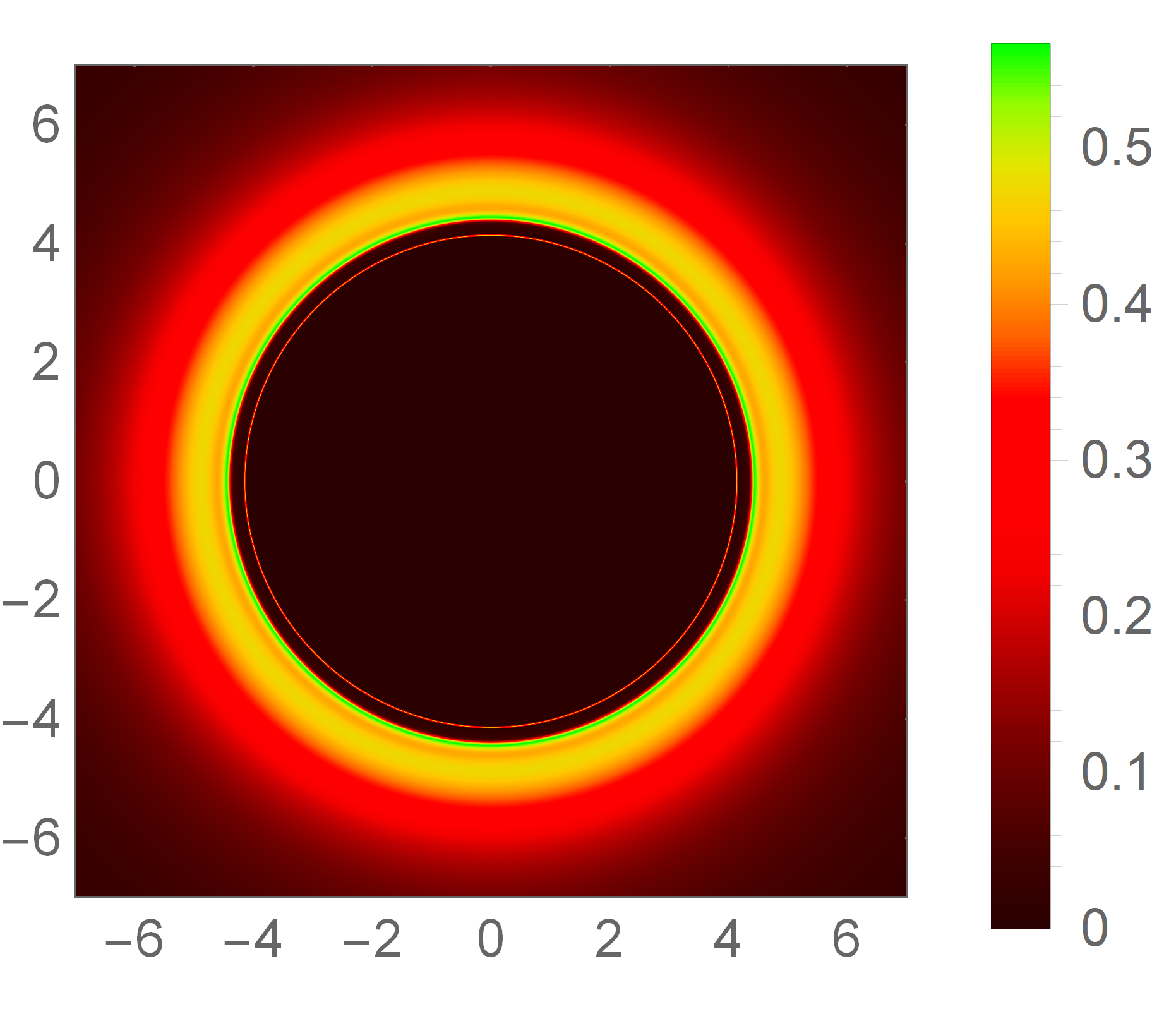}
\includegraphics[width=5.9cm,height=4.4cm]{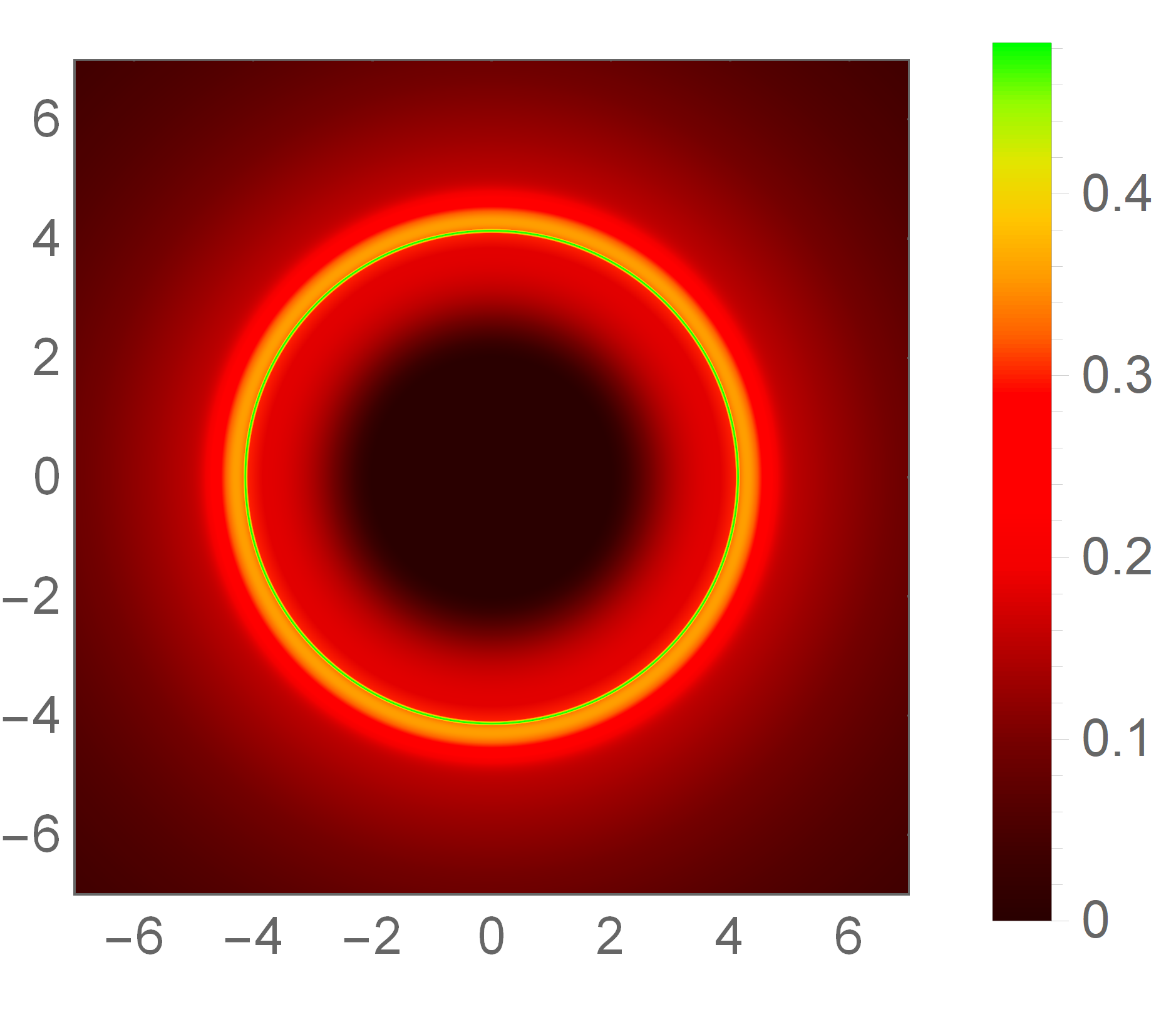}
\includegraphics[width=5.9cm,height=4.4cm]{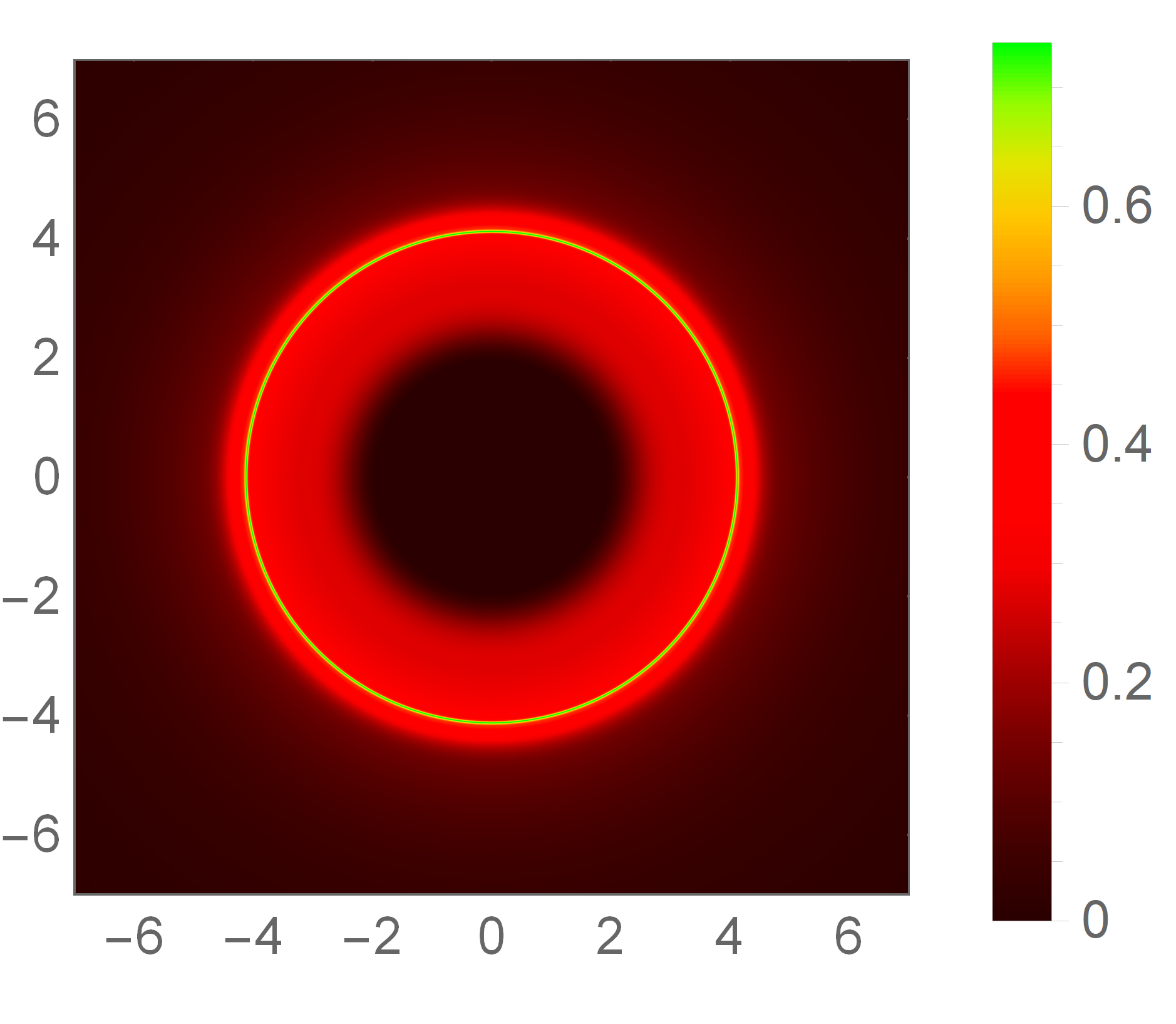}
\includegraphics[width=5.9cm,height=4.4cm]{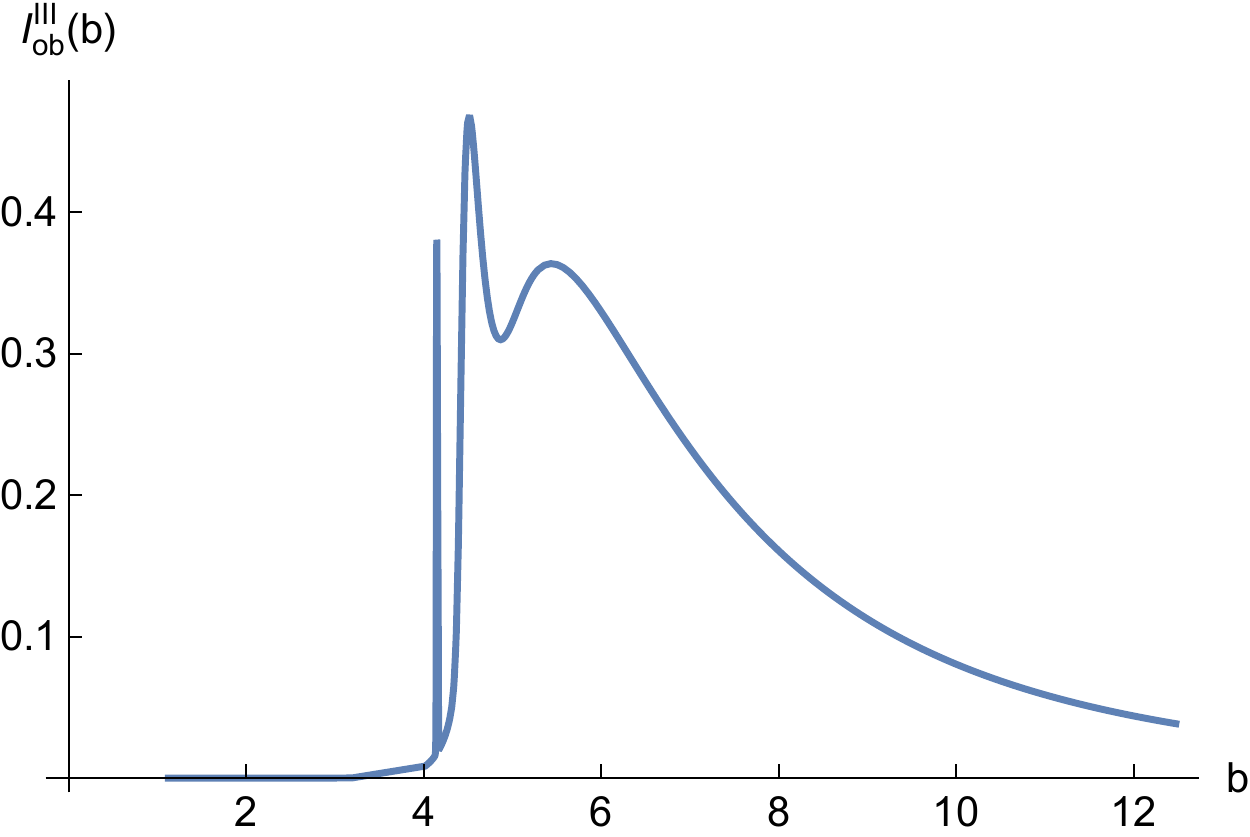}
\includegraphics[width=5.9cm,height=4.4cm]{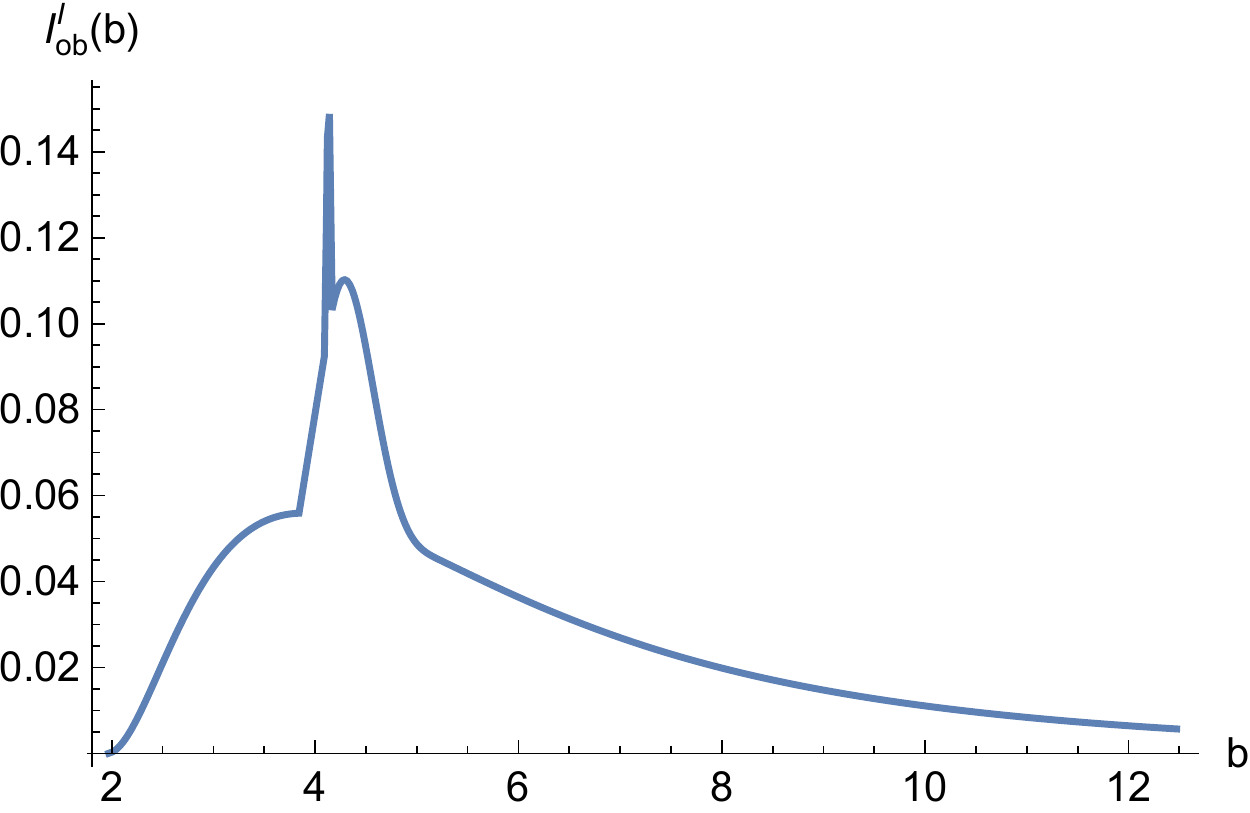}
\includegraphics[width=5.9cm,height=4.4cm]{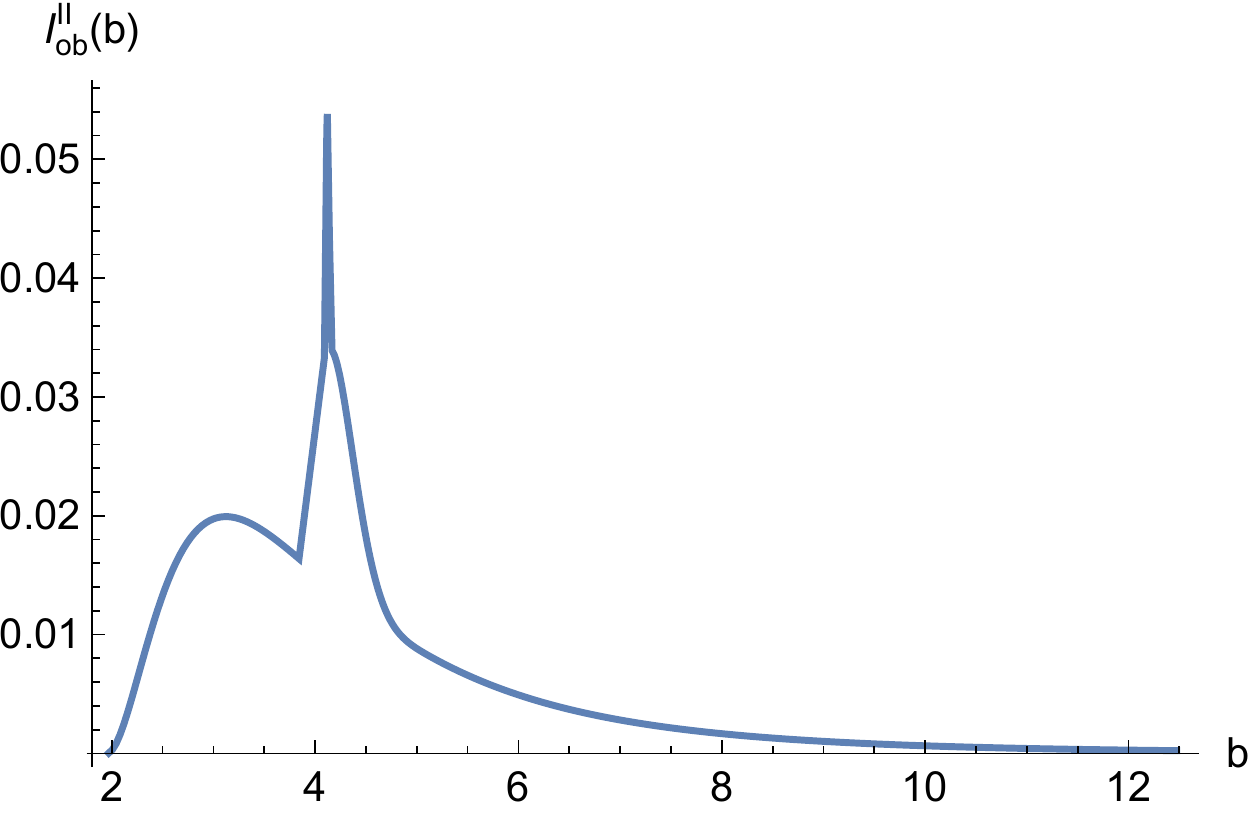}
\caption{The optical appearance of EiBI RN2 configurations (top panel)  in the GLM3 (\ref{eq:adGLM3}), GLM1 (\ref{eq:adGLM1}) and GLM2 (\ref{eq:adGLM2}) emission models, and the associated intensity for each of them (bottom panel), displaying the direct ($m=0$) and photon ring ($m=1,2$) emissions,  respectively.}
\label{fig:PRN2}
\end{center}
\end{figure*}

\begin{figure*}[t!]
\begin{center}
\includegraphics[width=5.9cm,height=4.4cm]{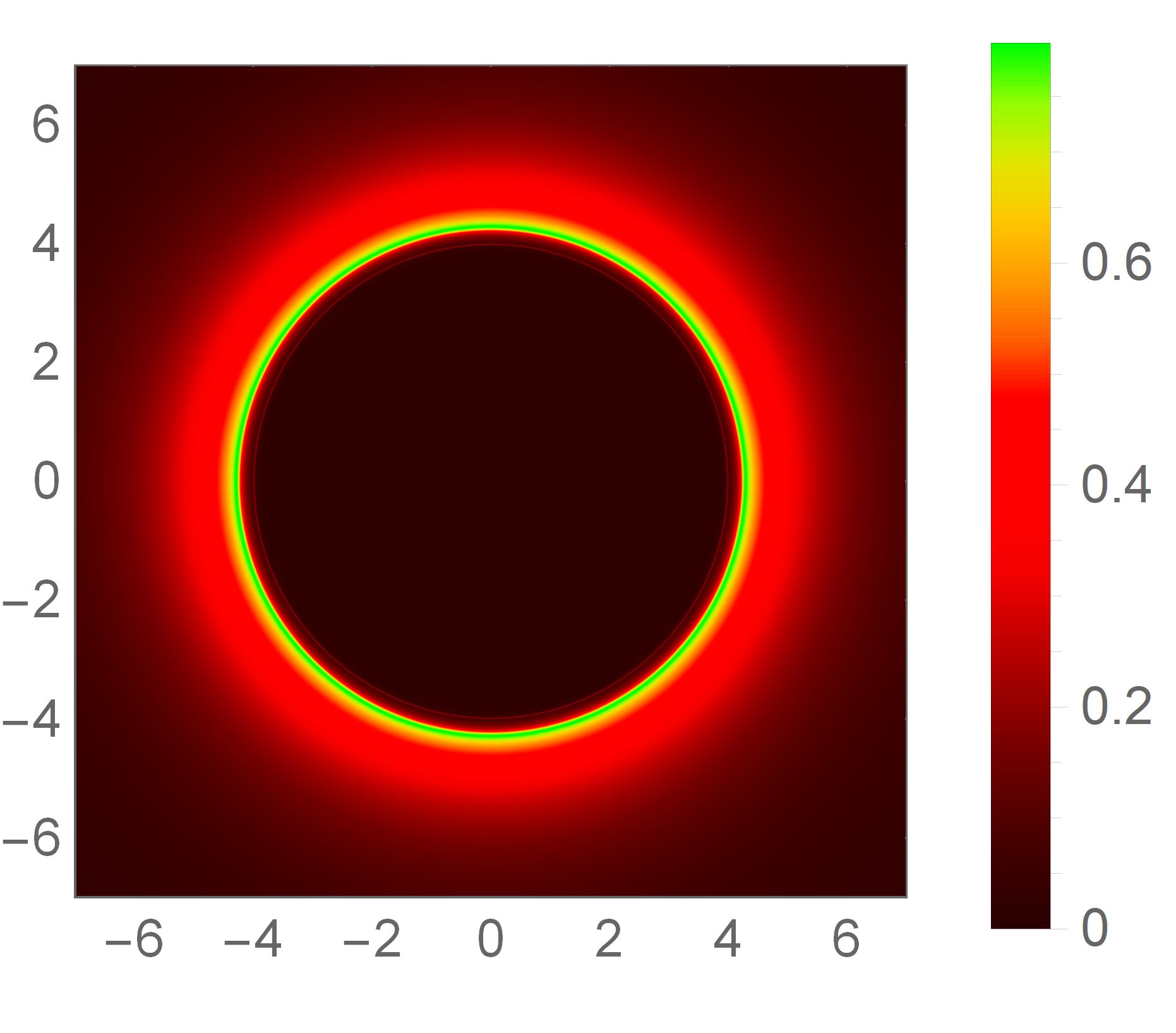}
\includegraphics[width=5.9cm,height=4.4cm]{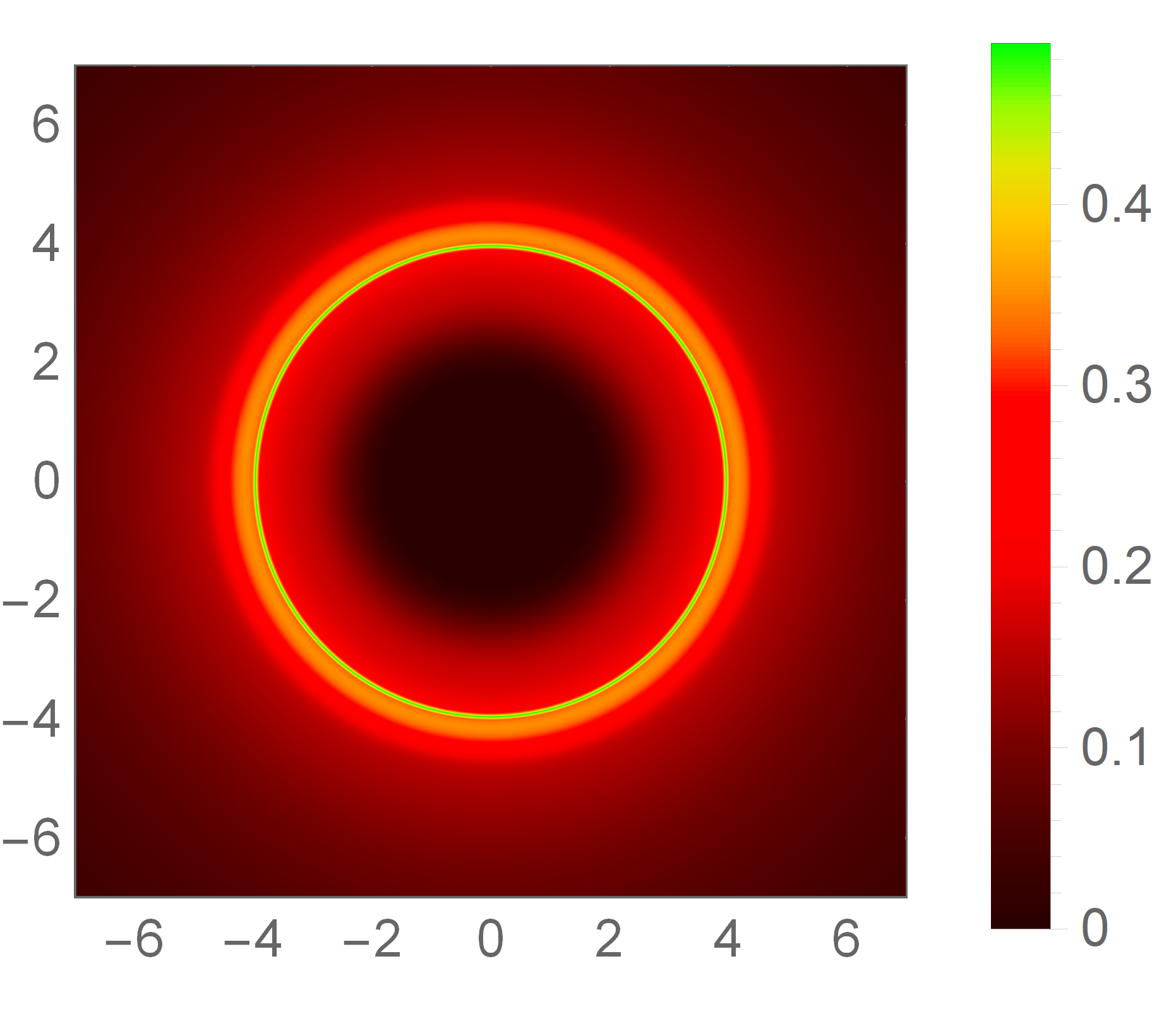}
\includegraphics[width=5.9cm,height=4.4cm]{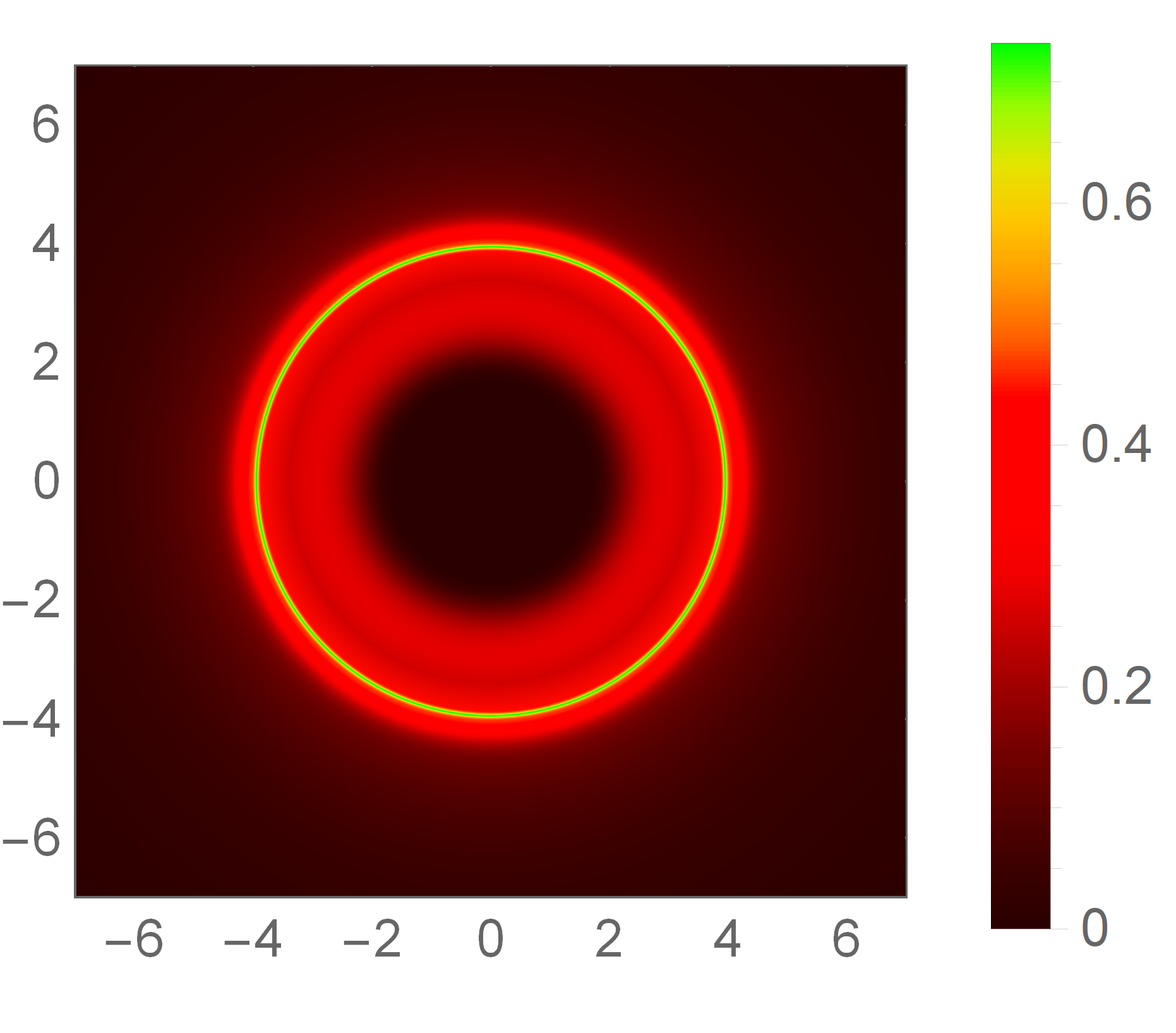}
\includegraphics[width=5.9cm,height=4.4cm]{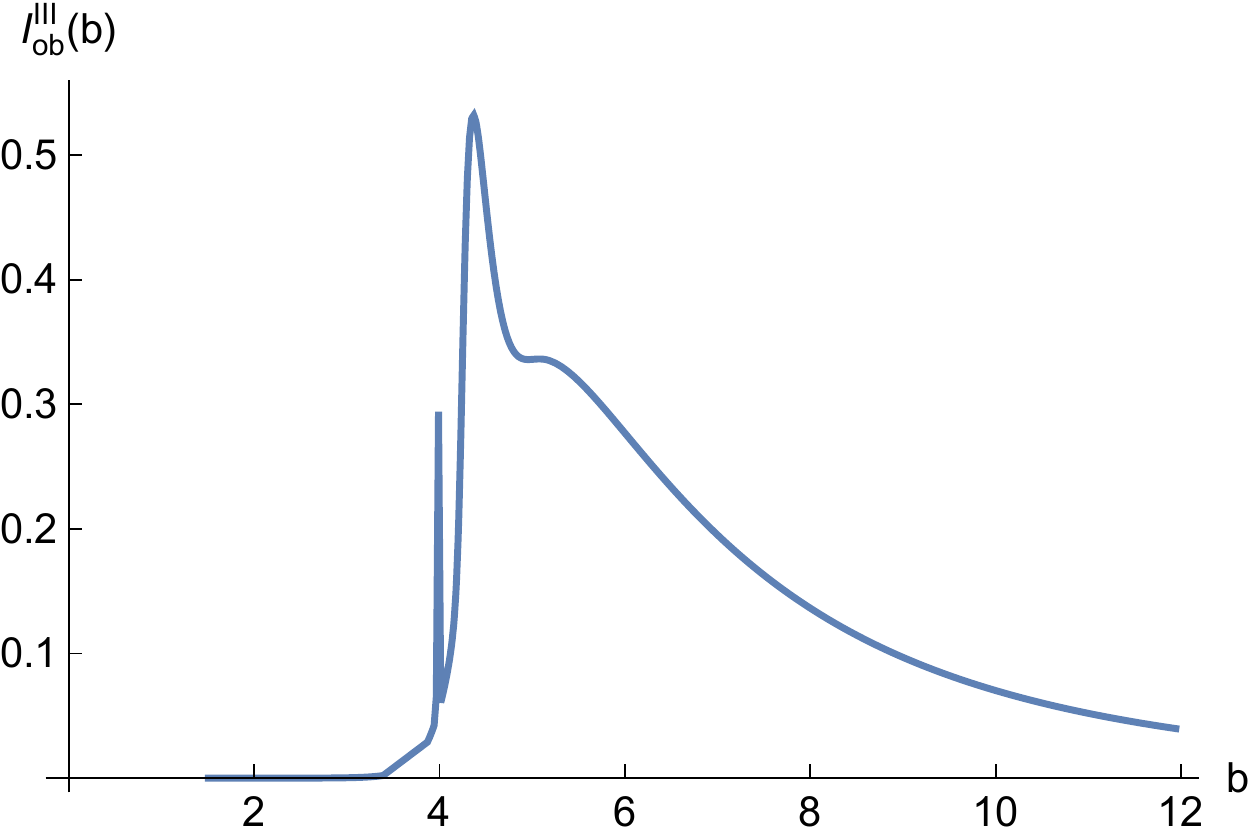}
\includegraphics[width=5.9cm,height=4.4cm]{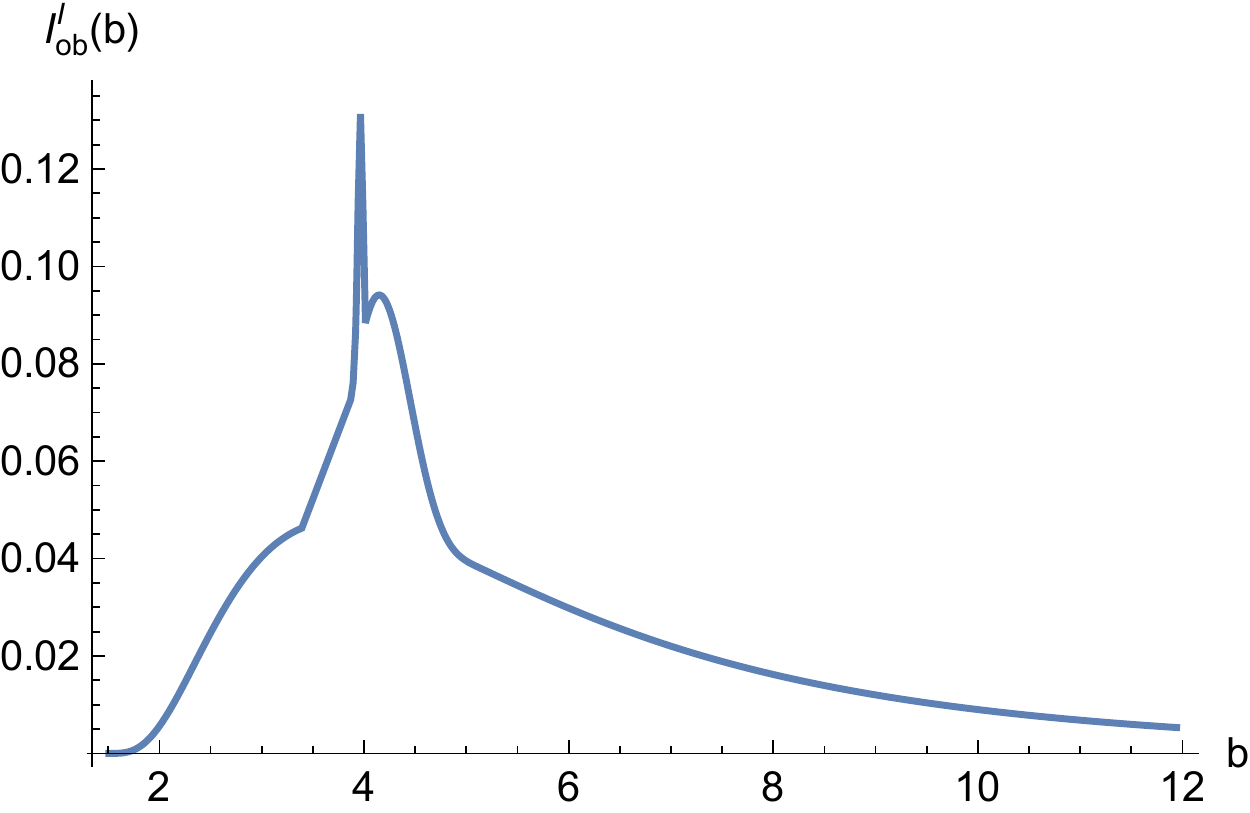}
\includegraphics[width=5.9cm,height=4.4cm]{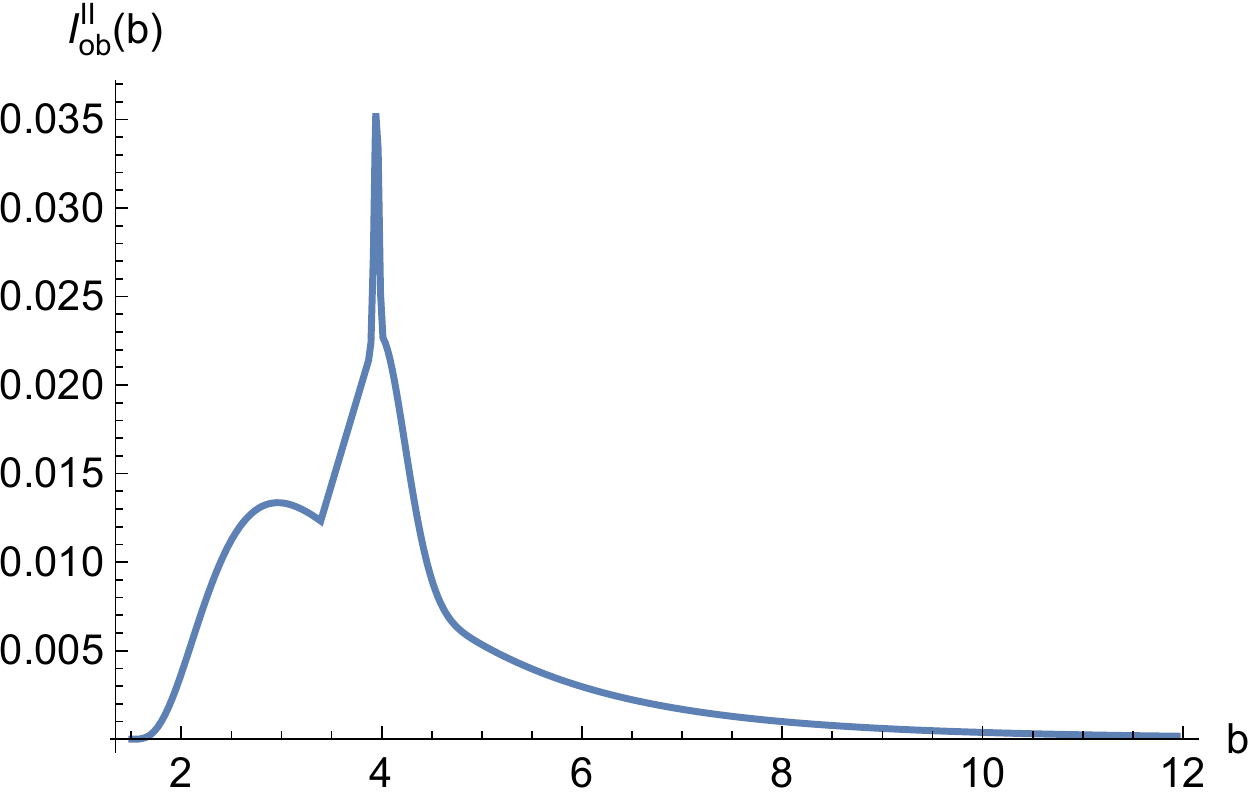}
\caption{The optical appearance of EiBI RN1 configurations (top panel)  in the GLM3 (\ref{eq:adGLM3}), GLM1 (\ref{eq:adGLM1}) and GLM2 (\ref{eq:adGLM2}) emission models, and the associated intensity for each of them (bottom panel), displaying the direct ($m=0$) and photon ring ($m=1,2$) emissions,  respectively.}
\label{fig:PRN1}
\end{center}
\end{figure*}

\begin{figure*}[t!]
\begin{center}
\includegraphics[width=5.9cm,height=4.4cm]{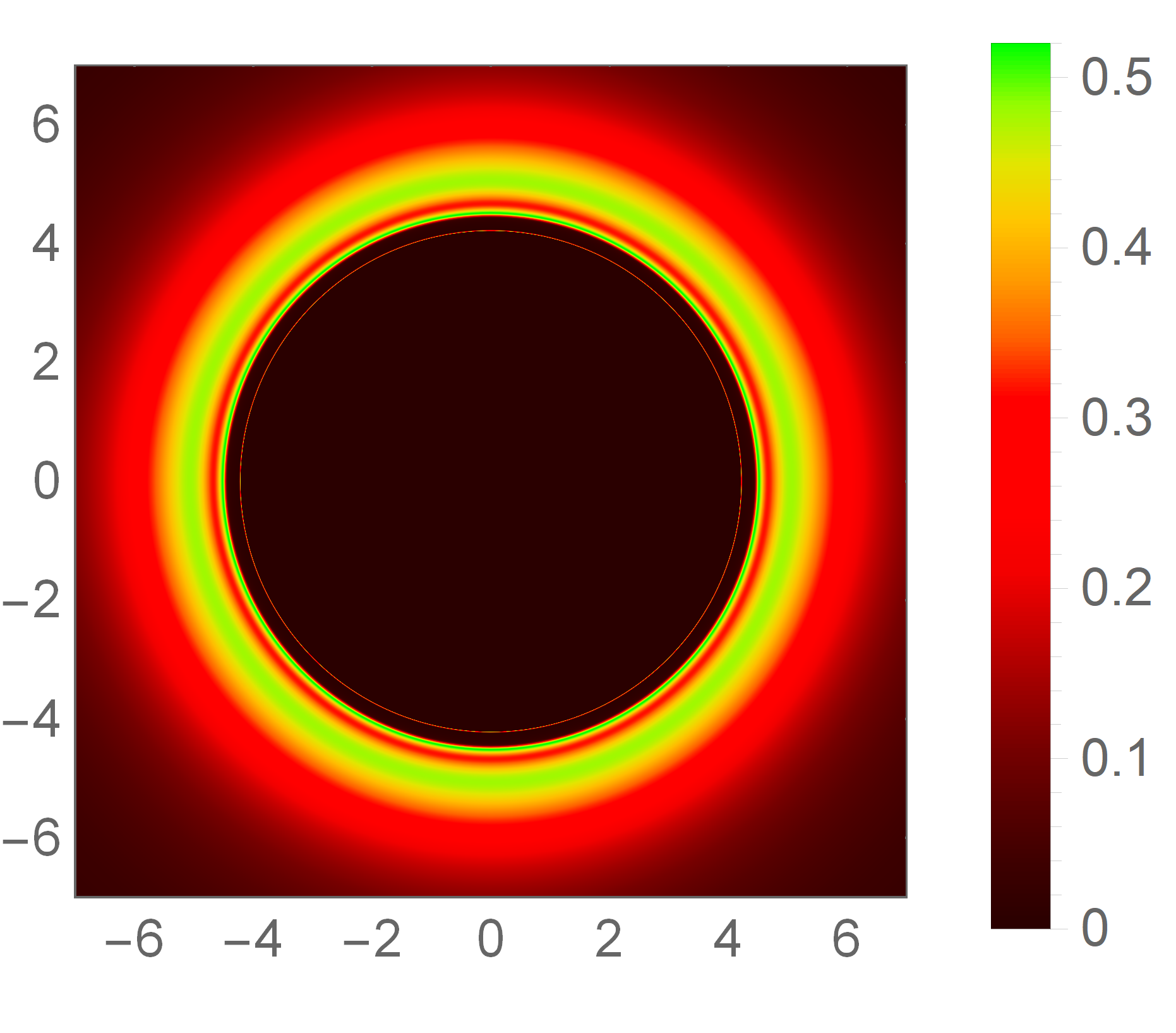}
\includegraphics[width=5.9cm,height=4.4cm]{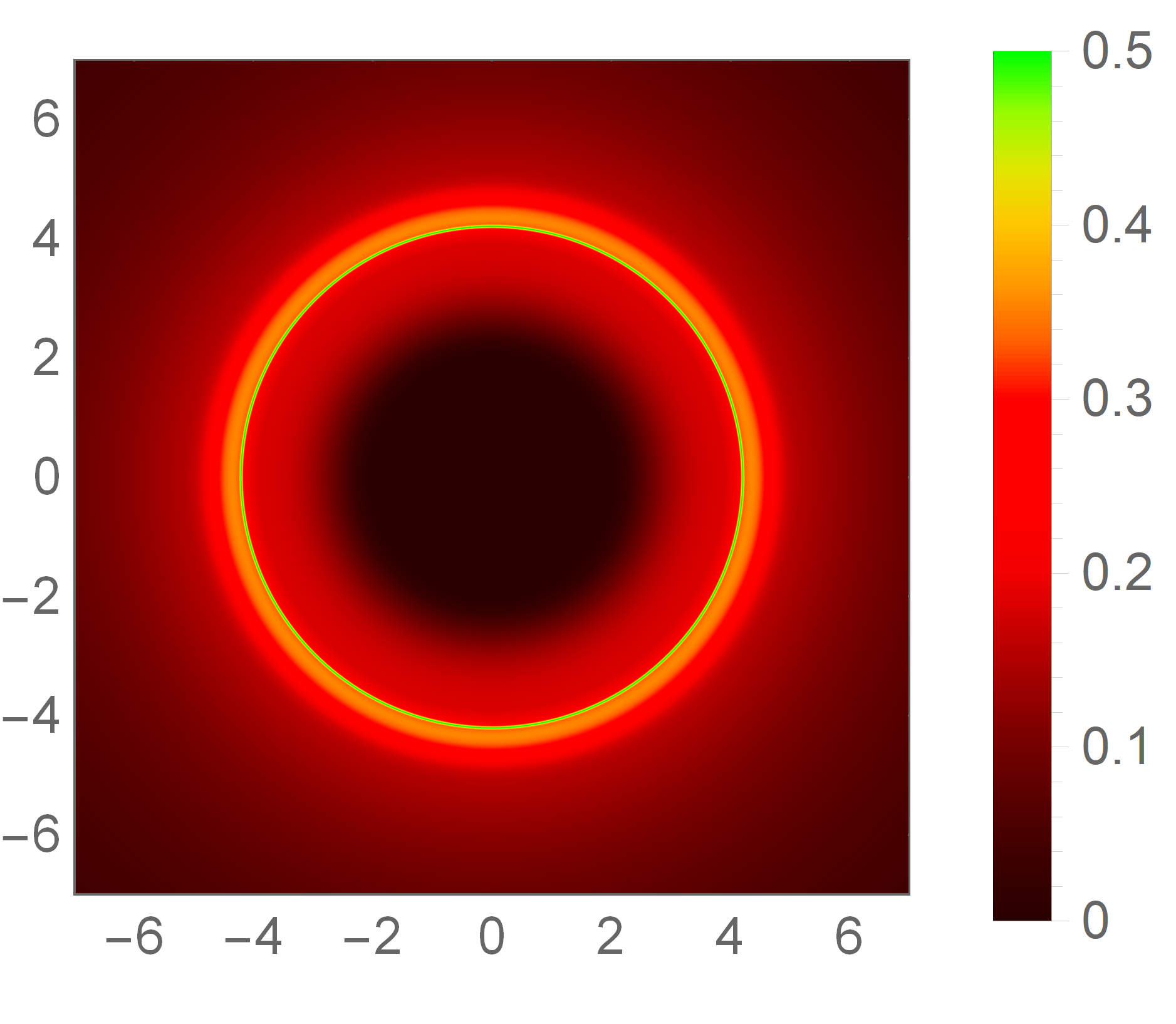}
\includegraphics[width=5.9cm,height=4.4cm]{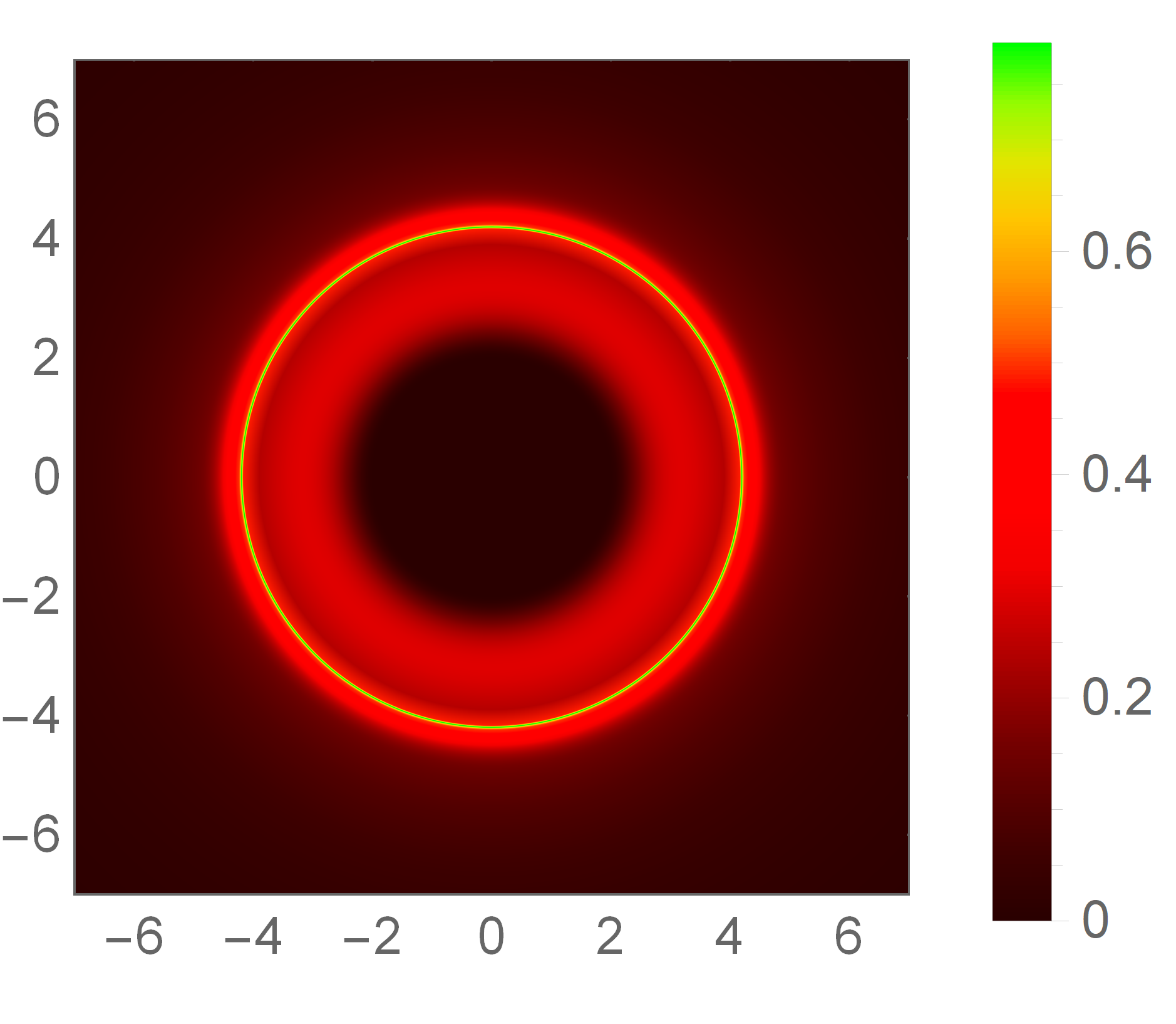}
\includegraphics[width=5.9cm,height=4.4cm]{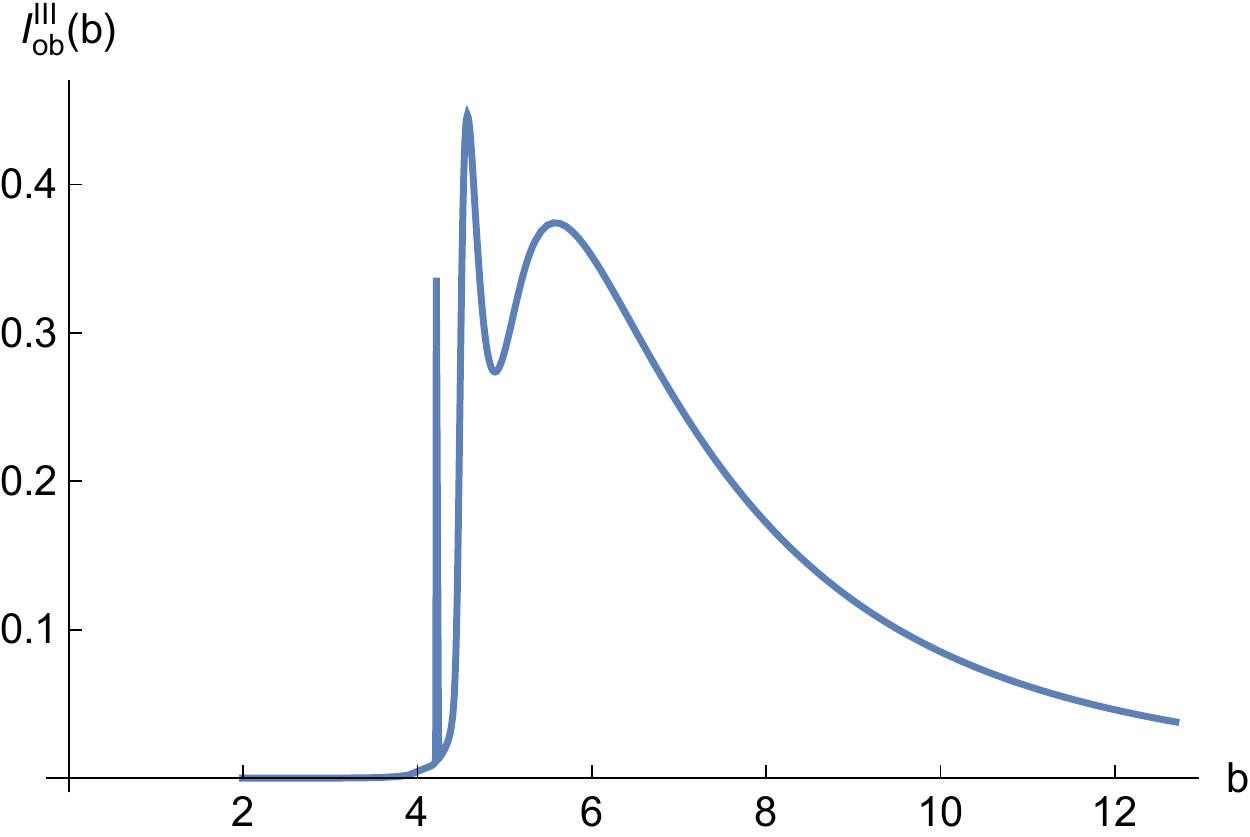}
\includegraphics[width=5.9cm,height=4.4cm]{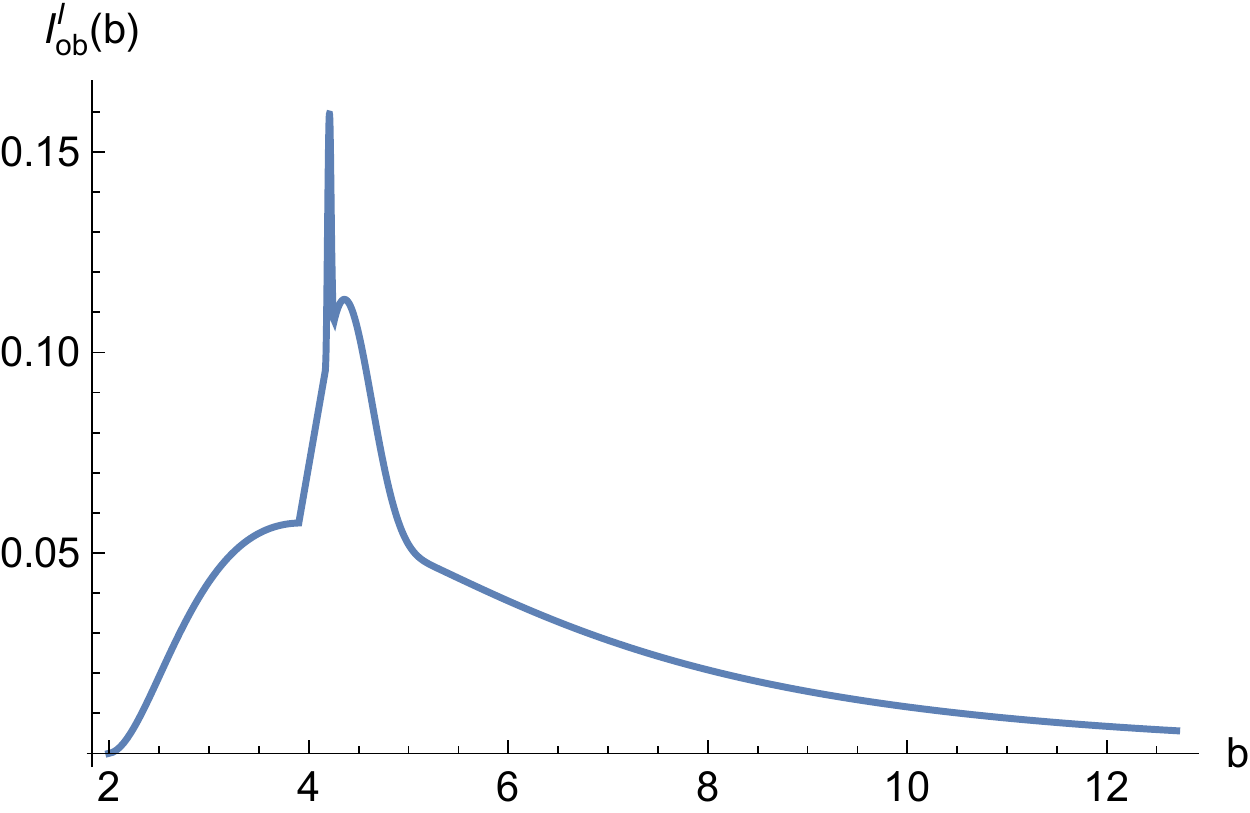}
\includegraphics[width=5.9cm,height=4.4cm]{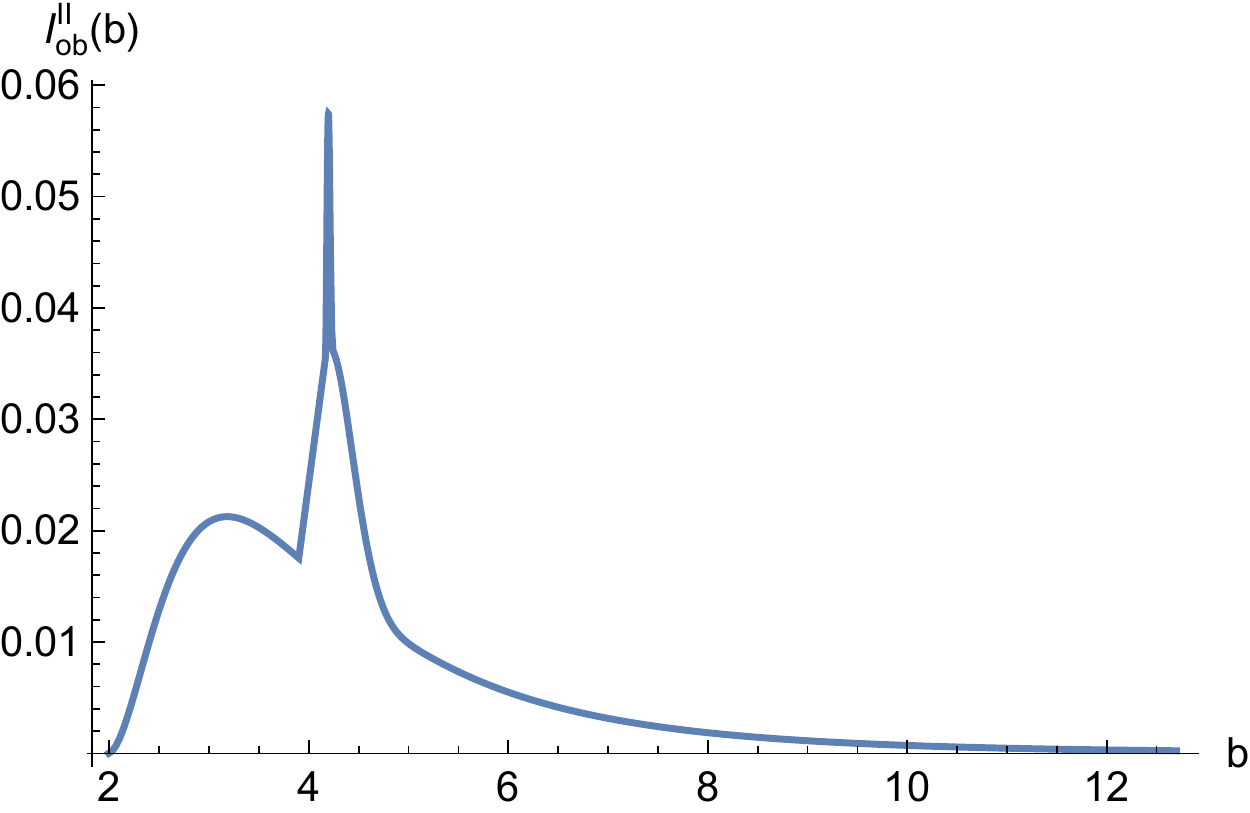}
\caption{The optical appearance of EiBI FC1 configurations (top panel)  in the GLM3 (\ref{eq:adGLM3}), GLM1 (\ref{eq:adGLM1}) and GLM2 (\ref{eq:adGLM2}) emission models, and the associated intensity for each of them (bottom panel), displaying the direct ($m=0$) and photon ring ($m=1,2$) emissions,  respectively.}
\label{fig:PMk1}
\end{center}
\end{figure*}

\begin{figure*}[t!]
\begin{center}
\includegraphics[width=5.9cm,height=4.4cm]{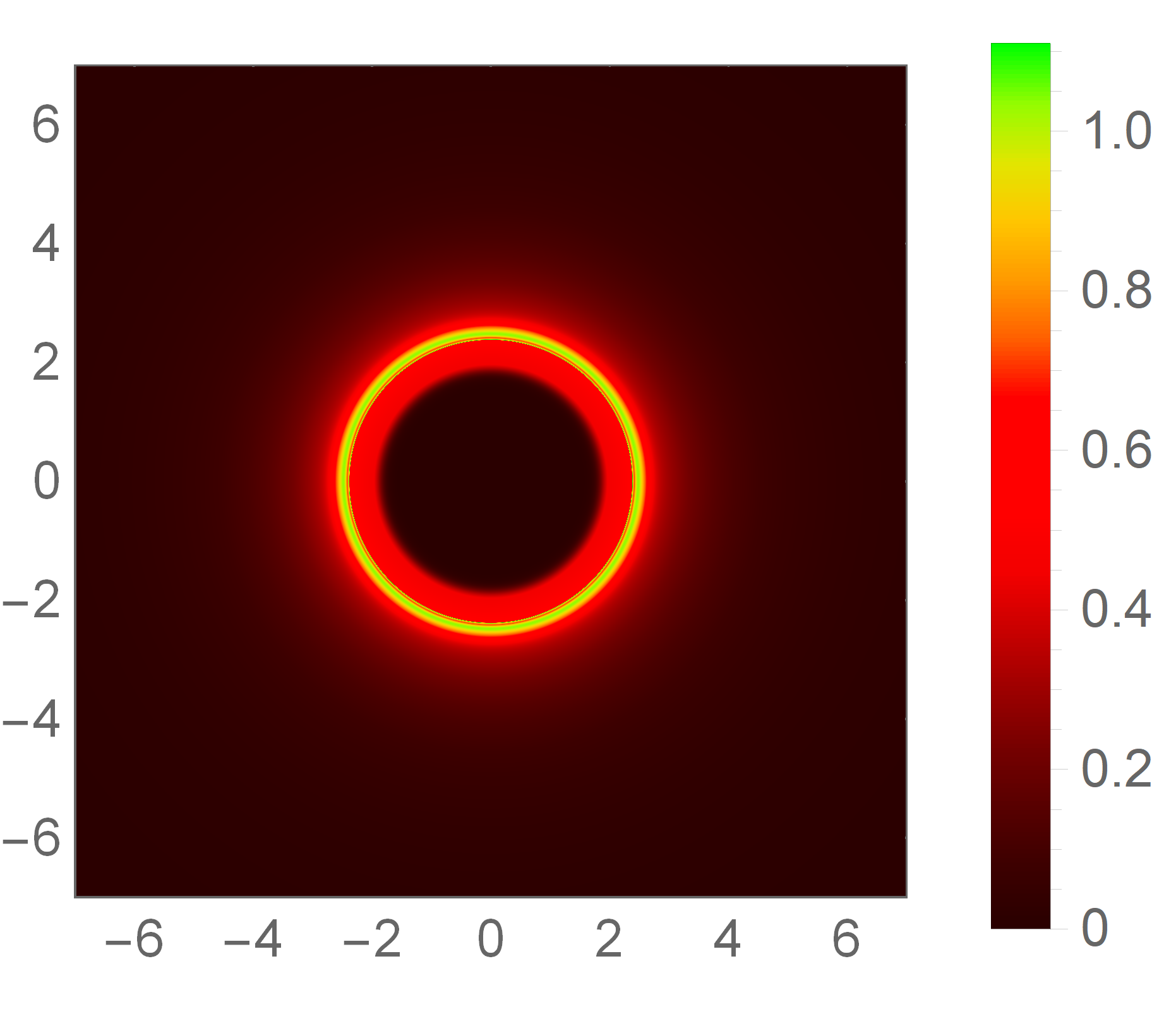}
\includegraphics[width=5.9cm,height=4.4cm]{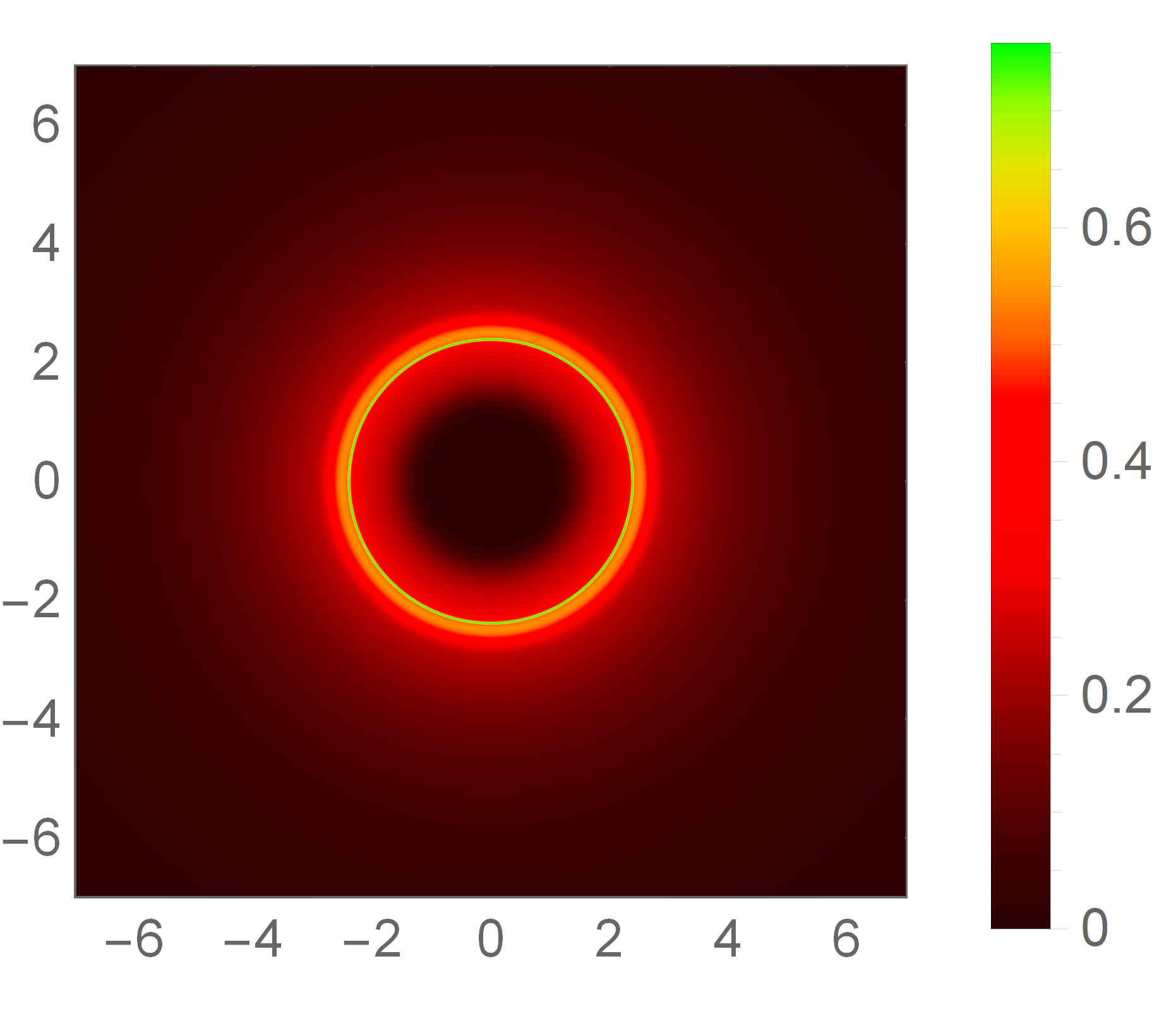}
\includegraphics[width=5.9cm,height=4.4cm]{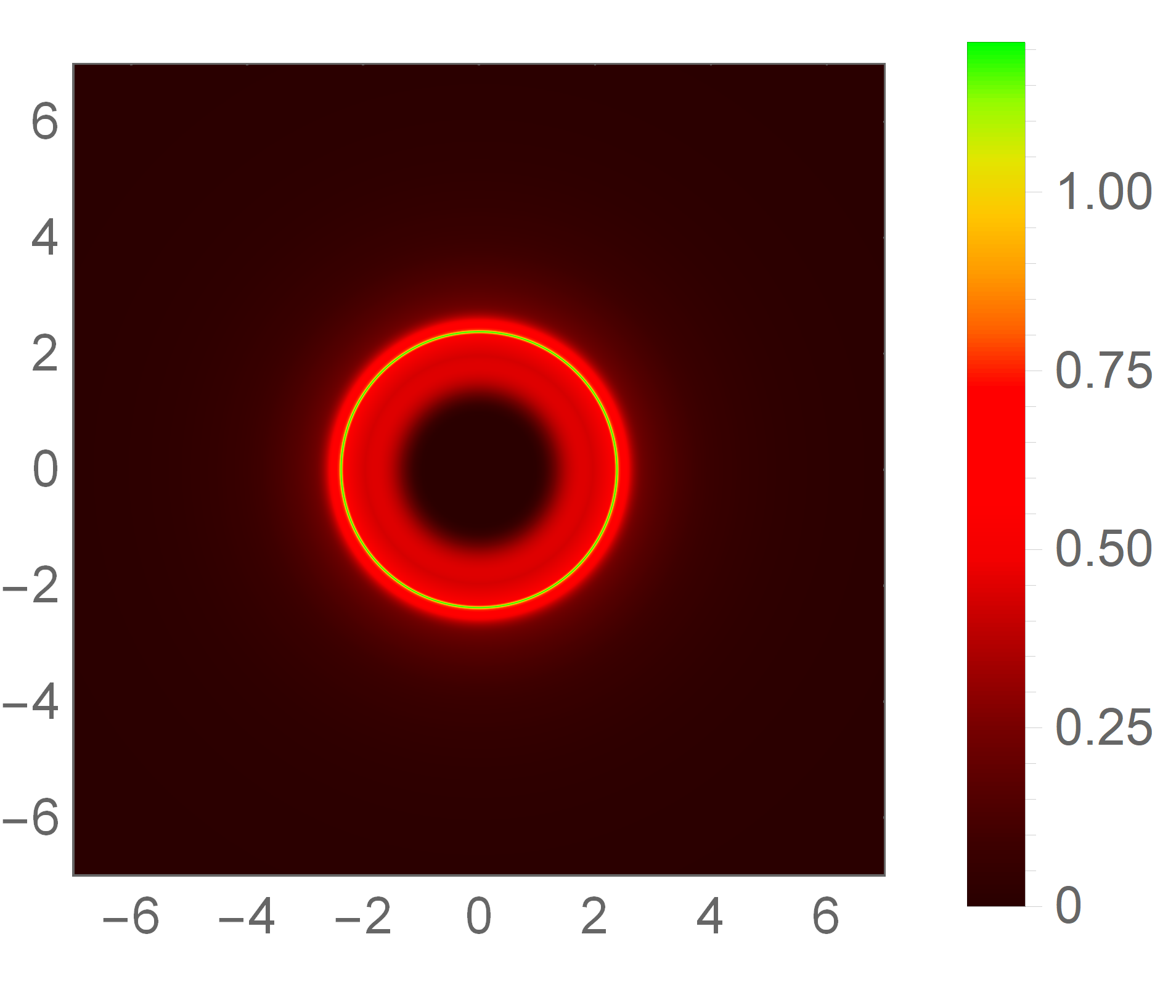}
\includegraphics[width=5.9cm,height=4.4cm]{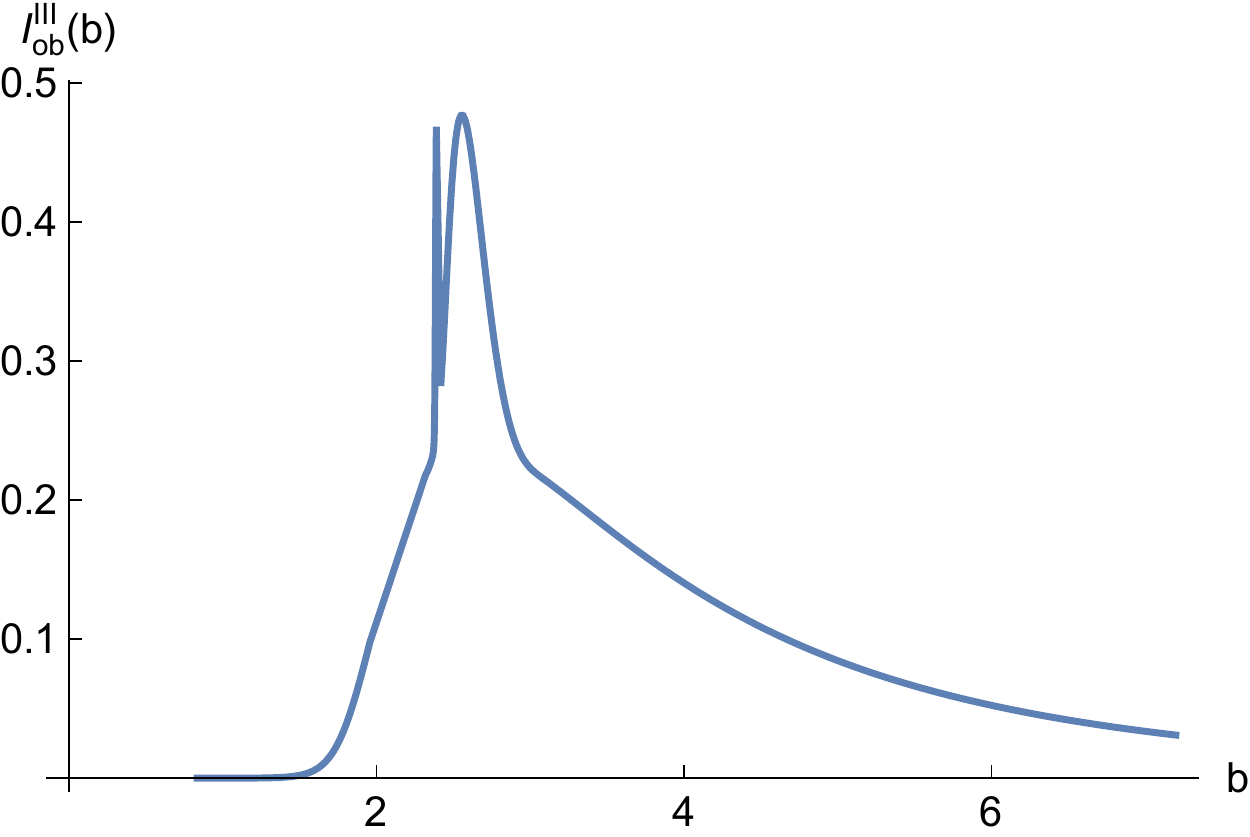}
\includegraphics[width=5.9cm,height=4.4cm]{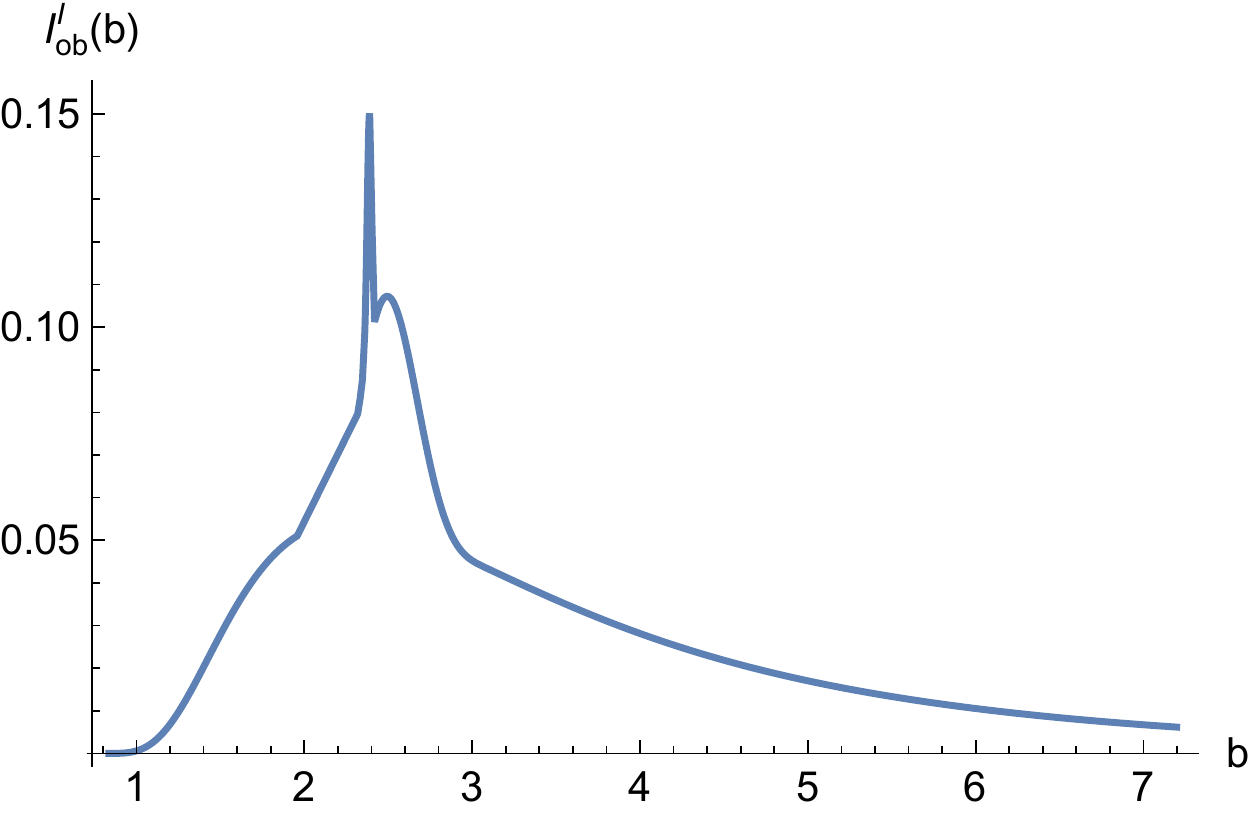}
\includegraphics[width=5.9cm,height=4.4cm]{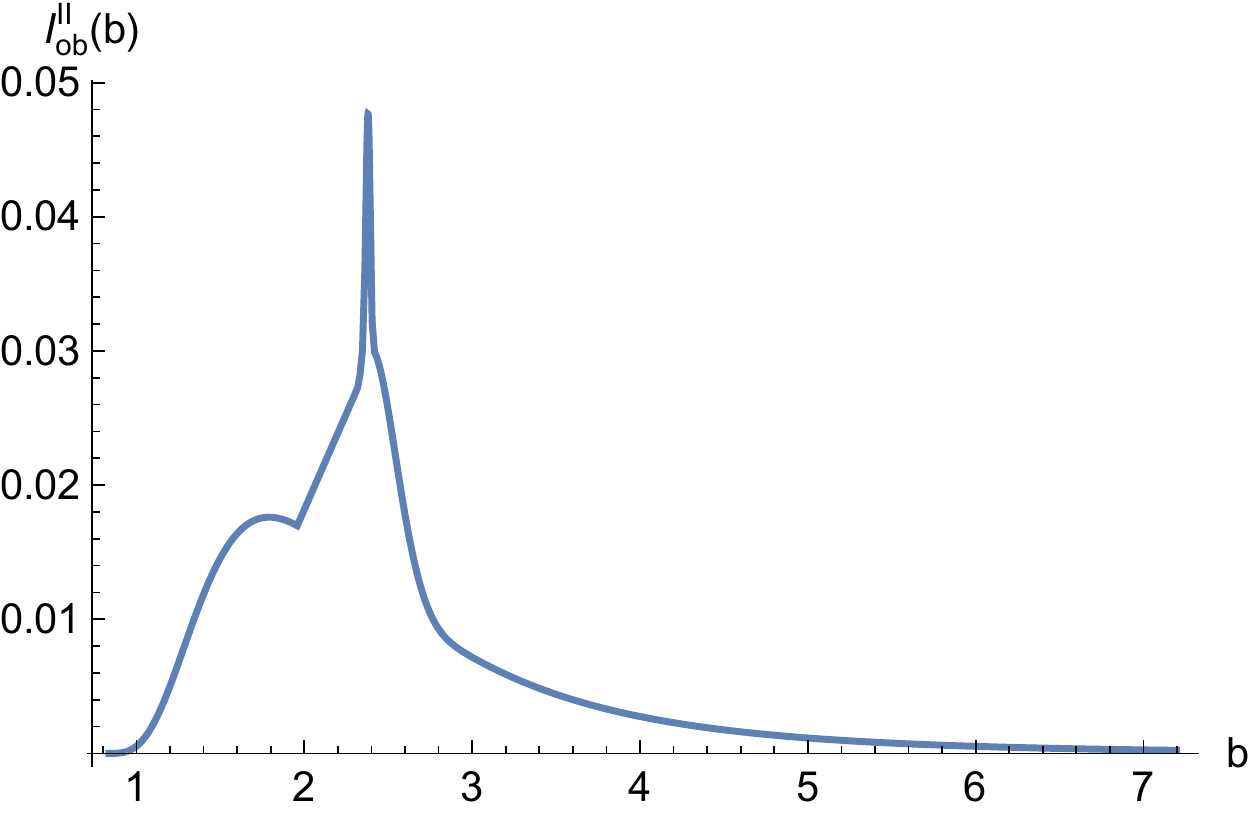}
\caption{The optical appearance of EiBI  FCt configurations (top panel)  in the GLM3 (\ref{eq:adGLM3}), GLM1 (\ref{eq:adGLM1}) and GLM2 (\ref{eq:adGLM2}) emission models, and the associated intensity for each of them (bottom panel), displaying the direct ($m=0$) and photon ring ($m=1,2$) emissions,  respectively.}
\label{fig:PMk1e}
\end{center}
\end{figure*}

Having our setting ready, we now proceed with the generation of images of our modified gravitational configurations. This is done in a two-step process: first the ray-tracing process outlined in Sec. \ref{sec:IV.A} integrates the geodesic equation
(\ref{eq:geoeqr}) to build the transfer function $r_m(b)$ into separated lists for every $m=0,1,2, \ldots$, and second this list is {\it illuminated} with the three functions of (\ref{eq:Ichoice}), each of them suitably normalized to their respective maximum of Fig. \ref{fig:disks}, to build the final distribution of luminosity as given by (\ref{eq:Iob}). The net result is a circular two-dimensional pattern of luminosity displaying the main ring of radiation together with the photon rings. In order to facilitate the comparison of images, we limit the displayed axis range of impact parameter in the images at $b \in (-7,7)$ for every configuration.

We thus systematically construct the optical appearance of the seven classes of configurations appearing in Table \ref{Table:I} based on the implementation of the three intensity profiles (\ref{eq:adGLM3}),  (\ref{eq:adGLM1}) and  (\ref{eq:adGLM2}). We proceed first with the black hole configurations: Sch, RN2, RN1, FC1 and FCt, which are depicted in the set of Figures \ref{fig:PSch}, \ref{fig:PRN2}, \ref{fig:PRN1}, \ref{fig:PMk1}, \ref{fig:PMk1e}, respectively. On the top panel we depict the (normalized) image for the (from left to right)  GLM3 (\ref{eq:adGLM3}), GLM1 (\ref{eq:adGLM1}) and GLM2 (\ref{eq:adGLM2}) emission models, while on the bottom panel we include the observed intensity profiles for the same models. There is nothing surprising in these plots as compared to our initial expectations and what we have in the Schwarzschild case of Fig. \ref{fig:Sch}: the intensity profiles of the GLM3 model display the typical lump of the direct emission, $m=0$, with its maximum located near the (gravitationally redshifted) inner edge of the emission, and that thanks to the fact that it is extended over a wide range of impact parameter values, it dominates the optical appearance of the object. Isolated (GLM3 model), or superimposed to it (GLM1/GLM2 models), there are the two narrow spikes associated to the $m=1,2$ photon rings (the higher the $m$ the inner the location in the image and the sharper the spike). The degree of overlapping between the $m=0$ and $m=1,2$ emissions depends both on the gravitational configuration and on the emission model, though the latter has a much larger influence on it. This can be seen by going left-to-right on the sequence of images for a given gravitational configuration and comparing between the different figures at fixed GLM model.

The image of each object is also affected by this degree of overlapping. Varying degrees of non-overlapping are maximized in the GLM3 model (left plots): this is clear in Fig. \ref{fig:PSch} for the Sch configuration, which is the most conservative extension of the usual GR-Schwarzschild/RN solution for these EiBI configurations, and which splits the three emissions into their individual constituents. This manifests in the image of the object as a wide ring of radiation enclosing two photon rings (the innermost of it, corresponding to $m=2$, barely visible at naked eye). The central brightness depression in this case thus fills all the central part of the image up to the $m=2$ photon ring (since we are dismissing as negligible the contributions of higher-order photon rings to the total luminosity, see however Section \ref{Sec:hopr}.). For other EiBI configurations, the two photon rings are still mostly visible, but the direct emission is overlapped with the $m=1$ ring or both, to some extend. Now, as we move forward towards the GLM1/GLM2 models, the direct emission completely dominates the image in all the range of impact parameter values, having inserting on it the two photon rings, and which manifest as the formation of a single spike in the GLM2 model, which is accompanied by a little bump at a slightly larger impact parameter value in the GLM1 model. This provides a way of distinguishing the GLM1/GLM2 models: in the latter the $m=1,2$ emissions blend to yield a single photon ring, while in the former there are two rings superimposed on top of the direct emission.

On the other hand, in these GLM1/GLM2 models the size of the shadow is strongly reduced to its inner counterpart in such a way that the non-zero brightness region lies well inside the critical curve. This size is essentially a gravitationally lensed image of the event horizon and, therefore, it has little to do with the GLM1/GML2 models but it is instead heavily associated to the gravitational configuration setting the location of the event horizon. This effect is also present in those EiBI configurations which qualitatively deviate heavily from the Schwarzschild one, as in the FCt of Fig. \ref{fig:PMk1e}, where we see some radiation leaking off the region inner to the photon ring. In all cases, the photon ring(s) appear in the optical appearance of the black holes as a boost of luminosity between the outer edge of the direct emission  and the inner shadow. Next we head on this very direction to track this question in more detail.

\subsection{Lyapunov exponents and photon rings relative luminosities}

In terms of luminosities, for GLM2 models the fact that the $m=1,2$ emissions blend into a single photon ring provides a non-negligible boost in the maximum luminosity of this hybrid photon ring. Nonetheless, our code allows us to track individually the properties of each of these individual photon rings: locations, impact parameters and, more importantly to our purposes here, their corresponding luminosities as compared to their RN counterparts and also to that of the direct emission itself.  A marker of this relevant feature is provided by the so-called Lyapunov exponents, which provide a measure of the instability scale for trajectories slightly displaced from the bound orbit, that is, $r=r_{ps}+\delta r_0$ with $\delta r_0 \ll r_{ps}$. This means that, in a spherically symmetric space-time of the type considered here, for a photon starting its trip at a certain $\delta r_0$ away from the critical curve, after a certain coordinate time $t$ has passed it will be located at a distance
\begin{equation}
\delta r \approx e^{\lambda t} \delta r_0
\end{equation}
where $\lambda$, measuring the growth rate of these radial perturbations in coordinate time, is dubbed in \cite{Cardoso:2014sna} as the {\it principal Lyapunov exponent}, and it can be expressed as 
\begin{equation} \label{eq:Lya}
\lambda= \sqrt{\frac{\mathcal{V}''(r)}{2\dot{t}}}
\end{equation}
which is dependent on the second derivative with respect to $r$ (indicated by a prime) of a modified potential  defined in our case as (recovering the one of GR for $\Omega_{+} \to 1$)
\begin{equation} \label{eq:effsecond}
\mathcal{V}(r(x))=\Omega_+^2 \left(E^2-\frac{AL^2}{r^2}\right)
\end{equation}
for a static spherically symmetric geometry of the form (\ref{eq:SSS}). Using the conserved quantities of the system and the definitions introduced so far for the EiBI geometries considered in this work, one can characterize this second derivative of the potential defined in (\ref{eq:effsecond}) in the $x$-representation as (here $A_{ps} \equiv A(r(x_{ps}))$)
\begin{equation} \label{eq:lyax}
\frac{d^2\mathcal{V}}{dx^2} \Big \vert_{x=x_{ps}}=\frac{L^2}{r^4(x_{ps})} \Omega_{+}^2(x_{ps})\left[2A_{ps} - r^2(x_{ps}) A_{ps}''\right]
\end{equation}
or, alternatively, in the $r$-representation, as
\begin{equation} \label{eq:lyax}
\frac{d^2\mathcal{V}}{dr^2} \Big \vert_{r=r_{ps}}=\frac{L^2}{r_{ps}} \Omega_{-}(r_{ps})\left[2A_{ps} - r_{ps}^2 A_{ps}''\right]
\end{equation}
where $x$-derivatives can be traded by $r$-derivatives by using both the photon sphere condition and the frame transformation (\ref{eq:dxdr}). Eq.(\ref{eq:lyax})  is in agreement with the formulae found in the literature \cite{Cardoso:2008bp} for general spherically symmetric space-times of the form (\ref{eq:SSS}), and which in our case entail modifications to the usual GR structure via both the explicit $\Omega_+$ factor but also via the new shape of the metric function $A(r(x))$.

\begin{center}
\begin{table}[]
\begin{center}
\begin{tabular}{|c|c|c|c|}
\hline
 & GR-Schwarzschild & GR-RN $Q=0.5$  & GR-RN $Q=1$    \\ \hline
$\gamma_{2/1}$  & 3.15075  & 3.05215  & 2.2593    \\ \hline
$E_{2/1}$  & 23.35  & 21.16  & 9.57   \\ \hline
$E_{III}^1$    &  21.00   & 19.29   & 12.14   \\ \hline
$E_{III}^2$    &  28.63   & 26.19   &  13.09 \\ \hline
$E_{I}^1$      &  12.56   & 11.64     &   7.34   \\ \hline
$E_{I}^2$      &  24.73   & 22.52   &   10.78 \\ \hline
$E_{II}^1$     &   9.79  & 8.99   &    5.15\\ \hline
$E_{II}^2$     &   23.42  &  21.25  &   9.50 \\ \hline
\end{tabular}
\caption{The Lyapunov exponent $\gamma_{2/1}$ (taking $M=1$) corresponding to the theoretical luminosity ratio of the $(m=2)/(m=1)$ photon rings, the extinction factor $E_{2/1}=e^{\gamma_{2/1}}$ associated to it, and the ratio of real luminosities between the $(m=1)/(m=0)$ and $(m=2)/(m=1)$, for the usual GR-Schwarzschild black hole, a charged GR-RN black hole, and the GR-RN extreme black hole.}
\label{Table:LyaGR}
\begin{tabular}{|c|c|c|c|c|c|}
\hline
 & Sch  & RN2  & RN1e  & FC1  & FCt  \\ \hline
$\gamma_{2/1}$              & 3.06288     & 2.56472   & 1.99167 & 2.62805 & 2.32797  \\ \hline
$E_{2/1}$  & 21.38  & 12.99  & 7.32 & 13.19 & 10.25   \\ \hline
$E_{III}^1$          & 19.37   & 12.80 & 11.44  & 13.38   & 9.85  \\ \hline
$E_{III}^2$         & 26.44 & 15.46 & 12.39  & 18.48  & 16.45  \\ \hline
$E_{I}^1$   & 11.73  & 8.33 &  7.41  & 8.53  & 8.10 \\ \hline
$E_{I}^2$     & 22.76 & 14.36  & 11.56  & 14.35  & 14.95 \\ \hline
$E_{II}^1$ & 9.06   & 6.13  & 5.24  & 6.27  & 6.00  \\ \hline
$E_{II}^2$ & 21.49 & 12.49 &  10.03  & 13.08  & 12.47  \\ \hline
\end{tabular}
\caption{The Lyapunov exponent $\gamma_{2/1}$ (taking $M=1$) corresponding to the theoretical luminosity ratio of the $(m=2)/(m=1)$ photon rings, the extinction factor $E_{2/1}=e^{\gamma_{2/1}}$ associated to it, and the ratio of real luminosities between the $(m=1)/(m=0)$ and $(m=2)/(m=1)$, for the five classes of EiBI black holes: Schwarzschild-like, the two-horizon and extreme RN-like, and the two single-horizon finite-curvature cases (FC1 and its limiting version FCt).}
\label{Table:LyaP}
\end{center}
\end{table}
\end{center}

The instability of the orbits is better understood (and better tuned to the potential observables that characterize the photon rings of the background geometry) when, instead of the coordinate time, it is expressed in terms of the increase in the number of half-orbits. Indeed, after $m$-half orbits, a given photon departing at a distance $\delta r_0$ from the critical curve will be located at a radius $\delta r$ away from it given by \cite{Pretorius:2007jn}
\begin{equation} \label{eq:Lya}
\delta r =e^{\gamma m} \delta r_0
\end{equation}
Since the Lyapunov exponent $\gamma$ controls the relative flux between successive rings, in the limit $m \to \infty$ it provides a universal quantifier of the contribution of a given spherically symmetric geometry to the properties of the image \cite{Johnson:2019ljv}, which is independent of the properties of the accretion disk. In particular, for the Schwarzschild case this exponent equals $\gamma=\pi$ so that this relative flux decays with successive rings as $e^{-\pi} \simeq 0.0432$ \cite{Luminet:1979nyg}: thus each of such ring would be $\sim 23.1$ times fainter than the previous one, which we call its extinction rate. For the sake of this work we shall concentrate on the Lyapunov exponent between the photon rings $m=1,2$ instead of the infinite $m$ limit, since these are the rings we hope to see in future interferometric projects \cite{Johnson:2019ljv}. Thus, we shall evaluate it according to
\begin{equation}
\gamma_{2/1} = \log(l_2/l_1)
\end{equation}
where $l_{1,2}$ are the locations of the $m=1,2$ photon rings with respect to the critical curve, that is, $l_{1,2}=r_{1,2}-r_{ps}$. In the Schwarzschild case, this number equals $\gamma_{2/1} \approx 3.15075$, which amounts to a $\sim 0.3 \%$ of difference with the exact number of the $m \to \infty$ limit.  In Table \ref{Table:LyaGR} we display this number (first row) for the reference GR black hole configurations: Schwarzschild black hole, two-horizons RN black hole, and extreme black hole, and also its associated extinction rate (second row) between successive rings (i.e. its exponential version, as given by Eq. (\ref{eq:Lya})). Note that due to the presence of the exponential, this rate decays quickly with $\gamma_{2/1}$: for instance, for a value $\gamma_{2/1} \approx 2.45$ we get a twice less severe extinction, and this will be typically the case for most configurations when we push the electric charge to large enough values. Similarly, in Table \ref{Table:LyaP} we display the same numbers for the five black hole configurations of the EiBI case. In this case the Sch and RN1 configurations can be put into direct correspondence with their GR-RN ($Q=0.5$ and $Q=1$, respectively) counterparts. We see that the Lyapunov exponent is larger in those EiBI cases which can be put in direct correspondence with their GR counterparts, so the photon rings of the former are expected to be less luminous than those of the former, an effect enhanced with an increase of the electric charge.

Nonetheless, the Lyapunov exponent is a theoretical number whose value must be compared with the real luminosity of each photon ring (as compared to either the direct emission or to each other) when specific emission models are switched on to illuminate the object. We find thus convenient to compute the extinction rate of the $m=1$ photon ring as compared to the direct emission ($m=0$), and the extinction rate of the $m=2$ photon ring as compared to the $m=1$ one (since it is a target of future interferometric projects \cite{Johnson:2019ljv}), i.e.
\begin{equation}
E^1=\frac{I_{m=1}}{I_{m=0}} \quad ; \quad E^2=\frac{I_{m=2}}{I_{m=1}}
\end{equation}
assuming that all three emission models (labelled by a subindex $\{III,I,II\}$) are normalized in their total intensity to unity. The first rate gives a measure of the degree of visibility of the first photon ring when superimposed with the direct emission, while the second rate provides a measure of how the Lyapunov exponent transforms into actual relative luminosity between successive photon rings. The combined analysis of both rates can actually probe regular geometries alternative to GR ones in a less dependent way of the accretion disk uncertainties, see e.g. \cite{Eichhorn:2022oma} for a discussion on this point. These observationally-relevant numbers are displayed in Table \ref{Table:LyaGR} for GR configurations and in Table \ref{Table:LyaP} for the EiBI ones. In both tables we see the comparison of the real extinction rates modelled by the emission profile of the disk as compared to the purely-geometrical (Lyapunov) calculations. Several observations are relevant from these rates:

\begin{itemize}

\item The extinction rates are reduced with the increase of charge in both GR and EiBI configurations, an effect that is greatly enhanced the closer we are to the critical charge limiting the existence of black hole configurations (recall that, taking $M=1$, this corresponds to $Q=1$ in the GR case, but $Q \approx 1.0511$ in the EiBI one). This was already expected on the grounds of the computation of the Lyapunov exponent.

\item The extinction rates, at equal mass and electric charge, are slightly higher in the EiBI case than in the GR one for every emission model, meaning that the corresponding rings are comparatively fainter. This could have been also foreseen from both the computation of the Lyapunov exponent and from the reduction of the shadow's size in the EiBI case.

\item Accretion disk models GLM3/GLM1/GLM2 infuse comparative larger luminosities to both the $m=1$ and $m=2$ photon rings at every value of charge in both the GR and EiBI configurations as we move in the sequence; indeed, the photon rings of the GLM2 model can be up to twice times more luminous than those of the GLM3 model.

\item The relevant extinction ratio $E^2$ between the $(m=2)/(m=1)$ photon rings is also consistent with the trend above: in the GR case this rate is lower in the GLM2 model than in the GLM1 one, and the latter is larger than in the GLM3 one. In all cases the EiBI values are higher than in the GR case, thus entailing a lower luminosity contrast between the photon rings than in the GR configurations.

\end{itemize}

The above numbers contain the tiny details of the set of figures \ref{fig:PSch}, \ref{fig:PRN2}, \ref{fig:PRN1}, \ref{fig:PMk1} and \ref{fig:PMk1e} extracted from the optical appearance (top) of each configuration (normalized intensity in each figure) for the emission models GLM3 (left), GLM1 (middle) and GLM2 (right) alongside the observed intensity profile (bottom). For low values of the electric charge, the GLM3 allows to see the peaks associated to each photon rings clearly isolated from the direct emission, though this neat separation is blurred as the electric charge is increased with the result that the photon rings end up overlapped with the direct emission in some cases. In the GLM1 and GLM2 models the direct and photon ring emissions are always overlapped, but this was already expected and assumed as an inherent consequence of the extension of the accretion disk to the event horizon of the black hole configurations. Because of this fact, the corresponding optical appearances of each geometry in combination with every emission model is different not only on the latter (which is the main actor in distinguishing between images) but also on the former, since there are obvious differences as we go deeper in the sequence Sch $\to$ RN2 $\to$ RN1 for a fixed emission model. In particular, photon rings are clearly more visible within this sequence, though this effect is already present in the original RN solutions. FC1/FCt geometries are worth of similar comments, though here we do not compare to their counterpart images in GR.

The bottom line of the discussion above is that the features of black hole images are highly entangled between the modifications to the background geometry by the EiBI corrections (at equal mass and electric charge) and the properties of the accretion disk itself, whose separation is a delicate and persistent issue for every black hole configuration  \cite{Lara:2021zth}. The single clean modification is that photon rings in the EiBI case are slightly less luminous at equal parameters than their GR counterparts, while at the same time yielding an enhancement of the shadow's size. We conjecture this effect to be the result of a relative weakening of the gravitational interaction in the EiBI case which is somewhat behind the ability of the model to remove the troubles with geodesic incompleteness in these solutions, while at the same time affecting the trademark of the curved space-time around them via the feature of its photon rings. 

Let us stress however, that the Lyapunov exponent is not a direct observable, while the relative fluxes between successive photon rings will be hard to achieve. Other potential observables of interest are the ring's shape (such as its width) \cite{Paugnat:2022qzy} and, more promising still, the so-called auto-correlation functions, namely, the correlations of observed brightness fluctuations across different times and locations  since they exhibit a universal structure \cite{Hadar:2020fda}, whose analysis goes far beyond the scope of the present work.

\subsection{Higher-order photon rings of horizonless compact objects} \label{Sec:hopr}

\begin{figure*}[t!]
\begin{center}
\includegraphics[width=5.9cm,height=4.4cm]{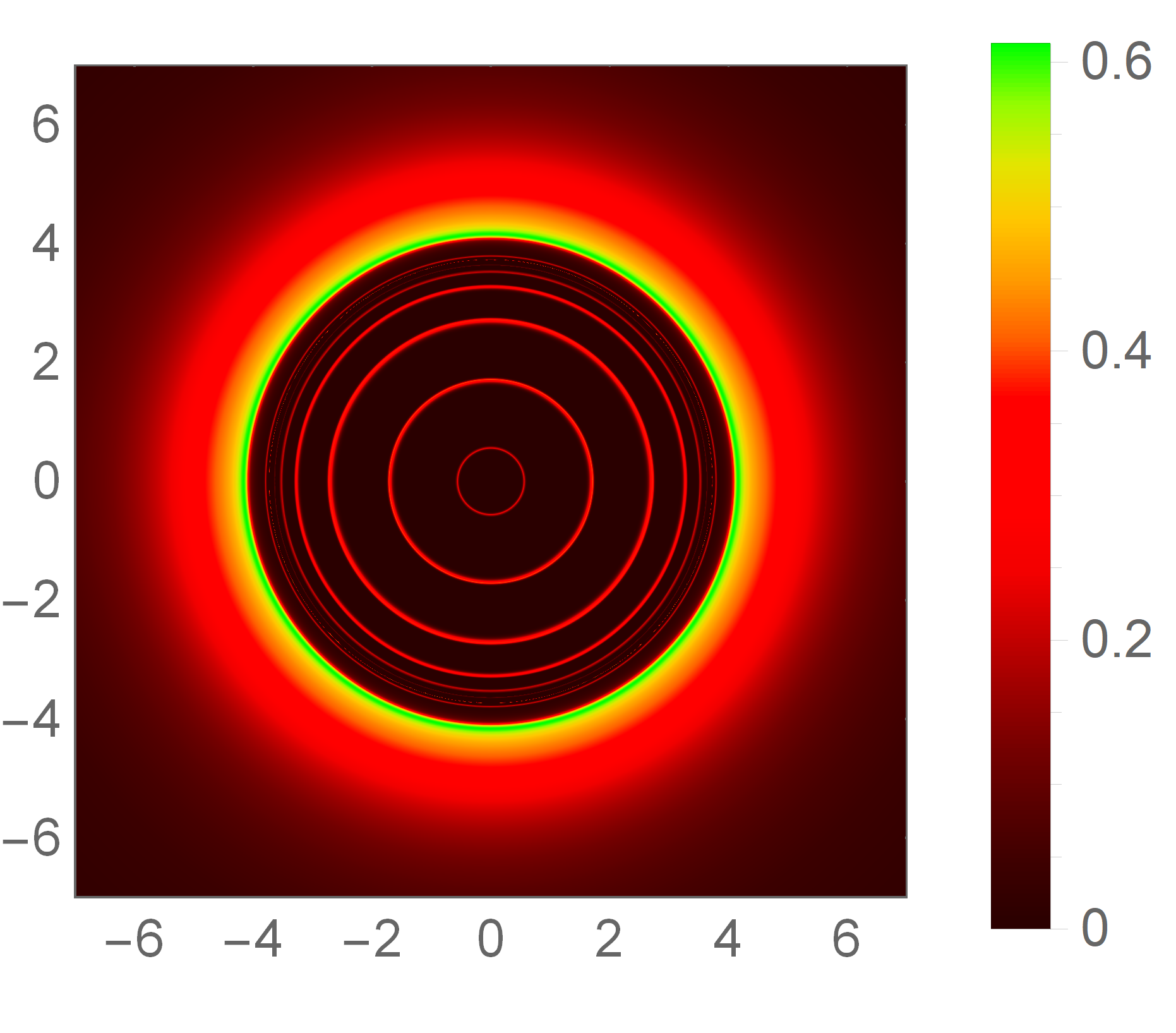}
\includegraphics[width=5.9cm,height=4.4cm]{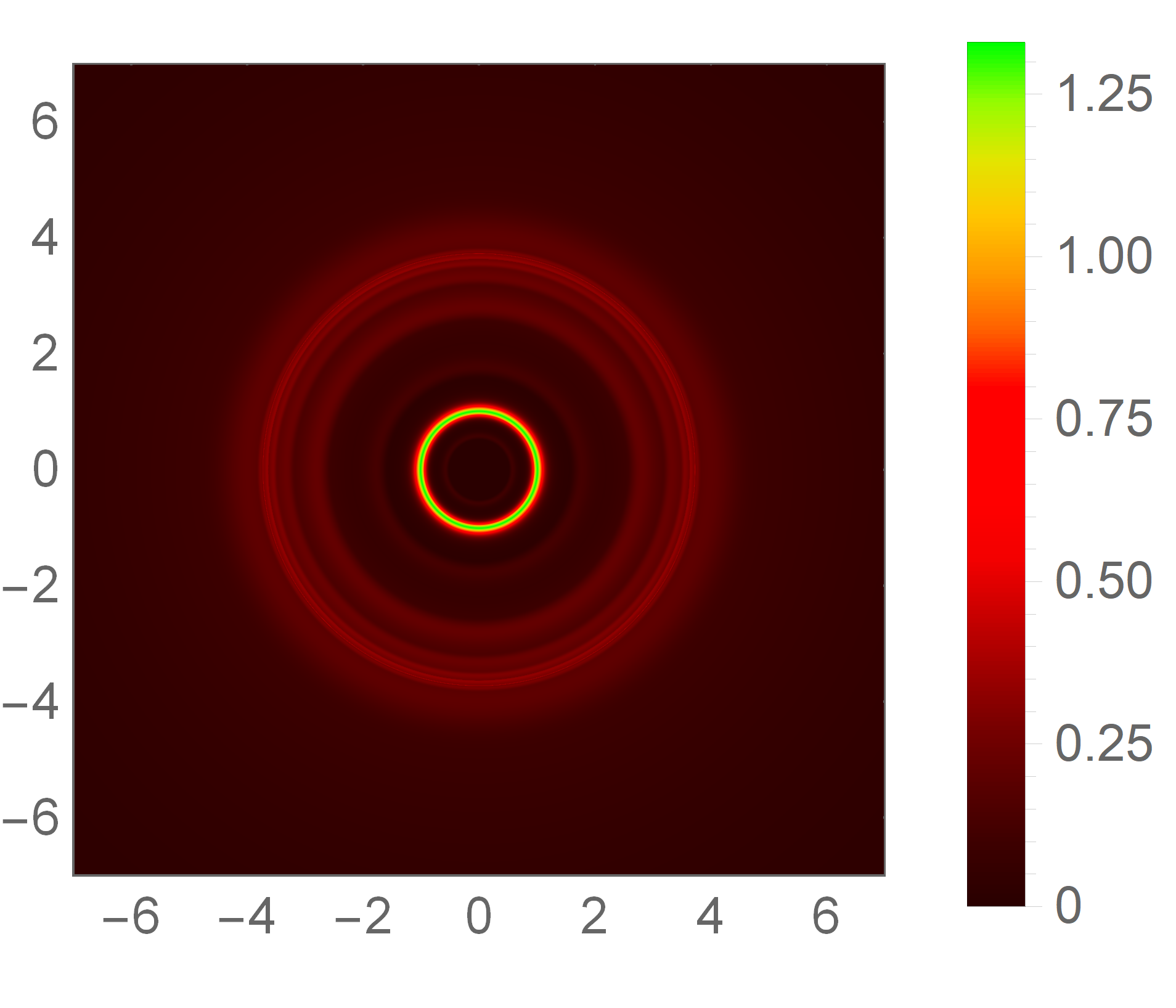}
\includegraphics[width=5.9cm,height=4.4cm]{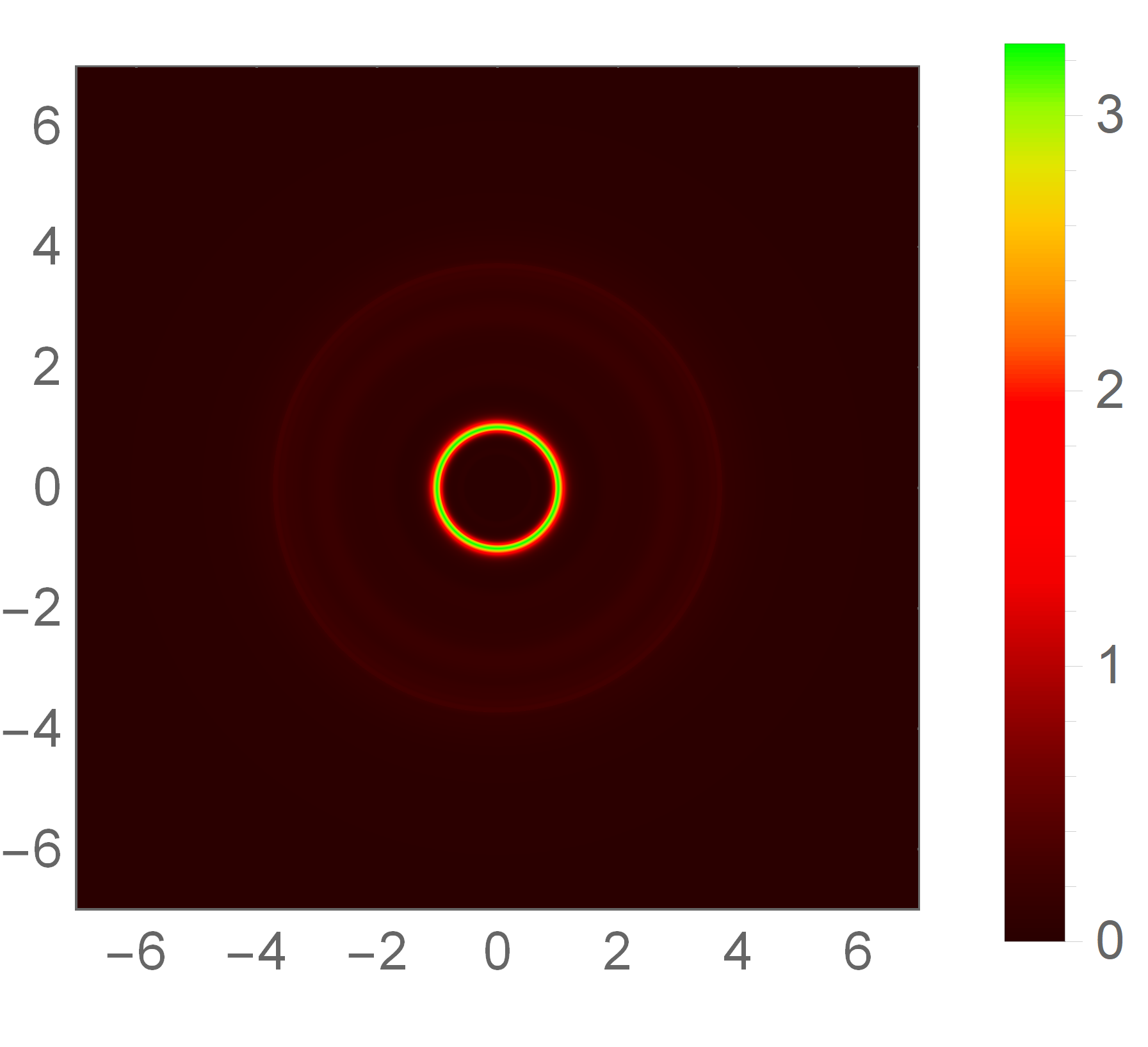}
\includegraphics[width=5.9cm,height=4.4cm]{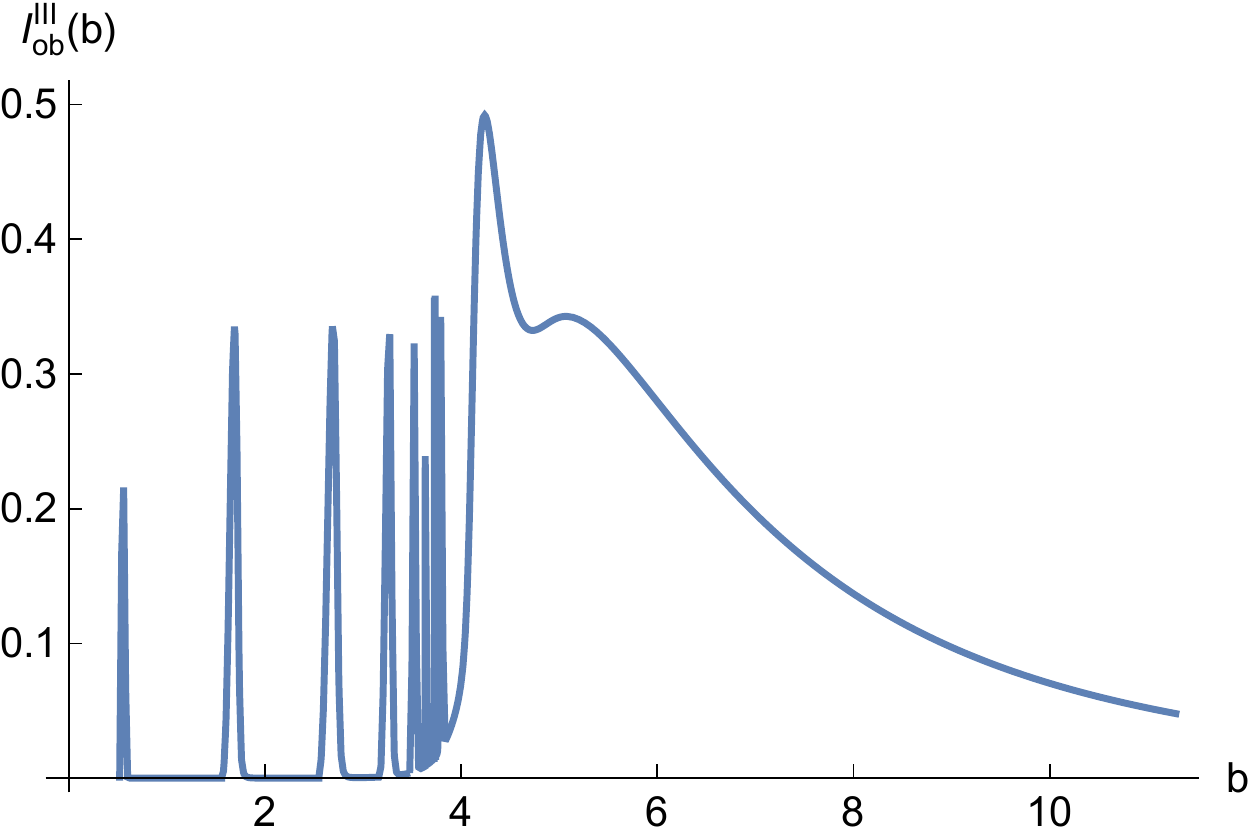}
\includegraphics[width=5.9cm,height=4.4cm]{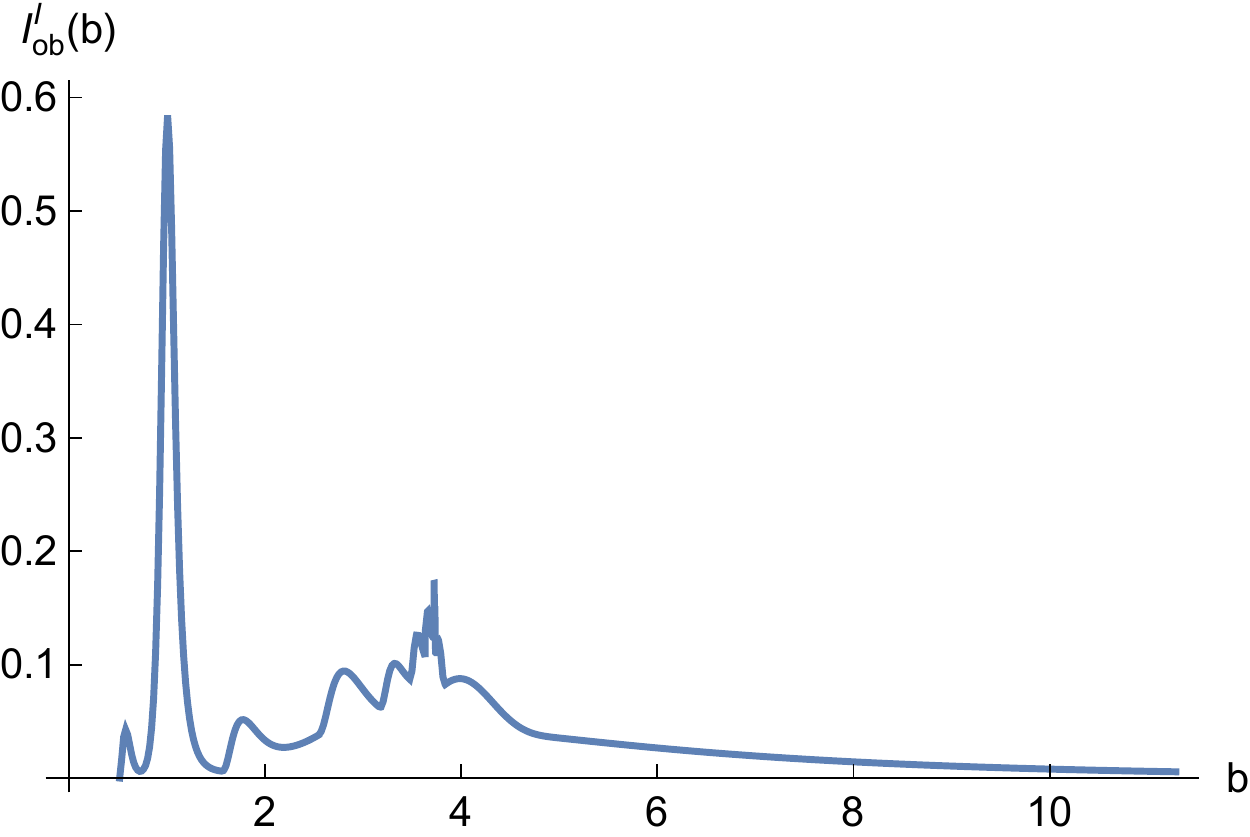}
\includegraphics[width=5.9cm,height=4.4cm]{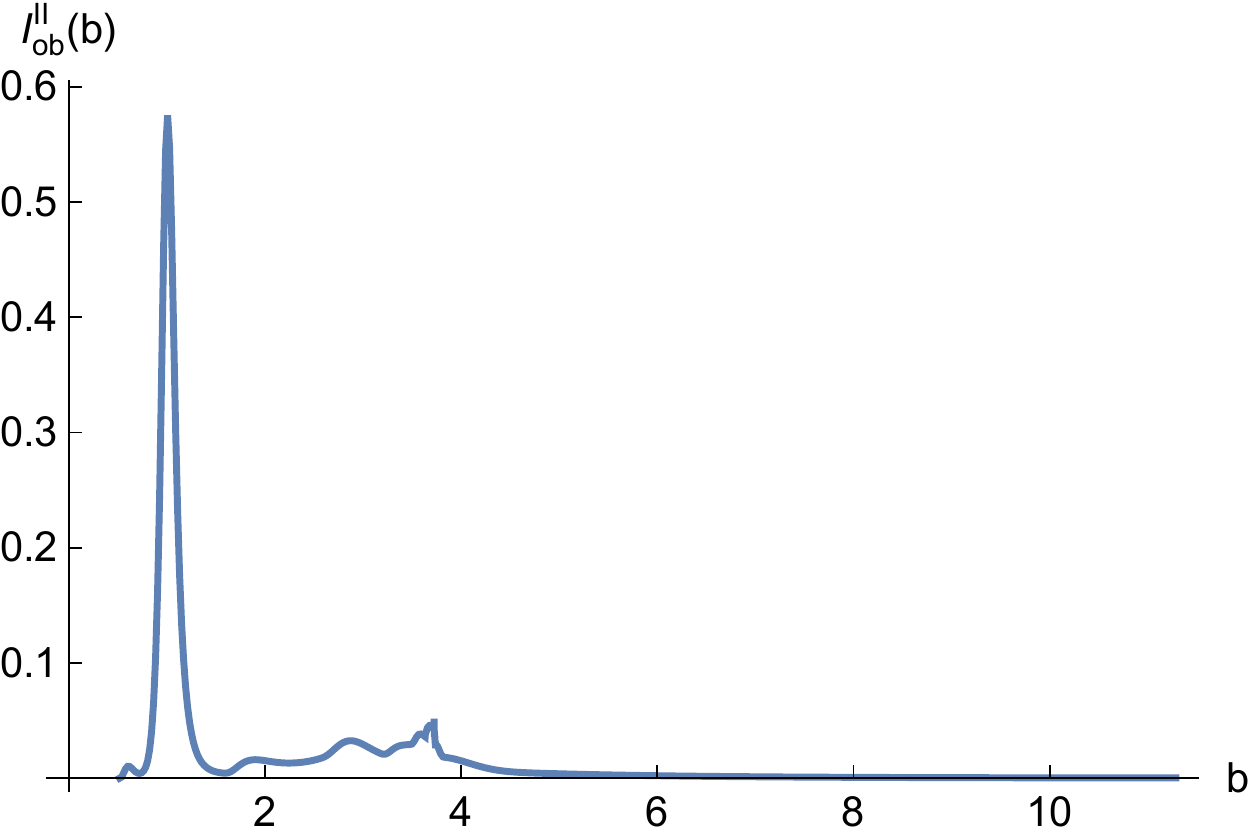}
\caption{The optical appearance of EiBI  RN0 configurations (top panel)  in the GLM3 (\ref{eq:adGLM3}), GLM1 (\ref{eq:adGLM1}) and GLM2 (\ref{eq:adGLM2}) emission models, and the associated intensity for each of them (bottom panel), displaying the direct ($m=0$) and photon ring ($m=1,2,3,4,5$) emissions,  respectively, via a number of peaks $M>m$.}
\label{fig:PRN0}
\end{center}
\end{figure*}

Since the presence of an event horizon prevents one from extracting more physical consequences of the regularity of these space-times, to tackle this challenge we consider next the two scenarios in which these configurations get rid of their event horizons, corresponding to the RN0/FC0 configurations. Horizonless compact objects have been long studied in the literature due to the conceptual and operational differences they pose as compared to black hole space-times, and which may have repercussions at the observational level via several avenues (a detailed review can be found in \cite{Cardoso:2019rvt}). In order to stand as firm contestants to the black hole hypothesis, such objects are required to satisfy a number of physical admissibility conditions, including suitable gravitational collapse mechanisms, stability against linear perturbations, and being supported by a physically sound theory of the gravitational and matter fields. While not all such horizonless objects need to be compact enough to have a critical curve, the ones that we are going to consider in this section, corresponding to the RN0 configurations, have it.

For the sake of shadow images the fact that in this case light trajectories can get access to the full internal part of the effective potential (given the absence of an event horizon) allows further contributions to photon ring emissions on top of those originated near the critical curve. Indeed, such new contributions are associated to those rays circulating between the location of the maximum of the effective potential, its minimum (dubbed as the {\it anti-photon sphere}) and the infinite potential slope [recall Fig. \ref{fig:potential}, green curve]. The net effect is that the corresponding object will show a {\it multi-ring structure}, namely, a cascade of additional non-negligible higher-order photon ring contributions (for an overly general analytic analysis of this problem we refer the reader to Refs. \cite{Bisnovatyi-Kogan:2022ujt,Tsupko:2022kwi}). These appear thanks to the enhancing effect of the region between the photon sphere and the internal part of the potential such that every photon with low enough impact parameter hits the infinite potential slope, allowing for further  transits of light trajectories around the photon sphere on its way back, this way collecting additional luminosities\footnote{A different, but somewhat related scenario corresponds to those compact objects having two accessible photon spheres, which can be realized in both black holes \cite{Guo:2022muy} and symmetric \cite{Tsukamoto:2022vkt} and asymmetric  \cite{Guerrero:2022qkh} wormholes. In such a case, the potential well between both photon spheres acts as source of additional higher-order trajectories.}. As a consequence, the expectations based on the theoretical Lyapunov exponent regarding an exponential suppression of the luminosity of successive photon rings with higher $m$ does not necessarily hold here, and one must study this problem on a case-by-case basis for the combination of background geometry and accretion disk's emission (for the canonical case of Kerr-Newman black holes see the analysis of its multi-ring structure of \cite{Hou:2022gge}).

In Fig. \ref{fig:PRN0} we depict the result of our simulations for the images of the RN0 configurations appearing in Table \ref{Table:I}, where we consider light trajectories going up to the $m=5$ intersection with the disk, since we verified this is the last trajectory which provides visible enough associated rings. As for the number of iterations in tracking the critical curve, however, we limit them to 1000, in order to shorten computation times and because here we are not that interested on precisely determining the corresponding luminosities but on seeing the general visual appearance. Note that in this case despite the fact that no horizon is present, these trajectories cannot get to the wormhole throat radius since the potential grows without bound for a larger radius than the throat one. The multi-ring structure is clearly visible (top figures), while the observed intensity (bottom figures) reveals that many new narrow peaks have appeared in the internal region of the impact parameter space, which are particularly visible in the GLM3 model. Indeed, the one-to-one correspondence between trajectories crossing the equatorial plane $m$ times and the number of photon rings $M$ is broken with $M>m$. This was expected on the grounds of the lessons learned from some toy-models having a potential ``well" in which these new rings can be originated \cite{Guo:2022muy,Guerrero:2022msp,Guo:2022ghl}, qualitatively similar as the one found in these configurations. These new rings appear as sharp images in the optical appearance (top figures) of the object, with a strong dependence on the emission model. Indeed, only in the GLM3 model most of these rings appear as separated images from the direct emission; in the GLM1/GLM2 models they end up instead overlapped with the direct emission (given the fact that the latter extends all the way down to the potential slope) in  quite a complex way, and we actually see a strong dependence on the luminosity infused to each of them depending of the decay of the intensity with the distance between each model. Despite their relative faintness, we recall that the sharpness of these rings grant their dominance in the Fourier spectrum of interferometric detections and, as such, they are potentially observable (should they happen to be there) in future probes \cite{Johnson:2019ljv}. On the other hand, the actual central brightness depression of the image is sharply reduced as compared to the black hole images, and it is bounded by the innermost of the higher-order photon rings. Therefore, the uncloaking of the wormhole throat unmistakably distinguishes these configurations from their black hole counterparts, both in the multi-ring structure and in the drastic reduction of the actual shadow's size.

It should be stressed that similar features as those discussed here can be found in the overcharged RN configurations of GR. However, in such a case one is faced with the insurmountable difficulty of the presence of a space-time singularity at its center as given by its incompleteness character, since purely radial ($b=0$) null geodesics can reach it in finite affine time without any possibility of further extension. In the present case the presence of the wormhole throat removes this theoretical problem but it changes little the optical appearance of the object since such GR incomplete geodesics contribute nothing to the image: the differences are again tiny in terms of relative luminosities between successive photon rings. Note also that in the present case the fact that the wormhole throat lies beyond the infinite potential slope means that the throat cannot actually be reached by any null geodesics (save also by purely radial ones) since each of them find a turning point and, from that point of view, this wormhole configuration could be regarded as non-traversable.

This kind of objects having both a photon and an anti-photon sphere has been strongly criticised in the literature on the grounds that the stable light ring (the minimum of the effective potential) may trigger the development of a instability by trapping massless modes that eventually grow large enough so as to back-react upon the background geometry and destabilize the whole configuration \cite{Keir:2014oka,Cardoso:2014sna}. This is case, for instance, of  a family of compact objects satisfying the above conditions that has been widely studied in the literature, namely, ultra-compact bosonic stars. The recent results of \cite{Cunha:2023xrt} using full non-linear numerical simulations of two families of such boson stars with both a photon and anti-photon spheres show that the final fate of such instability is that such objects either collapse into a black hole or migrate to configurations without critical curves, i.e., without photon rings in their images. In our case, the analysis of this problem would entail to possibly consider both pieces of the wormhole (i.e. in the regions $x>0$ and $x<0$) and taking its non-trivial topology into account, something beyond the scope of this work.

\subsection{A finite-curvature traversable wormhole}

\begin{figure*}[t!]
\begin{center}
\includegraphics[width=5.9cm,height=4.4cm]{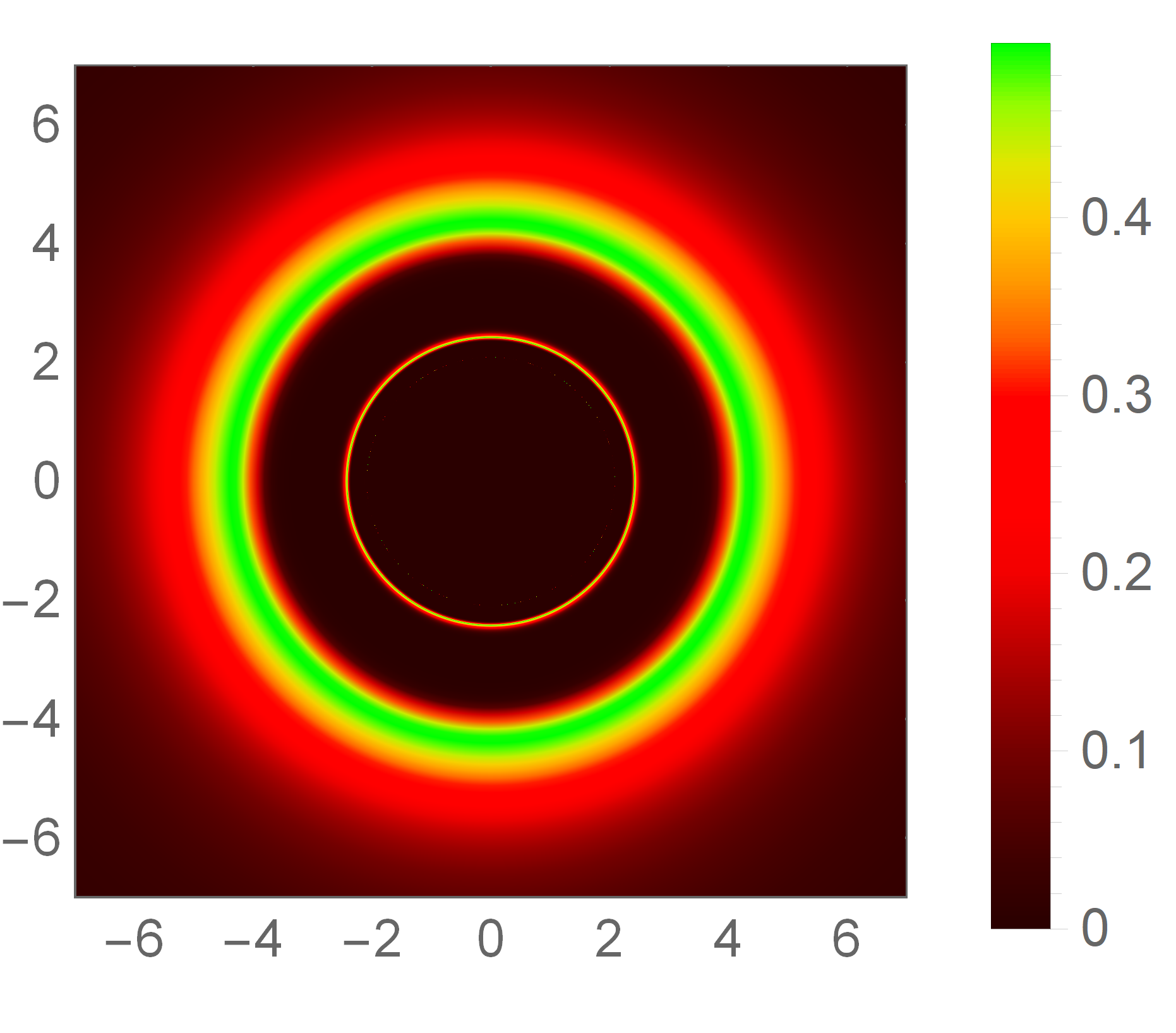}
\includegraphics[width=5.9cm,height=4.4cm]{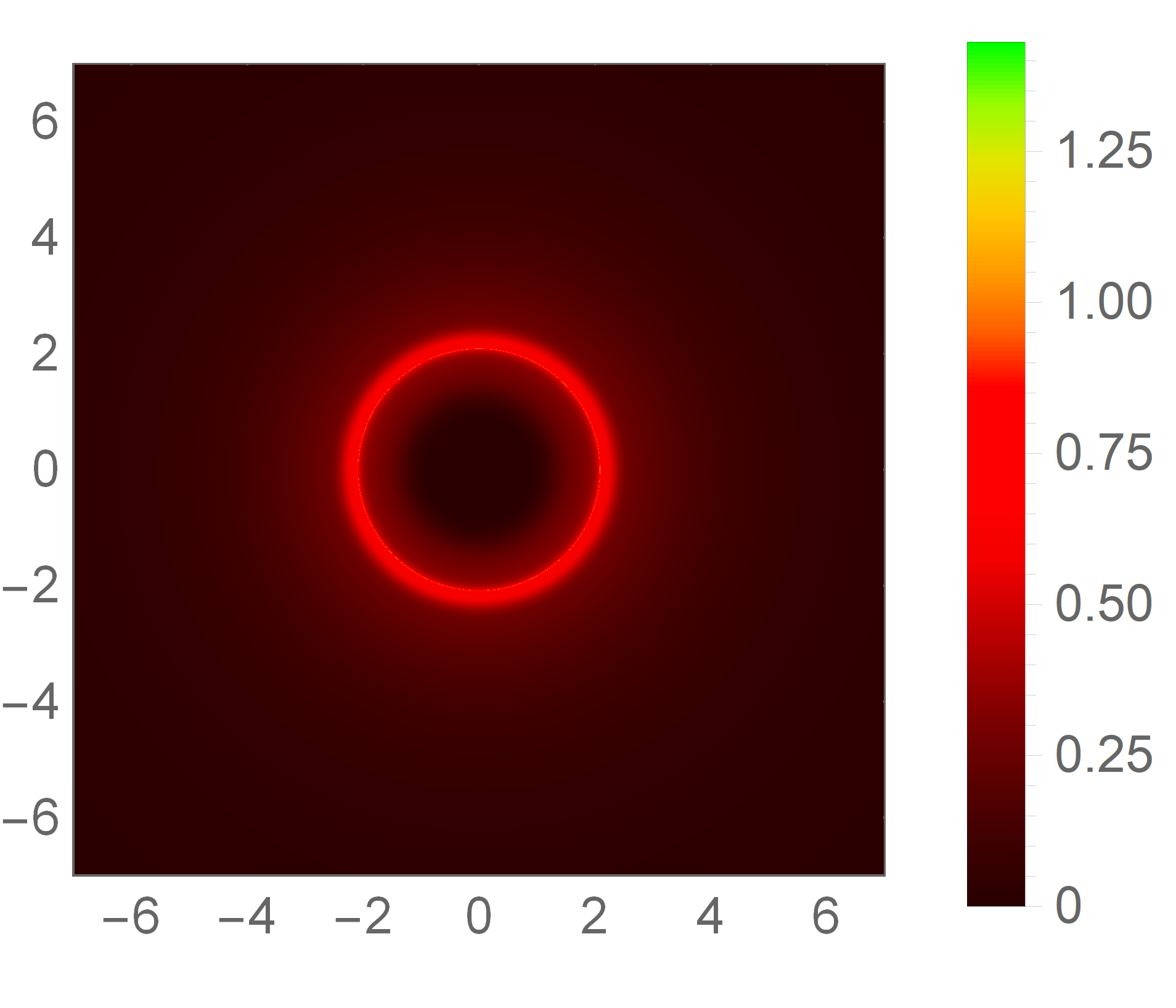}
\includegraphics[width=5.9cm,height=4.4cm]{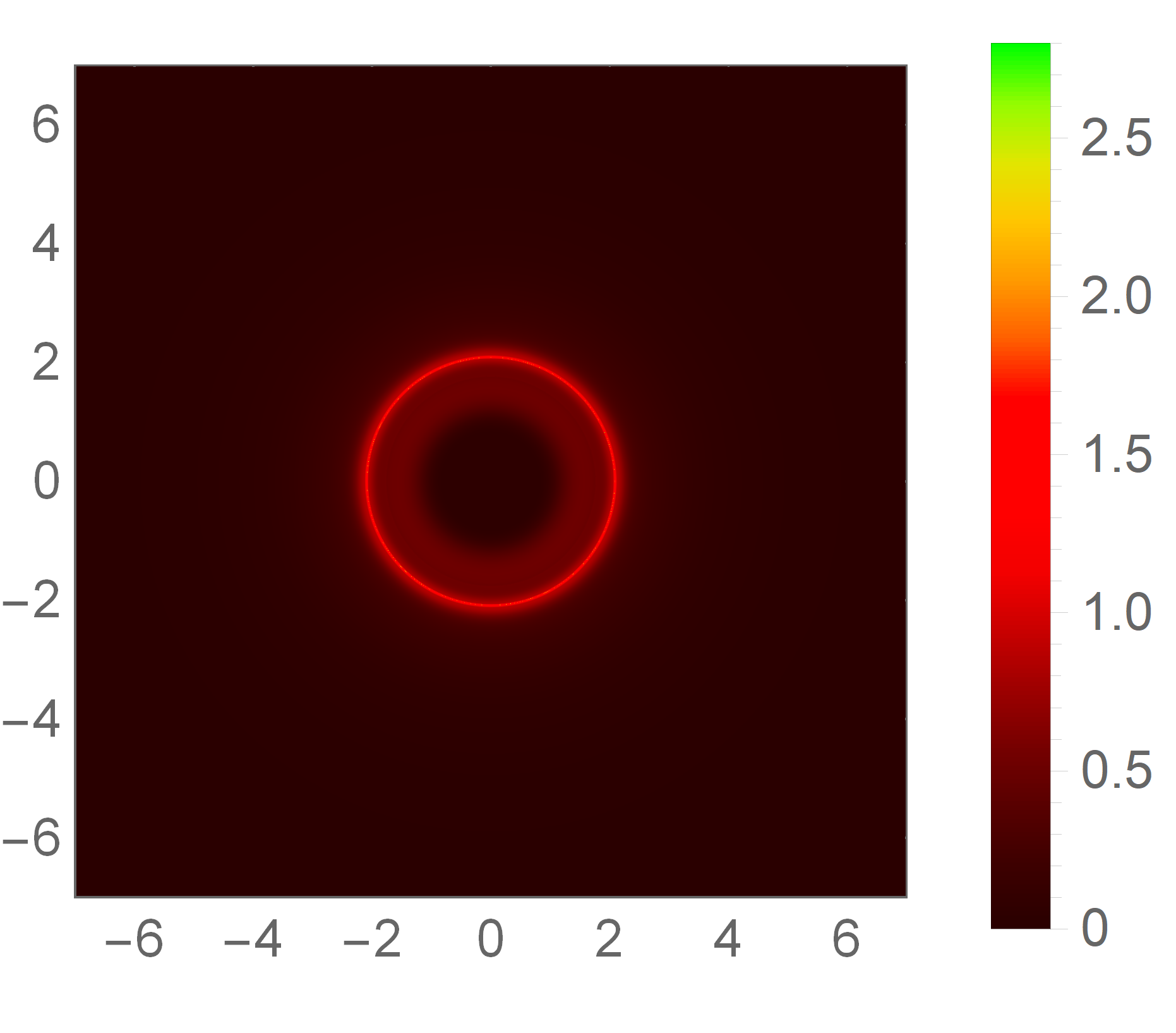}
\includegraphics[width=5.9cm,height=4.4cm]{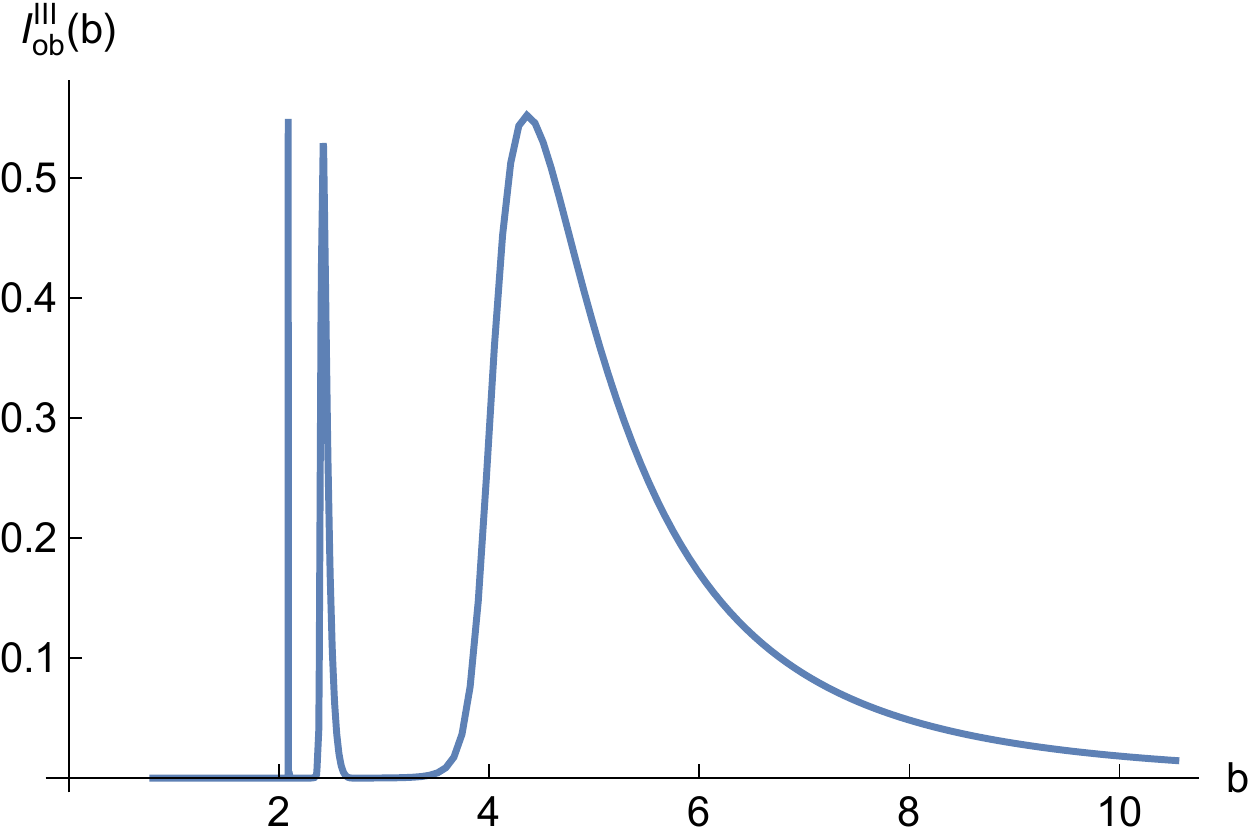}
\includegraphics[width=5.9cm,height=4.4cm]{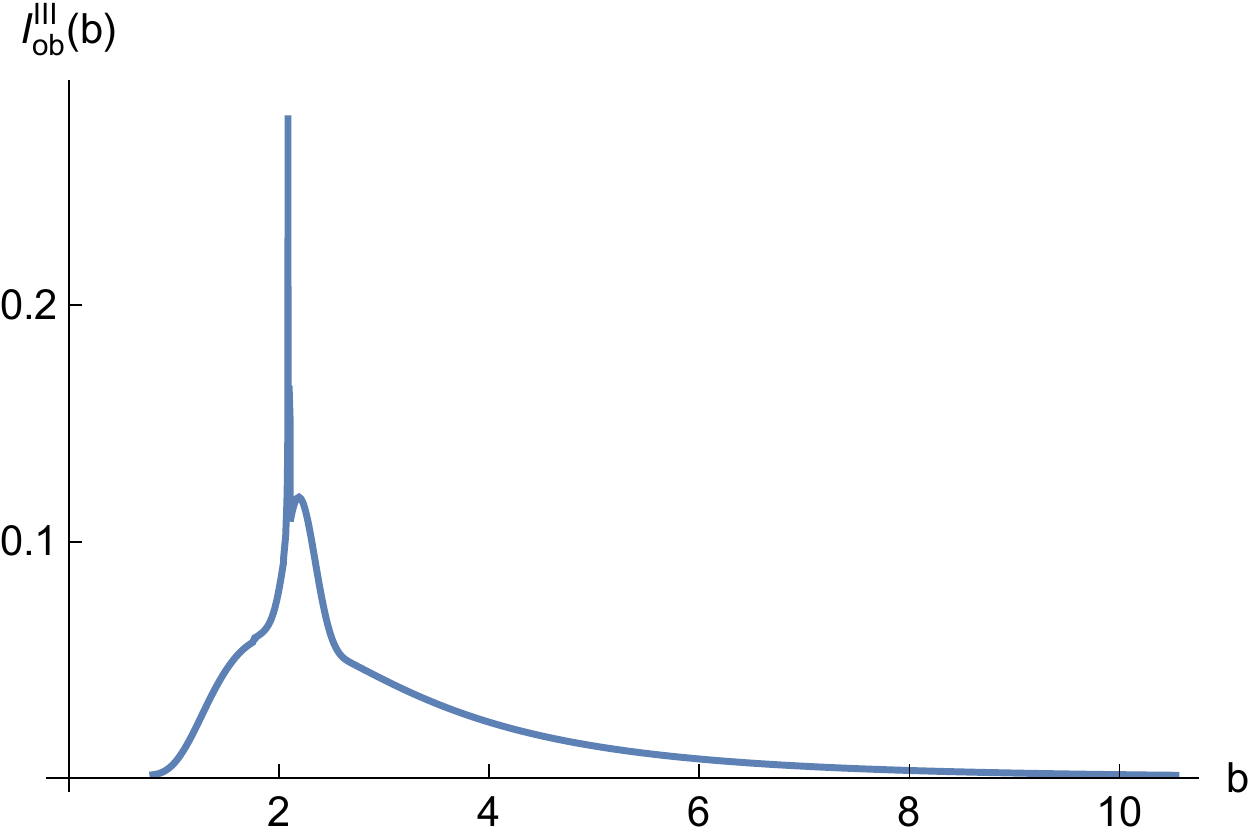}
\includegraphics[width=5.9cm,height=4.4cm]{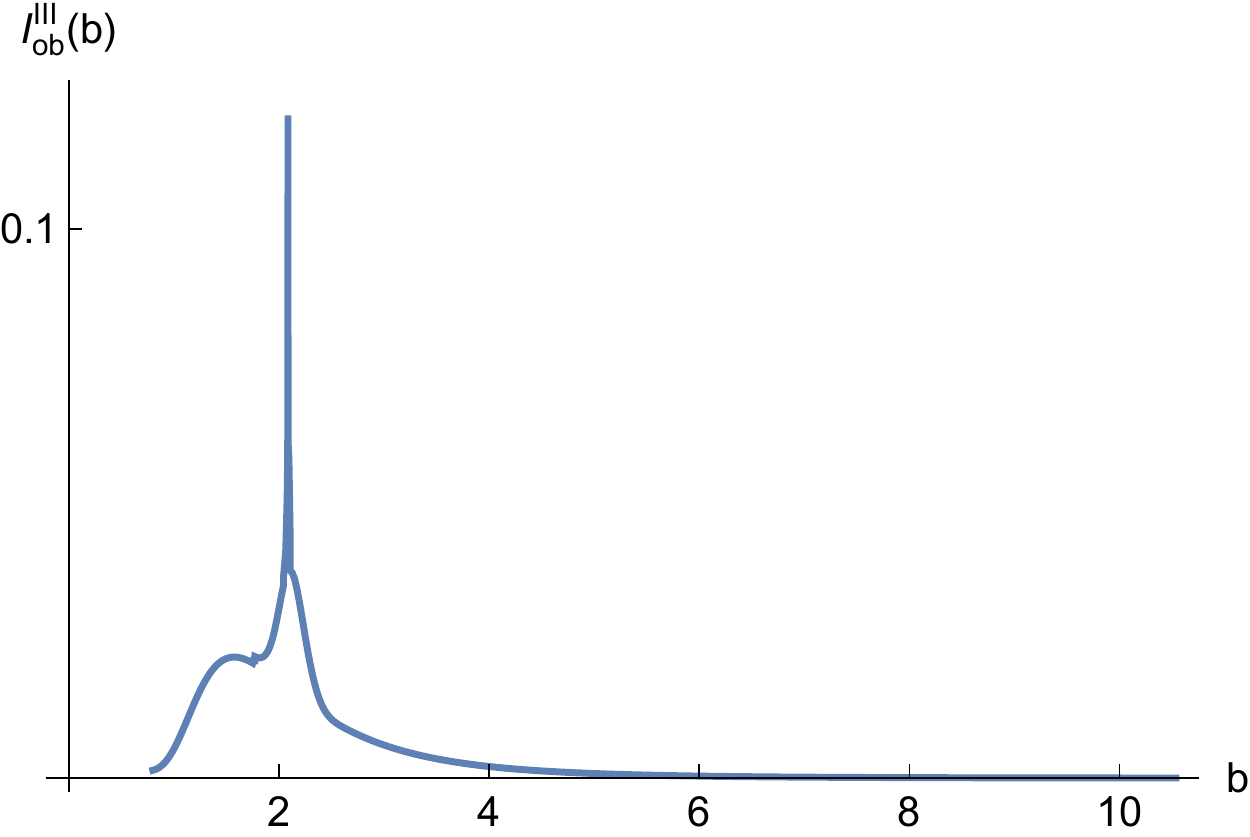}
\caption{The optical appearance of EiBI  FC0 configurations (top panel)  in the GLM3 (\ref{eq:adGLM3}), GLM1 (\ref{eq:adGLM1}) and GLM2 (\ref{eq:adGLM2}) emission models, and the associated intensity for each of them (bottom panel), displaying the direct ($m=0$) and photon ring ($m=1,2$) emissions,  respectively.}
\label{fig:FC0}
\end{center}
\end{figure*}

In the FC0 configurations the effective potential is everywhere finite while also displaying a critical curve [recall Fig. \ref{fig:potential}, cyan curve]. But now the wormhole throat is not only uncloaked by also reachable by any light ray with small enough impact parameter. However, the process of generation of images is very similar to those of the black hole ones since every trajectory passing over the maximum of the effective potential will be swallowed up by the wormhole throat instead of the event horizon, so the optical appearance and observed intensity are also very similar, see top and bottom panels of Fig. \ref{fig:FC0}. The results would be very different should we consider radiation flowing from the other side of the wormhole throat to our local space-time patch, since in such a case there could be trajectories with impact parameters lower than the critical one coming to color the otherwise black central region, and which would constitute strongly different shadow images (we shall address this upgrade in forthcoming works on the subject).

This way, this kind of object where the stable critical curve is absent could circumvent the potentially pathological long-lived modes associated to it \cite{Cardoso:2014sna}. Furthermore, it is not covered by the theorem exposed in Ref. \cite{Cunha:2017qtt} by which critical curves (photon spheres) in horizonless ultra-compact objects always come in pairs (i.e. one unstable and another stable) via a simple argument by which such every regular object must have the same topological charge associated to critical curves (with opposite signs for stable/unstable such curves) as Minkowski space-time, namely, zero, in order to be created by a dynamical formation process. The two features that conspire to produce this effect for these configurations are i) the topologically non-trivial character of the space-time via the presence of a wormhole throat (see the discussion at the end of \cite{Cunha:2023xrt}) and ii) the fact that the metric at the throat is finite and larger than one (which has the side-effect of removing the presence of curvature divergences). Admittedly these configurations are somewhat ``artificial" since any small modification to the cubic charge-to-quadratic mass ratio (\ref{eq:R1}) would make them to suddenly ``leap" into any of the other families of configurations in which the second effect is absent. Nonetheless, these configurations are also very suggestive since they naturally implement Wheeler's {\it geons} idea \cite{Wheeler:1955zz}, namely, self-sustained sourceless gravito-electromagnetic entities with both their electric charge and mass realized from a topological origin \cite{Olmo:2013gqa}. This topic deserves a much deeper analysis, which we leave for future work as it also goes beyond the scope of this paper.

\section{Conclusion and discussion} \label{sec:V}

In this work we have studied the optical appearance of several families of modified black hole and horizonless configurations when illuminated by a thin accretion disk. These geometries are found within a theoretically well-motivated and observationally viable extension of General Relativity dubbed as Eddington-inspired Born-Infeld gravity, formulated in a metric-affine context and coupled to an standard electromagnetic (Maxwell) field. As such, its counterpart in GR would be the Kerr-Newman solution, downgraded to the Reissner-Nordstr\"om solution in the spherically symmetric limit, the case of interest for the sake of this work. These families of configurations can be split into seven: five of black hole type and two of horizonless one, depending on two ratios between their masses and charges. This way, setting a mild value for the gravitational's theory parameter in the branch for which all configurations develop a wormhole throat on their innermost region (hidden behind the event horizon in the black hole cases)  we have selected representative samples of each  family of configurations.

With this in hand, we have proceeded to the analysis of their shadow and photon ring images. First we have developed the null geodesic formalism, suitably adapted to the peculiarities of these geometries, with particular attention to light trajectories that can turn several (half-)times around each configuration. When the disk is transparent to its own radiation and furthermore geometrically thin, this creates a series of self-similar photon rings (characterized by an integer number $m \geq 1$ counting the number of intersection with the equatorial plane of the disk) on top of the direct emission of the disk itself ($m=0$), and with exponentially-decreasing contributions to the total luminosity of the object. We have characterized such rings according to the theoretical luminosity provided by the Lyapunov exponent associated to the instability scale of trajectories near the the bound orbits, and the real luminosity after switching on three models of monochromatic emission characterized by a single distance-dependent function, previously introduced in the literature. One of them is designed to probe the structure of such photon rings (limited to $m=1,2$, which are the only ones that provide visible contributions to the optical appearance of all black hole configurations) when they are (mostly) separated from the direct emission, while the other two extend all the way down to the horizon with a difference dependence on the radial distance, and which mix the two photon rings over the direct emission of the disk.

The main conclusion of our analysis is that photon ring and shadow images of these modified black hole configurations closely resemble their GR counterparts, with the properties of the accretion disk emission being the major actor casting these images. Two main qualitative differences arises though: the relative luminosity of the photon rings, particularly on the ratio between the $m=2$ and $m=1$ ones, is slightly fainter than  their GR counterparts, an effect that is enhanced as larger values of the electric charge are considered. Secondly, the inferred shadow's radius of the Sgr A$^*$ from the EHT Collaboration according to some calibration factors suggests that the shadow's radii of the present configurations are also slightly larger than their GR counterparts. As for the two horizonless (traversable wormhole) configurations, one of them creates a stable photon sphere on its effective potential in addition to the unstable one, allowing for a multi-ring structure towards the $m=3,4,5$ trajectories and which yields an optical appearance of the corresponding object that strongly departs from their GR-images. Criticisms on these configurations can be made at the potential instability that the stable photon sphere may trigger, which requires a deeper analysis. The second configuration resembles GR-images, but it has some theoretical niceties regarding the absence of stable photon sphere (and its apparent run-away from the hypotheses of \cite{Cunha:2017qtt}) and its interpretation as geonic objects. In any case, our analysis confirms the differences in the optical appearance of modified black holes and horizonless compact objects found in the literature, namely, the presence of a multi-ring structure and a reduced size of the shadow's radius of the latter as compared to the former, two main observables in shadow observations and which can effectively act as discriminators between these two categories of objects.

This work must be seen as a first step towards a systematic analysis of shadow and photon ring  images of black holes and ultra-compact configurations within a metric-affine framework of gravitational extensions of GR, where the singularity-removal of the configurations of the latter can be achieved by two different mechanisms (see \cite{Bejarano:2017fgz} for a discussion on this point). Since the shape of the modified gravitational configurations introduces several additional difficulties both in the analytic analysis of their properties as well as in the numerical integration of their light trajectories (in particular via the corresponding computational times), the disk's modeling is quite simplified. In order to obtain more realistic images, though, our analysis must be upgraded in various directions:

\begin{itemize}

\item Real accretion disks' geometry is neither infinitely thin nor completely spherical (see however \cite{Narayan:2019imo,Bauer:2021atk}), so an upgrade to disks of some non-negligible thickness would be appropriate.

\item The optical thinness of the disk needs not be so at all frequencies, something disregarded in our analysis, which assumes a monochromatic, everywhere thin emission. Furthermore our assumption of a single frequency in the disk's frame contrasts with the EHT observations assumptions, which are made at a constant $1.3$mm frequency on the observer's frame \cite{EventHorizonTelescope:2022xqj}. As for the choice of the $I(r)$ function, we supported ourselves on a family of profiles (the GLM ones) extracted from the matching between semi-analytic models and the outcomes of GRMHD simulations, but this is a fast-evolving field in which there is plenty of room for improvement.

\item Here we consider only spherically symmetric black holes which, despite being known in analytic form, posed troubles of several kinds due to its much larger complexity than its GR counterpart in order to find its optical signatures. Note in this sense that for a black hole rotating at full speed the shadow's size is up to a $\sim 7 \%$ smaller than its spherically symmetric counterpart \cite{Psaltis:2018xkc}. While this small difference could be ignored by arguing that there are other aspects of the modelling that influence the images more, the fact is that the photon sphere degenerates into a photon shell of unstable geodesics such that light rays can now perform more involved trajectories in the radial and angular directions, which will also leave their imprints in the corresponding images (via shapes and luminosities of their photon rings). Now, the finding of rotating black hole configurations  (at least under analytical form) is a major problem in every modified theory of gravity, which in the present case can be shortcut via the application of a {\it mapping method} allowing to find infinite classes of rotating black holes with reasonable ease \cite{Guerrero:2020azx}: the rotating counterpart of the spherically symmetric configurations discussed in this paper was actually found in the work \cite{Afonso:2021pga}. Whether the integrability of the geodesic equation in such a case still holds is a major question of interest to carry out the analysis of their images, something we shall tackle in the future. Furthermore, from a practical point of view, even if this is possible, computational times may become problematic.

\item Our current images are displayed at face-on orientation, while the EHT observations of M87 display an inclination of the rotation axis with the observer's line of sight of about 17 degrees \cite{EventHorizonTelescope:2020qrl}. Nonetheless one needs to go to larger inclinations in order to find strong deviations in the shape of photon rings as compared to the face-on case, basically via a bending of this shape to cross a section of the plane of the image.  The implementation of such a feature in the current configurations is problematic due to computational times. As mentioned, the transition from GR configurations to EiBI ones may increase this time by a factor $\sim 20-50$ due to the presence of special functions in the definition of the background geometry metric components (something that may worsen as we move towards other metric-affine configurations found in the literature), upon which the incorporation of inclination would further enlarge such times.

\end{itemize}

Individually, every such upgrade requires hard work beyond what we have at the moment. We can roughly split them into two kinds: modifications to the background geometry and to the ray-tracing code. For the former, the community has accumulated a good deal of knowledge on metric-affine geometries, particularly on the spherically symmetric case, and we also recently hinted on how to generalize these geometries to the axially symmetric case \cite{Guerrero:2020azx} via the development of new powerful methods \cite{Afonso:2018bpv}.  For the sake of this work we chose perhaps the most workable model of metric-affine gravity found so far, but there are others available in the literature such as $f(R)$ \cite{Bejarano:2017fgz} or quadratic gravity \cite{Olmo:2012nx}, coupled to both non-linear electromagnetic fields and to different classes of (anisotropic) fluids, many of which are also successful at resolving space-time singularities. For the latter, its implementation has varying degrees of difficulty: the extension of the geometry of the disk to be completely spherical and the implementation of the inclination of the disk are reasonable enough, while more refined intensity profiles are certainly possible: thick (but not completely spherical) disks and light trajectories in axially symmetric background geometry is a daunting challenge.

To conclude, space-time singularities (despite being shielded from the asymptotic observers by the present of an event horizon in black hole space-times) are typically seen as an abhorrent feature of the classical description provided by GR. If such singularities are to solved by an upgraded description of the gravitational field at certain scales, metric-affine gravities could potentially live from that job. This work can thus be seen as the first one towards the understanding of what modifications can be expected in shadow and photon ring images of singularity-free geometries of metric-affine type, which can be potentially searched for using very long-baseline interferometry \cite{Carballo-Rubio:2022aed}.  In implementing this programme we have already faced the widespread and well known problem that the role played by the background geometry and the geometrical, optical, and emission properties of the accretion disk are highly entangled in determining the properties of such images. Most geometrical modifications to GR predictions tend to be too mild to be easily recognizable via characterization of their images (as in the black hole cases considered here), while those introducing qualitatively new features (such as the RN0 horizonless configurations) may be in outright contradiction with observed images. The ``contamination" enacted by the poorly understood physics of the disk could be potentially disentangled via potential observational discriminators in the shape, size and luminosity of the photon rings (mostly the $m=2$ one) as well as in the properties of the shadow region between GR and any of its competing approaches. This question has been thoroughly explored in the literature \cite{Gralla:2020pra,Volkel:2020xlc,Kocherlakota:2022jnz,Ayzenberg:2022twz}, and it is the most relevant long-term goal of our collaboration on this subject to find what metric-affine gravities have to say about it.

\section*{Acknowledgements}

JLR acknowledges the European Regional Development Fund and the programme Mobilitas Pluss for financial support through Project No.~MOBJD647, and project No.~2021/43/P/ST2/02141 co-funded by the Polish National Science Centre and the European Union Framework Programme for Research and Innovation Horizon 2020 under the Marie Sklodowska-Curie grant agreement No. 94533. DRG is funded by the {\it Atracci\'on de Talento Investigador} programme of the Comunidad de Madrid (Spain) No. 2018-T1/TIC-10431. This work is supported by the Spanish National Grants FIS2017-84440-C2-1-P, PID2019-108485GB-I00, PID2020-116567GB-C21 and PID2020-117301GA-I00 funded by MCIN/AEI/10.13039/501100011033 (``ERDF A way of making Europe" and ``PGC Generaci\'on de Conocimiento"), and the project PROMETEO/2020/079 (Generalitat Valenciana). This article is based upon work from COST Actions CA18108 and CA21136, supported by COST (European Cooperation in Science and Technology).

\end{document}